\documentclass[12pt]{article}
\usepackage{cancel}

\usepackage[utf8]{inputenc} 
\usepackage[T1]{fontenc} 
\usepackage{amsmath, amsfonts, amssymb, amsthm, mathtools} 
\usepackage{geometry} 
\usepackage{multirow} 
\usepackage{xr} 
\usepackage{float}
\usepackage{graphicx} 
\graphicspath{{images/}}
\usepackage{tikz} 
\usetikzlibrary{angles, quotes}
\usepackage[normalem]{ulem}
\usepackage[ruled,vlined, linesnumbered]{algorithm2e} 
\usepackage{algorithmic} 

\usepackage[authoryear]{natbib} 

\numberwithin{equation}{section}

\usepackage{latexsym} 
\usepackage{enumerate} 
\usepackage{enumitem}
\usepackage{setspace} 
\usepackage[toc,title,titletoc,header]{appendix} 
\usepackage[bottom]{footmisc} 
\usepackage{cases}
\usepackage[pdfborder={0 0 0}]{hyperref} 
\hypersetup{
    colorlinks=true,
    linkcolor=magenta,
    citecolor=blue,
    filecolor=magenta,
}

\usepackage{libertine} 
\usepackage[libertine]{newtxmath} 

\usepackage{comment}

\newtheorem{theorem}{Theorem}
\newtheorem{corollary}{Corollary}
\newtheorem{lemma}{Lemma}
\newtheorem{proposition}{Proposition}
\newtheorem{assumption}{Assumption}
\newtheorem{definition}{Definition}
\newtheorem{example}{Example}
\newtheorem{remark}{Remark}

\newcommand{\prob}{\mathbb{P}}

\newcommand{\expect}{\mathbb{E}}
\newcommand{\ind}{\mathbb{I}}

\renewcommand{\vec}[1]{\mathbf{#1}}

\DeclareMathAlphabet{\mathbmit}{OML}{cmm}{b}{it}
\renewcommand{\vec}[1]{\mathbmit{#1}}

\makeatletter
\@tfor\next:=abcdefghijklmnopqrstuvwxyz\do
 {\begingroup\edef\x{\endgroup
    \noexpand\@namedef{v\next}{\noexpand\vec{\next}}%
  }\x}

\begin{document}
\relax
\hypersetup{pageanchor=false}
\hypersetup{pageanchor=true}

\title{\vspace{0cm}\textbf{Partially Identified Rankings from Pairwise Interactions}\thanks{We are grateful to Ivan Canay for his guidance in this project. We are also thankful to Eric Auerbach, Federico Bugni, Joel Horowitz, Charles Manski, and the participants of the Econometrics Reading Group at Northwestern University for their comments and suggestions. We thank Kurt Lavetti for having shared his code. All errors are our own.}}

\author{
\begin{minipage}{.45\textwidth}
    \centering
    Federico Crippa\\
    Department of Economics\\
    University of California, Berkeley \\
    \footnotesize \url{federico.crippa@berkeley.edu}
\end{minipage}
\hfill
\begin{minipage}{.45\textwidth}
    \centering
    Danil Fedchenko\\
    Department of Economics\\
    The University of Melbourne\\
    \footnotesize \url{danil.fedchenko@unimelb.edu.au}
\end{minipage}
}

\date{\vspace{1cm} September 10, 2026}
\maketitle

\vspace{-1cm}
\thispagestyle{empty}

\begin{spacing}{1.1}
    \begin{abstract}
        This paper considers the problem of ranking objects based on their latent merits using data from pairwise interactions. We allow for incomplete observation of these interactions and study what can be inferred about rankings in such settings. First, we show that identification of the ranking depends on a tradeoff between the tournament graph and the link function: in parametric models, such as the Bradley-Terry-Luce, rankings are point identified even with sparse graphs, whereas nonparametric link models require dense graphs. Second, moving beyond point identification, we characterize the identified set in the nonparametric link model under any tournament structure and represent it through moment inequalities. Finally, we propose a likelihood-based statistic to test whether a ranking belongs to the identified set. We study two testing procedures: one is finite sample valid but computationally intensive; the other is easy to implement and pointwise asymptotically valid. We illustrate our results using Brazilian employer-employee data and reject the hypothesis that the PageRank firm ordering constructed from job-to-job transitions belongs to the identified set.
    \end{abstract}
\end{spacing}



\thispagestyle{empty}

\newpage
\hypersetup{pageanchor=true}
\setcounter{page}{1}

\section{Introduction}

Consider the problem of ranking $q$ objects by their latent merits, when only pairwise interaction data are observed. Sports competitions provide a natural example: teams have unobservable strengths that influence the observable outcomes of games. These outcomes contain information about the teams' relative strengths and are typically used to rank teams by merit. In this paper, we ask whether the true ranking can be identified from binary pairwise comparison data under any given tournament structure.

This problem extends beyond sports competitions, and applications using pairwise interaction data to rank interacting objects can be found in the social sciences, with several recent examples also in economics. Job-to-job worker transitions are used to rank firms based on employee willingness to work for them \citep{sorkin2018ranking,corradini2023collective,lagos2024union}. Cross-journal citations serve as a method to rank academic journals according to their influence within the field \citep{stigler1994citation, palacios2004measurement, gu2022ranking}. In marketing, conjoint analysis leverages pairwise comparisons from surveys to construct consumer preference rankings and investigate the factors driving these preferences \citep{baier1998optimal,vcubric2019assessment}.

We show that whether the true ranking can be identified depends on two factors: the link function and the structure of the tournament graph. The link function defines the relationship between latent merits and pairwise interaction outcomes, while the tournament graph is an undirected graph in which vertices represent teams and edges connect teams that interact. Together, assumptions on the link function and the structure of the tournament graph determine what can be inferred about the true ranking. We also develop a procedure to test whether a particular ranking is compatible with the observed pairwise data. Because the true probabilities of the observed outcomes are unknown, the procedure accounts for the possibility that differences in observed ranking positions arise from sampling uncertainty rather than actual differences in latent merits.

Our first contribution is to show that, when a parametric form for the link function is assumed, as in the popular Bradley--Terry--Luce (BTL) model \citep{bradley1952rank,luce1959individual}, the ranking is point identified if and only if the tournament graph is connected. In a connected tournament graph, knowledge of the DGP of the observed interactions makes it possible to rank any pair of teams, even when they correspond to distant vertices. This result is driven by the assumption that the probability of each outcome is a known function of the difference between latent merits. Although this assumption provides strong identifying power, we show that informative conclusions about the ranking remain possible when the parametric restriction is relaxed. This gives researchers greater flexibility when there is no clear rationale for imposing a specific functional form.

Our second contribution is to relax the parametric restriction on the link function and consider a nonparametric link model that ensures the existence of a ranking. Under this weaker restriction, connectedness alone generally yields only partial identification of the ranking. We provide a sharp characterization of its identified set. Point identification requires a denser tournament graph: each team must either compete directly against every other team or share a common opponent with it. Our characterization shows that pairwise interactions can still provide informative conclusions under more credible assumptions, even when point identification is not possible. The nonparametric link model relies on the assumption of strong stochastic transitivity and coincides with the model considered by \citet{shah2016stochastically} and \citet{chatterjee2019estimation}.

Our third contribution is to propose two statistical tests of whether a particular ranking belongs to the identified set under the nonparametric link model. The first test is valid in finite samples for any tournament graph and any number of games, although its implementation can be computationally demanding in large tournaments. We therefore also propose a second test that is pointwise asymptotically valid and consistent against alternative rankings outside the closure of the identifying restrictions. The latter is particularly well suited to settings in which the number of games is large.

Both tests are based on a likelihood ratio statistic that compares restricted and unrestricted estimators of the outcome probabilities for each pairwise interaction. The unrestricted estimator is the vector of sample average outcomes, whereas the restricted estimator imposes the closed version of the restrictions implied by our sharp characterization of the identified set. Intuitively, a large difference between the two estimators indicates that the tested ranking imposes restrictions that are not supported by the data, leading to rejection of the null hypothesis that the ranking belongs to the identified set. The two procedures differ in how they compute the p-value associated with the test statistic. The first computes the p-value for the least favorable distribution, which requires solving an optimization problem specific to each application. The second test relies on a pivotal distribution that is least favorable asymptotically, as the number of repetitions becomes large.

As an illustration, we apply our procedure to assess whether the firm ranking constructed from job-to-job transitions is supported by the data. We use the \textit{Rela\c{c}\~{a}o Anual de Informa\c{c}\~{o}es Sociais} (RAIS), an administrative census of formal-sector employment in Brazil, and follow the empirical literature in constructing the PageRank firm ranking from employer-to-employer transitions \citep{sorkin2018ranking,corradini2023collective,lagos2024union}. We then test whether this PageRank ordering belongs to the identified set under the nonparametric link model. The test rejects the ranking, and this conclusion is robust across a range of alternative specifications. Descriptive evidence of Condorcet cycles in the transition data further suggests that the incompatibility is not specific to the ranking procedure. The illustration demonstrates how our method can be used to assess whether rankings constructed from pairwise interactions are supported by the underlying data.

\subsection{Related Literature}

Models considering rankings from pairwise interaction data have a long history in statistics and social sciences, dating back to seminal works by \cite{thurstone1927three}, \cite{bradley1952rank}, and \cite{luce1959individual}\footnote{Comprehensive reviews on the applications, estimation, computation, and extensions of these models can be found in \cite{fligner1993probability}, \cite{marden1996analyzing}, and \cite{cattelan2012models}.}. Both the BTL and Thurstone models are parametric, assuming a specific functional form for the link function that connects the latent merits to the probabilities of the outcomes. More recently, \cite{shah2016stochastically}, \cite{shah2018simple}, \cite{chatterjee2019estimation}, and \cite{rastogi2022two} studied paired comparison models that relax these parametric restrictions. These papers are closely related to the nonparametric link model considered here: in particular, they consider stochastic transitivity restrictions without imposing a BTL or Thurstone functional form.

Our analysis differs from this literature in two main respects. First, we allow for deterministic incompleteness in the tournament graph. In \cite{shah2016stochastically}, \cite{shah2018simple}, and \cite{chatterjee2019estimation}, missing pairwise comparisons are typically modeled as part of the sampling design: each comparison is observed with a positive probability, so the pattern of observed comparisons is random and every pair can in principle be observed. In contrast, we take the tournament graph $G$ as fixed and allow arbitrary patterns of missing comparisons, including pairs that are never observed. This distinction is important in many applications, where the absence of some pairwise interactions is structural rather than random. We show that such deterministic incompleteness can lead to a loss of point identification. This result is related to the conclusions of \cite{pananjady2020worst}, who also highlight the impossibility of consistent estimation under certain patterns of missing interactions. However, to the best of our knowledge, explicitly framing the issue as a loss of point identification is new to the literature.

Second, our object of interest is the identified set for the ranking, together with valid inference on this set. By contrast, \cite{shah2016stochastically} and \cite{chatterjee2019estimation} focus primarily on estimation of the matrix of pairwise outcome probabilities. In this respect, our paper contributes to the recent econometric literature studying rankings as objects of identification and inference \citep{bazylik2021finite, mogstad2024inference, gu2023invidious}, as well as to applied work in which rankings are central empirical objects \citep{sorkin2018ranking, corradini2023collective, lagos2024union}.

In contrast to \cite{bazylik2021finite}, \cite{mogstad2024inference}, and \cite{gu2023invidious}, who study cases where objects are ranked based on potentially noisy measurements of their merits, we study the problem of ranking based on entirely unobservable, latent merits. We show how a researcher can utilize information from pairwise interactions to conduct inference on the ranking according to these latent merits.

Recent applied contributions \citep{sorkin2018ranking, corradini2023collective, lagos2024union} rely on this approach to construct rankings of firms using job-to-job transition data (see \cite{mas2025non} for a recent review). In their estimation of the rankings, these authors employ the PageRank algorithm \citep{page1999pagerank}, which is closely related to the parametric BTL model \citep{negahban2017rank, selby2024pagerank}. In the empirical illustration, we test and reject the hypothesis that the ranking thus constructed belongs to the identified set under the nonparametric link model.

The inference problem we consider has similarities with the literature on empirical testing of stochastic choice and random utility models. Each interaction between two objects can be viewed as an outcome of a stochastic choice made by an individual. While we impose similar stochastic regularities on the probabilities as those found in the stochastic choice literature \citep{oliveira2018new, blavatskyy2018fechner, Strzalecki_2024}, our primary focus is different. Specifically, the literature on empirical testing of stochastic choice models typically aims to test whether the model's assumptions align with observed choices. For instance, \cite{kitamura2018nonparametric} demonstrate how to test the rationality of agents under weak assumptions about their preferences, and \cite{regenwetter2010testing} discuss testing for strong stochastic transitivity of preferences. In contrast, our work focuses on the identification of and inference on rankings. However, the characterization of the identified set can be used to test whether this set is empty, which would imply that no ranking can rationalize the pairwise probabilities under strong stochastic transitivity. We do not develop such a test here, though we return to the question when discussing the empirical illustration in Section \ref{sec:empirical_illustration}.

On the technical side, our work relates to the literature on constrained statistical inference (see \cite{robertson1988order} for an excellent textbook treatment and references). Specifically, we use the constrained log-likelihood as a test statistic. By borrowing from and extending results in this literature, particularly from \cite{robertson1988order} and \cite{hu1997maximum}, we derive the asymptotic distribution of the test statistic. This allows us to test whether a particular ranking belongs to the identified set.

It is finally important to note that, although the sharp characterization of the identified set for the ranking that we derive takes the form of moment inequalities, the structure of our data generally differs from what is typically assumed in the literature on testing moment inequalities (see \citealp{canay2017practical}). As a result, the tools developed in that literature do not directly apply to our context. We discuss this issue further in Appendix \ref{appendix:moment_inequ}.

The rest of the paper is structured as follows. Section \ref{sec:model} introduces the model. Section \ref{sec:identification} derives conditions for point identification in the linear parametric model and characterizes the identified set for the nonparametric link model. In Section \ref{sec:inference}, we propose the statistical test for a ranking belonging to the identified set. Section \ref{sec:mc_simulations} studies the finite sample performance of the test through Monte Carlo simulations. Section \ref{sec:empirical_illustration} illustrates the procedure by testing a PageRank firm ordering constructed from Brazilian job-to-job transitions. Section \ref{sec:conclusion} concludes. Proofs and some additional results are relegated to the Appendix.

\textbf{Notation} In the rest of the paper, we use bold characters to indicate vectors ($\boldsymbol{\theta}$, $\vr$), and denote their elements by $\theta_\ell$, $r_\ell$; we denote by $[q]$ the set of all integers from 1 to $q$, i.e. $[q]:= \{1, 2, \dots, q\}$; in what follows we use the sign function, that is defined as: $\text{sign}(x):= I\{x \ge 0 \} - I\{x \le 0\}.$

\section{Model}\label{sec:model}

We introduce the model using the sports competition setting where \textit{teams} (the objects to be ranked) interact by playing \textit{games}. This concrete setting aids in the exposition, and we provide examples at the end of the section to illustrate how applications in economics and marketing fit within this framework.

Let $q \in \mathbb{N}$ represent the number of teams. Each team $\ell$ is characterized by a latent merit $\theta_\ell$, where a smaller $\theta_\ell$ indicates a higher merit. There may be ties: teams $\ell$ and $k$ for which $\theta_\ell = \theta_k$. Merits are collected in a vector $\boldsymbol{\theta}_0 \in \mathbb{R}^q$. The parameter of interest is a weak ordering (a strongly complete and transitive binary relation \citep{roberts1985measurement}) of $\boldsymbol{\theta}_0$, which we refer to as the ranking, formally defined in the next paragraph. First, we need to introduce the definition of ranks.

\begin{definition}
    {\normalfont (Ranks)}
    For a vector of latent merits $\boldsymbol{\theta} \in \mathbb{R}^q$,
    rank of team $\ell \in [q]$ defined as:
    \begin{align*}
        r_{\ell}(\boldsymbol{\theta}):= 1 + \sum_{i \in [q]} I \left\{ \theta_{\ell} > \theta_i \right\}.
    \end{align*}
\end{definition}
In words, the rank of team $\ell$ is 1 plus the number of teams that are better than team $\ell$\footnote{Alternatively, the rank for team $\ell$ could be defined as the total number of teams minus the number of teams that are worse than team $\ell$. The two definitions coincide when no teams have equal merits; otherwise, the rank according to our definition is no larger than the rank according to the alternative definition. As a result, the literature typically refers to our definition as the ``lower rank'' and the alternative as the ``upper rank''. In what follows, we focus on identifying the lower rank, which we will simply call \textit{rank}. The exposition would be similar for the upper rank.}. We denote the collection of ranks $(r_{\ell}(\boldsymbol{\theta}))_{\ell \in [q]}$ by $\vr(\boldsymbol{\theta}):= (r_{\ell}(\boldsymbol{\theta}))_{\ell \in [q]}$ and refer to it as the \textit{ranking}. The ranking is a $q$-dimensional vector of positive integers. It is useful to define the set of all logically possible rankings of $q$ teams (excluding, for example, rankings as $(1,2,4)'$):
\begin{gather*}
    \mathsf{R}:= \{\boldsymbol{r}:=(r_1, \dots, r_q)' \in \mathbb{N}^q:\ \forall\ \ell \in [q],\ r_{\ell} = 1 + \sum_{k \in [q]} I\{ r_k < r_{\ell} \} \}
\end{gather*}

We say that two teams interact if they played against each other. Any pair of teams $(\ell, k) \in [q] \times [q]$ with $\ell \neq k$ may interact. The tournament graph represents the collection of these interactions.

\begin{definition}
    {\normalfont (Tournament Graph)}
    Let $E$ be the collection of unordered pairs $(\ell, k)$, $\ell, k \in [q]$, $\ell \neq k$ such that the interaction between teams $\ell$ and $k$ occurs. Call a graph $G = ([q], E)$ the tournament graph.
\end{definition}

Whenever a game between $\ell$ and $k$ occurs, the outcome $Y_{\ell, k}$ is realized, representing a win ($Y_{\ell, k} = 1$) or a loss ($Y_{\ell, k} = 0$) for team $\ell$ against team $k$. Let $\prob_{F, \theta}$ be the distribution of $Y_{\ell, k}$, which depends on the link function $F$ and merits $\boldsymbol{\theta} \in \mathbb{R}^q$ as follows:
\begin{align}\label{eq:measure}
    \prob_{F, \boldsymbol{\theta}}(Y_{\ell, k} = 1) = F(\theta_\ell, \theta_k).
\end{align}

We will assume that the link function $F$ belongs to a collection of functions $\mathcal{F}$, which satisfy certain restrictions to ensure $\prob_{F, \boldsymbol{\theta}}$ is well-defined according to \eqref{eq:measure}. The family $\mathcal{F}$ is known to the econometrician and will determine several identification results based on the assumptions made about it.

For each interaction, there can be multiple games (repetitions), with independent and possibly different outcomes. Denote by $\vn \coloneqq \{ n_{\ell,k} \}_{(\ell,k) \in E}$ the collection of the number of games for each interaction $(\ell,k) \in E$. The tournament graph $G$ and the vector $\vn$ are treated as fixed features of the data generating process. They are not subject to restrictions and may be related to the teams' latent merits; for example, teams with similar merits may play more often. However, they are determined before the realizations of the outcomes considered in the sample and do not depend on the realized values of $\{Y_{\ell,k,i}\}$. For each $(\ell, k) \in E$, a researcher observes $n_{\ell, k}$ i.i.d. realizations of $Y_{\ell, k}$. Realizations of $Y_{\ell,k}$ and $Y_{i,j}$, for $(\ell,k) \neq (i,j)$, are independent and need not be identically distributed. The collected data are hence $\boldsymbol{Y}_{\boldsymbol{n}}:= (\{Y_{\ell, k, i}\}_{i=1}^{n_{\ell, k}} )_{(\ell, k) \in E}$. The following assumption summarizes the data generating process.

\begin{assumption} \label{as:dgp}
    {\normalfont (DGP)} There exist $F_0 \in \mathcal{F}$ and $\boldsymbol{\theta}_0 \in \mathbb{R}^q$ such that
    \begin{gather*}
        \boldsymbol{Y}_{\vn} = \left( \{Y_{\ell,k,i}\}_{i=1}^{n_{\ell,k}} \right)_{(\ell,k)\in E}
    \end{gather*}
    is generated according to \eqref{eq:measure}. The random variables $Y_{\ell,k,i}$ are mutually independent across all $(\ell,k,i)$, and, for each fixed $(\ell,k)\in E$, $Y_{\ell,k,1},\ldots,Y_{\ell,k,n_{\ell,k}}$ are identically distributed.
\end{assumption}

Intuitively, the model works as follows. First, nature chooses the number of teams $q$, the vector of their latent merits $\boldsymbol{\theta}_0$, a link function $F_0 \in \mathcal{F}$, the tournament graph $G$, and the number of games per edge $\boldsymbol{n}$. These objects are treated as fixed and may be linked arbitrarily. For example, teams with similar merits may play more often, and some pairs may never play. Next, conditional on them, the data are generated according to \eqref{eq:measure}. The framework therefore allows arbitrary deterministic incompleteness in $G$, but it does not allow $G$ or $\boldsymbol{n}$ to be selected as a function of outcomes in the sample. Sequential elimination tournaments, such as playoff brackets in which later matchups depend on previous wins, are therefore not covered by the baseline assumption without an additional model for the selection of observed comparisons.

The researcher knows the family $\mathcal{F}$ and the relation in \eqref{eq:measure}, and observes the graph $G$, the number of games $\boldsymbol{n}$, and, for each interacting pair $(\ell, k) \in E$, the outcomes $\{Y_{\ell, k, i}\}_{i=1}^{n_{\ell, k}}$. These are independent Bernoulli draws with unknown success probabilities $F_0(\theta_\ell, \theta_k)$. The goal is to recover the true ranking $\vr(\boldsymbol{\theta}_0)$.

Our aim is to study the minimal assumptions under which the ranking is recoverable and to isolate the identifying power of each model component. This is useful when common theoretical restrictions are hard to defend, including parametric assumptions on the link function and conditions that take the tournament graph as complete or incomplete at random (for example, \cite{shah2016stochastically}, \cite{oliveira2018new}, and \cite{chatterjee2019estimation} assume that each interaction is observed independently with some probability). Even with this minimal structure, meaningful inference about the ranking is possible. The analysis reveals a tradeoff between the richness of $\mathcal{F}$ and the structure of $G$: when the graph has few edges, parametric assumptions on the link function are needed; when the graph is denser, strong stochastic transitivity alone can recover the ranking.

The model is simple, but it provides a useful foundation that can be extended in several directions. The assumption of i.i.d. outcomes within pairs, for instance, rules out features such as home-field advantage. Describing team merits with a single dimension cannot reflect temporal or strategic variation, and focusing only on binary outcomes may seem too restrictive in settings where richer information is available. For example, moving beyond binary outcomes would allow the framework to incorporate draws or ordered measures of performance, such as goal differences in sports applications. Even so, the model is rich enough to capture the tradeoff we want to emphasize and flexible enough to accommodate further structure. Extending it to include additional features remains an interesting direction for future research.

The following examples illustrate how the framework can be applied to model rankings from pairwise interactions in contexts beyond team tournaments, such as in labor economics and marketing.

\begin{example}
    {\normalfont (Ranking Firms using Revealed Preference)}
    \cite{sorkin2018ranking} consider pairwise interactions of firms to construct a ranking based on workers' preferences. Each firm is characterized by a latent value that represents the willingness of workers to work for that firm. Although the econometrician cannot directly measure this value, they can observe workers transitioning from one firm to another. The probability of the transition from firm $\ell$ to firm $k$ is modeled as a function of the latent values $\theta_\ell$ and $\theta_k$. Interactions between firms $\ell$ and $k$ occur when one worker moves between the two firms and can be repeated $n_{\ell,k}$ times, where $n_{\ell,k}$ is the number of employer-to-employer transitions from firm $\ell$ to firm $k$, or from firm $k$ to firm $\ell$. In this case, the probability of choosing firm $\ell$ over firm $k$ can be interpreted as the probability of each worker choosing firm $\ell$ over $k$ or as the fraction of workers who prefer firm $\ell$ to firm $k$. Workers' revealed preferences are thus used to derive the firms' ranking. This approach has also been utilized in \cite{corradini2023collective} and \cite{lagos2024union}.
\end{example}

\begin{example}
    {\normalfont (Ranking Journals by Influence)} \cite{stigler1994citation} proposes a model of journal influence based on pairwise citation counts. Citations from journal $\ell$ to papers in journal $k$ are interpreted as a sign of journal $k$'s influence on journal $\ell$. While the econometrician cannot directly observe these influences, they can observe the pattern of cross-journal citations. The probability that journal $\ell$ cites or is cited by journal $k$ is modeled as a function of their latent influences, $\theta_\ell$ and $\theta_k$. These cross-citations are then used to rank the influence of the journals.
\end{example}

\begin{example}
    {\normalfont (Conjoint Analysis)}
    Conjoint analysis is a marketing research technique used to understand how customers value different products. It involves asking consumers to rank products based on their perceived quality in an experimental setting. In full-profile conjoint analysis, products are presented as fixed combinations of attributes, and the responses construct a ranking based on latent qualities \citep{rao2010conjoint}. A common approach is to use pairwise comparisons instead of ranking multiple products simultaneously. This method simplifies the decision-making process, reducing cognitive load and decision fatigue, as respondents only choose between two options at a time. The probability of choosing product $\ell$ over product $k$ is modeled as a function of latent qualities $\theta_\ell$ and $\theta_k$. Interactions occur if at least one respondent is presented with the choice set including products $\ell$ and $k$, with $n_{\ell,k}$ being the number of times the comparison is presented to respondents.
\end{example}

\section{Identification} \label{sec:identification}

\subsection{Definition of Identification}

We begin by formalizing the notion of identification used in this paper. Intuitively, the ranking is identified if the distribution of the observed data allows the researcher to infer the true ranking without ambiguity, given the known parameters $q$, $G$, $\mathcal{F}$, and $\boldsymbol{n}$. We define identification with respect to the tournament graph $G$ and the function family $\mathcal{F}$, keeping the dependence on $q$ and $\boldsymbol{n}$ implicit.

Let $\mathbf{P}(G, \mathcal{F})$ be the collection of permissible distributions that can be generated by the model described in Section \ref{sec:model}.
\begin{definition}
    {\normalfont (Permissible Distributions)}
    For a tournament graph $G$, and a family of functions $\mathcal{F}$, let the set of permissible distributions be defined as:
    \begin{align*}
        \mathbf{P}(G, \mathcal{F}):= \{(\prob_{\boldsymbol{\theta}, F}(Y_{\ell, k}))_{(\ell, k) \in E}:\ \boldsymbol{\theta} \in \mathbb{R}^q,\ F \in \mathcal{F} \}.
    \end{align*}
\end{definition}
When Assumption \ref{as:dgp} is satisfied, the distribution of observed data $P$ is a member of $\mathbf{P}(G, \mathcal{F})$. Different latent merits combined with different $F \in \mathcal{F}$ may give rise to the same distribution of observed data. We denote by $\Theta(G, \mathcal{F}, P)$ the set of observationally equivalent merits.

\begin{definition}
    {\normalfont (Observationally Equivalent Merits)}
    For a tournament graph $G$, a family of functions $\mathcal{F}$, and a distribution of observed data $P$, let the set of observationally equivalent merits be defined as:
    \begin{align*}
        \Theta(G, \mathcal{F}, P):= \{\boldsymbol{\theta} \in \mathbb{R}^q:\ \exists\ F \in \mathcal{F}:\ \prob_{\boldsymbol{\theta}, F}(Y_{\ell, k}) = P(Y_{\ell, k}), \forall\ (\ell, k) \in E \}.
    \end{align*}
\end{definition}

Likewise, different rankings may give rise to the same distribution of observed data. We denote by $R(G, \mathcal{F}, P)$ the set of observationally equivalent rankings, a subset of the set $\mathsf{R}$ of logically possible rankings.
\begin{definition}
    {\normalfont (Observationally Equivalent Rankings)}
    For a tournament graph $G$, a family of functions $\mathcal{F}$, and a distribution of observed data $P$ let the set of observationally equivalent rankings be defined as:
    \begin{align*}
        R(G, \mathcal{F}, P):= \{(r_{\ell}(\boldsymbol{\theta}))_{\ell \in [q]}:\ \boldsymbol{\theta} \in \Theta(G, \mathcal{F}, P) \}.
    \end{align*}
\end{definition}

Following the literature on identification, we say that a ranking is point identified if any permissible distribution of the observed data corresponds to a unique ranking.
\begin{definition}
    {\normalfont (Point identification of the Ranking)}
    For tournament graph $G$, and family of functions $\mathcal{F}$, the ranking is \textbf{point identified} if for all $P \in \mathbf{P}(G, \mathcal{F})$, $R(G, \mathcal{F}, P)$ is a singleton.
\end{definition}

When multiple rankings are consistent with the observed data, we say that the ranking is partially identified.

\begin{definition}
    {\normalfont (Partial identification)}
    For a tournament graph $G$ and a family of functions $\mathcal{F}$, the ranking is \textbf{partially identified} if it is not point identified. Further, $R(G, \mathcal{F}, P) $ is called \textbf{the identified set} for the ranking.
\end{definition}

\subsection{Point Identification in the Linear Parametric Model}

In this section, we assume the link function is known and linear. Hence $\mathcal{F} = \{F_0\}$ is a singleton with $F_0 \in \mathbf{F}_L$, where $\mathbf{F}_L$ is defined as follows.
\begin{definition}\label{def:linear}
    {\normalfont (Family $\mathbf{F}_L$)}
    Let $\mathbf{F}_L$ be a family of functions $F: \mathbb{R} \times \mathbb{R} \mapsto (0, 1)$ such that $\exists\ f:\mathbb{R} \mapsto (0, 1)$ that satisfies:
    \begin{enumerate}
        \item $F(x, y) \equiv f(y - x),\ \forall\ (x, y) \in \mathbb{R}^2$ \label{itm1_lin}
        \item $f(x) = 1 - f(-x),\ \forall\ x\ \in \mathbb{R}$\label{itm2_lin}
        \item $f$ is strictly increasing and continuous. \label{itm3_lin}
    \end{enumerate}
\end{definition}
The family $\mathbf{F}_L$ requires that the win probability depends only on the difference in latent merits, which implies a cardinal structure among them\footnote{In the stochastic choice literature, the family $\mathbf{F}_L$ is referred to as Fechnerian \citep{blavatskyy2018fechner}.}. The BTL model, widely used in applications \citep{stigler1994citation,sorkin2018ranking,corradini2023collective,lagos2024union}, is an example of a linear parametric model, with link function $f_{BTL}(x) \equiv \frac{\exp(x)}{1+\exp(x)}$, where $x$ is the difference in merits.

In the linear parametric model, point identification of the ranking is equivalent to having a connected tournament graph, as formally stated in the following theorem.

\begin{theorem}\label{th:ident_param}
    {\normalfont (Point Identification in the Linear Parametric Model)}
    Let Assumption \ref{as:dgp} hold with $F_0 \in \mathbf{F}_L$ known. For a tournament graph $G$ and $F_0$, the ranking is point identified if and only if $G$ is connected (that is, for any pair of teams there is a path in the graph connecting those two teams).
\end{theorem}

Theorem \ref{th:ident_param} establishes that the connectedness of the tournament graph is a necessary and sufficient condition for the point identification of rankings when the link function is linear and known. The necessity is straightforward: if in the graph there are disconnected components, it is impossible to compare the merits of teams across them.

To build intuition for the sufficiency result, note that the model assumes
\begin{gather*}
    f(\theta_{k} - \theta_\ell) = P(Y_{\ell, k} = 1),\ \forall\ (\ell, k) \in E,
\end{gather*}
and since $f$ is known and strictly monotone, this can be rewritten as
\begin{gather*}
    \theta_{k} - \theta_\ell = f^{-1}(P(Y_{\ell, k} = 1)),\ \forall\ (\ell, k) \in E.
\end{gather*}
This reformulation produces a system of linear equations in $\boldsymbol{\theta}$, where the data identifies the right-hand sides. If the tournament graph is connected, this system contains at least $q-1$ equations for $q$ unknowns: to fix the values, a normalization is needed. As the ordering is invariant to this normalization, solving the system provides a unique ranking of $\boldsymbol{\theta}$. Thus, connectedness is sufficient for point identification: if the link function is linear and known, even with a minimally connected tournament graph, a tree, one achieves point identification.

The sufficiency argument relies critically on the assumption that $f$ is both linear and known. Its identifying power comes from the fact that each observed win probability pins down the difference in merits between the two teams involved, so that these differences can be chained along any path in the graph. Once the link function is unknown, this chaining is no longer available: result \ref{itm3_ranking_general} of Theorem \ref{th:sharp_ranking_general} below shows that connectedness is then insufficient for point identification, and the counterexample in its proof is constructed with a link function that is itself linear. Assuming a known linear parametric form for $F_0$, as done in previous literature, therefore provides significant identifying power. In the next section, we relax both assumptions and study the nonparametric link model.

\subsection{Identification in the Nonparametric Link Model}

In this section, we study identification under a general family of link functions $\mathbf{F}$, introduced in Definition \ref{def:func} below. We refer to this specification as the nonparametric link model: the link function $F$ is left unspecified except for symmetry and monotonicity restrictions that encode strong stochastic transitivity. Related work uses the term nonparametric for pairwise comparison models based on strong stochastic transitivity; see, for example, \cite{chatterjee2015matrix}.

\begin{definition} \label{def:func}
    {\normalfont (Family $\mathbf{F}$)} Let $\mathbf{F}$ be a family of functions $F: \mathbb{R} \times \mathbb{R} \mapsto (0, 1)$ that satisfy the following conditions: \begin{enumerate}
        \item $F(x, y) = 1 - F(y, x),\ \forall\ (x, y) \in \mathbb{R}^2$. \label{itm1}
        \item $F$ is strictly decreasing in the first argument (keeping the second fixed), and strictly increasing in the second argument (keeping the first fixed). \label{itm2}
    \end{enumerate}
\end{definition}
Condition \ref{itm1} guarantees that $\prob_{F, \boldsymbol{\theta}}(Y_{\ell, k} = 1) = \prob_{F, \boldsymbol{\theta}}(Y_{k, \ell} = 0) = 1 - \prob_{F, \boldsymbol{\theta}}(Y_{k, \ell} = 1)$. Condition \ref{itm2} formalizes the idea that, all else being equal, the probability of winning against a stronger (weaker) opponent is strictly smaller (larger). Notice that Condition \ref{itm1} implies $F(\theta_{\ell}, \theta_{\ell}) = 1/2$, $\forall\ \theta_{\ell} \in \mathbb{R},$ so that whenever two teams with the same latent merits play, the outcome is a fair coin flip. The family $\mathbf{F}$ imposes an ordinal relation between the team merits, requiring that whenever $\theta_\ell < \theta_k$, team $\ell$ wins against team $k$ with a probability greater than 0.5\footnote{This concept appears in the literature on stochastic choice, giving rise to what is known as a simple scalability representation of random utility (Definition 3.26 in \cite{Strzalecki_2024}).}.

Recall that $\mathsf{R}$ is the set of all logically possible rankings for $q$ teams. The following theorem characterizes the identified set $R(G, \mathbf{F}, P)$ (result \ref{itm1_ranking_general}), and determines conditions under which the ranking is point identified (result \ref{itm3_ranking_general}).

\begin{theorem}\label{th:sharp_ranking_general}
    {\normalfont (Identification in the Nonparametric Link Model)}
    Let Assumption \ref{as:dgp} hold with $F_0 \in \mathbf{F}$, where $F_0$ is unknown, and consider some $\boldsymbol{r}:=(r_1, \dots, r_q)' \in \mathsf{R}$. Then
    \begin{enumerate}[label=(\alph*)]
        \item \label{itm1_ranking_general}
        $\boldsymbol{r} \in R(G, \mathbf{F}, P)$ if and only if the following three conditions hold:
        \begin{enumerate}[label=(\roman*)]
            \item
            For all $(\ell,k)\in E$,
            \begin{equation}
                \text{sign}\Big(\expect_P\big[Y_{\ell,k}-\tfrac12\big]\Big)
                =
                \text{sign}(r_k - r_\ell).
                \label{eq:dir_games}
            \end{equation}
            \item
            For all $(i,k),(j,\ell)\in E$ with $r_i = r_j$,
            \begin{equation}
                \text{sign}\Big(\expect_P[Y_{i,k}-Y_{j,\ell}]\Big)
                =
                \text{sign}(r_k - r_\ell).
                \label{eq:indir_games_tie}
            \end{equation}
            \item
            For all $(i,k),(j,\ell)\in E$ with $r_i > r_j$,
            \begin{equation}
                r_\ell \ge r_k
                \quad\Longrightarrow\quad
                \expect_P[Y_{i,k}] < \expect_P[Y_{j,\ell}].
                \label{eq:indir_games}
            \end{equation}
        \end{enumerate}
        \item The ranking is point identified with respect to $G$ and $\mathbf{F}$ if and only if there is a path of length at most two between any two teams in the tournament graph. \label{itm3_ranking_general}
    \end{enumerate}
\end{theorem}

Result~\ref{itm1_ranking_general} provides necessary and sufficient conditions for a ranking to belong to the identified set, so $R(G,\mathbf{F},P)$ is sharply characterized by \eqref{eq:dir_games}--\eqref{eq:indir_games}. We give an intuitive explanation for necessity; sufficiency is proven in the Appendix. Because every $F\in\mathbf{F}$ is strictly stochastically transitive, win probabilities must respect the order of latent merits, and hence of ranks. Condition~\eqref{eq:dir_games} is the direct implication: a weaker team beats a stronger one with probability below $1/2$. Condition~\eqref{eq:indir_games_tie} extends this when $r_i=r_j$, including the case $i=j$: when $i$ and $j$ face opponents $k$ and $\ell$, the side facing the stronger opponent must have the lower win probability. In particular, if $i=j$, a team's win probability is strictly decreasing in the strength of its opponent. Condition~\eqref{eq:indir_games} is the indirect restriction when $i$ is ranked strictly worse than $j$: whenever $k$ is at least as strong as $\ell$, the weaker team $i$ must be strictly less likely to beat $k$ than the stronger team $j$ is to beat $\ell$.

When there are no ties, these conditions are equivalent to a simple restriction on a probability matrix ordered by $\boldsymbol{r}$:

\begin{corollary} \label{cor:matrix_representation}
    {\normalfont (Matrix Representation of Identification Conditions)}
    Under the conditions of Theorem \ref{th:sharp_ranking_general}, fix
    $\boldsymbol{r}\in\mathsf{R}$ without ties, and form the $q\times q$
    matrix $A_{\boldsymbol{r}}$ whose rows and columns are ordered from best
    to worst under $\boldsymbol{r}$. Set every diagonal entry equal to
    $1/2$. For $i\neq j$, set
    $a_{i,j}=\expect_P[Y_{\ell,k}]$ if $(\ell,k)\in E$ and
    $(r_\ell,r_k)=(i,j)$, and leave $a_{i,j}$ empty otherwise. Then
    $\boldsymbol{r}\in R(G,\mathbf{F},P)$ if and only if, for all nonempty
    entries,
    \begin{align*}
        i\geq i',
        \quad
        j\leq j',
        \quad
        (i,j)\neq(i',j')
        \quad\Longrightarrow\quad
        a_{i,j}<a_{i',j'}.
    \end{align*}
\end{corollary}

When ties in the ranking occur, if teams $\ell$ and $k$ share the same rank, their corresponding entries in the probability matrix must be equal.
Therefore, the assumption that $\boldsymbol{r}$ has no ties in Corollary \ref{cor:matrix_representation} is made without loss of generality. If multiple teams share the same rank, the tournament graph $G$ can be modified by merging the teams with identical ranks into a single node and adjusting the set of edges accordingly.

The following example uses the representation described in Corollary \ref{cor:matrix_representation} to illustrate partial identification, and to compare results of Theorems \ref{th:ident_param} and \ref{th:sharp_ranking_general}.

\begin{example} \label{example:partial_identification}
    {\normalfont (Partial Identification for the Ranking)}
    Consider the tournament graph illustrated in Figure \ref{fig:tournament_partial_identification}. There are four teams ($A$, $B$, $C$, and $D$) and three interactions. The numbers on the edges represent the probabilities of the team on the left winning over the team on the right. For example, $\expect[Y_{A,B}] = 0.75$.
    \begin{figure}[H]
        \centering
        \begin{tikzpicture}
            \node[shape=circle,draw=black] (A) at (0,0) {A};
            \node[shape=circle,draw=black] (B) at (2,0) {B};
            \node[shape=circle,draw=black] (C) at (4,0) {C};
            \node[shape=circle,draw=black] (D) at (6,0) {D};
            \path [-] (A) edge node[anchor=south] {$0.75$} (B);
            \path [-] (B) edge node[anchor=south] {$0.7$} (C);
            \path [-] (C) edge node[anchor=south] {$0.2$} (D);
        \end{tikzpicture}
        \caption{Tournament graph for Example \ref{example:partial_identification}.}
        \label{fig:tournament_partial_identification}
    \end{figure}

    In this example, the identified set for the nonparametric link model contains exactly two rankings without ties, corresponding to the orderings $\theta_A < \theta_D < \theta_B < \theta_C$ and $\theta_D < \theta_A < \theta_B < \theta_C$. The associated probability matrices (rows/columns ordered from best to worst under each ranking) are:
    \begin{gather*}
        \underbrace{
        \begin{bmatrix}
            0.5 & & 0.75 & \\
            & 0.5 & & 0.8\\
            0.25 & & 0.5 & 0.7\\
            & 0.2 & 0.3 & 0.5
        \end{bmatrix}
        }_{\theta_A < \theta_D < \theta_B < \theta_C}
        \text{ and }
        \underbrace{
        \begin{bmatrix}
            0.5 & & & 0.8\\
            & 0.5 & 0.75 & \\
            & 0.25 & 0.5 & 0.7\\
            0.2 & & 0.3 & 0.5
        \end{bmatrix}
        }_{\theta_D < \theta_A < \theta_B < \theta_C}
    \end{gather*}
    In both matrices, each entry is greater than all entries southwest of it: Theorem \ref{th:sharp_ranking_general} then establishes that both rankings belong to the identified set. With the nonparametric link model, hence, it is impossible to determine the order relation between teams $A$ and $D$. In contrast, Theorem \ref{th:ident_param} demonstrates that, when we assume a known linear functional form for the link function, it becomes possible to retrieve this order.

    Consider, for example, the BTL model. It imposes the following system of linear equations:
    \begin{gather*}
        \begin{aligned}
            \begin{cases}
                f_{BTL}(\theta_B - \theta_A) = 0.75 \\
                f_{BTL}(\theta_C - \theta_B) = 0.7 \\
                f_{BTL}(\theta_D - \theta_C) = 0.2
            \end{cases}
            \quad \longrightarrow \quad
            \begin{cases}
                \theta_B - \theta_A = f_{BTL}^{-1}(0.75) = 1.099 \\
                \theta_C - \theta_B = f_{BTL}^{-1}(0.7) = 0.847 \\
                \theta_D - \theta_C = f_{BTL}^{-1}(0.2) = -1.386
            \end{cases}
        \end{aligned}
    \end{gather*}
    By normalizing $\theta_A = 1$, the system is solved by $\theta_B = 2.099$, $\theta_C = 2.946$, and $\theta_D = 1.559$, meaning that, with the BTL model, team A is stronger than team D, as $\theta_A < \theta_D$.

    Some applications involve sparse tournament graphs with a diameter greater than two, similar to the graph illustrated in Figure \ref{fig:tournament_partial_identification}. This example demonstrates that in such contexts the nonparametric link model remains informative, as the identified set contains two of the 24 rankings without ties, but it also acknowledges the impossibility of establishing an order for all team pairs when interactions are limited. In contrast, the linear parametric model can always establish a complete ranking in a connected graph, thanks to the identification power of the linear parametric assumption.
\end{example}

Result~\ref{itm3_ranking_general} of Theorem~\ref{th:sharp_ranking_general} establishes that, to guarantee point identification when the link function is unknown but known to belong to the family $\mathbf{F}$, the tournament graph must contain sufficiently many edges. More precisely, its diameter must be at most $2$\footnote{A sharp edge-count condition that guarantees this property uniformly over all graphs with $q$ vertices is
\begin{gather*}
    |E| \geq \binom{q-1}{2}+1 = \frac{(q-1)(q-2)}{2}+1.
\end{gather*}
Indeed, with only $\binom{q-1}{2}$ edges, the graph could consist of a complete graph on $q-1$ vertices and one isolated vertex, in which case the diameter condition fails.}. Compared to Theorem \ref{th:ident_param}, when the link function $F_0$ is assumed to be known and belongs to the family $\mathbf{F}_L$, the graph is required to have more edges to ensure the ranking is point identified. The key takeaway from Theorems \ref{th:ident_param} and \ref{th:sharp_ranking_general} is the tradeoff between the richness of the link function family and the tournament graph structure required to unambiguously recover the ranking. With few edges, parametric assumptions become necessary for identification, even when the data distribution is entirely known.

Theorem \ref{th:sharp_ranking_general} sharply characterizes the identified set $R(G, \mathbf{F}, P)$ for the ranking. $R(G,\mathbf{F},P)$ is generally a set of $q$-dimensional vectors of rankings consistent with the distribution of observed data $P$. This object can be difficult to visualize, especially when the number of teams $q$ is large. However, once $R(G,\mathbf{F},P)$ is obtained, the identified set for the individual rank of each team can be determined by projecting $R(G,\mathbf{F},P)$. This is formally stated in the following Corollary.
\begin{corollary}
    {\normalfont (Identified Set for the Rank of team $\ell$)}
    For any team $\ell \in [q]$, the sharp identified set for its rank can be constructed as
    \begin{gather*}
        R_{\ell}(G, \mathbf{F}, P):= \left\{ 1_{\ell}' \cdot \boldsymbol{r}:\ \boldsymbol{r} \in R(G, \mathbf{F}, P) \right\}
    \end{gather*}
    where $1_{\ell}$ is a $q$-dimensional vector with $1$ at $\ell$-th position, and 0 everywhere else.
\end{corollary}

\begin{remark}
    {\normalfont (Shape of the Identified Set)}
    $R_{\ell}(G, \mathbf{F}, P)$ represents the set of integer ranks for team $\ell$ that are consistent with the data generating process $P$. However, note that if $\underline{r}_{\ell} = \min\ R_{\ell}(G, \mathbf{F}, P)$ and $\overline{r}_{\ell} = \max\ R_{\ell}(G, \mathbf{F}, P)$, the set $R_{\ell}(G, \mathbf{F}, P)$ does not necessarily contain all integers between $\underline{r}_{\ell}$ and $\overline{r}_{\ell}$. This can occur, for instance, when the ranking order between teams A and B, or A and C, cannot be identified, but the ranking order between teams B and C can, and the data indicate that they have the same rank. In this case, team A could be ranked better than both B and C, causing its rank to improve by 2 positions, or worse than both, causing its rank to drop by 2 positions. However, its rank cannot change by only 1.
\end{remark}

\begin{remark}
    {\normalfont (Sharpness from Transitivity)}
    In general, the sharp identified set for the full ranking must be derived to obtain the sharp identified set for individual ranks. When the tournament graph is complete, the rank of team $\ell$ can be characterized using only the probabilities of comparisons involving $\ell$: in the absence of ties, $r_\ell=1+\sum_{i\neq \ell} I\{p_{i,\ell}>1/2\}$. In incomplete graphs, however, these probabilities provide only direct information about $r_\ell$. Other comparisons, including comparisons that do not involve $\ell$, may still restrict the possible values of $r_\ell$ through transitivity. This also has consequences for testing: methods such as \cite{mogstad2024inference}, which build confidence sets for ranks directly, can be easier to apply, but in this setting they generally target an outer set rather than the sharp identified set.

\end{remark}

\section{Inference}\label{sec:inference}

Theorem \ref{th:sharp_ranking_general} shows that even without strong parametric assumptions, the nonparametric link model can be informative about the ranking of merits, including cases where point identification does not hold. Because the model is stochastic, it is important to develop procedures that account for randomness when assessing whether the data support or reject a given ranking. Theorem \ref{th:sharp_ranking_general} provides inequalities that characterize the identified set for the ranking in the nonparametric link model. In this section, we study a procedure to test these inequalities against the observed data distribution.
The discussion of testing a specific ranking is an important starting point and can serve as a basis for procedures designed for more targeted research questions.

For instance, if a test for $\boldsymbol{r} \in R(G, \mathbf{F}, P)$ is available, one can build a confidence set for the ranking using the duality between hypothesis testing and confidence set construction. Once such a set for $R(G, \mathbf{F}, P)$ is obtained, projection methods can deliver confidence sets for individual ranks, or the procedure can be adapted to develop inference tools suited to specific research questions. A key limitation of this approach is computational. Constructing the full confidence set for $R(G, \mathbf{F}, P)$ by test inversion requires checking all rankings in $\mathsf{R}$, whose cardinality grows faster than $q!$, which is infeasible. This difficulty does not arise from our partial identification results. Even under point identification, estimating the ranking is computationally challenging. The computer science literature has studied this problem extensively, proposing algorithms that approximate the optimal solution to an otherwise infeasible task; see for example \cite{shah2016stochastically} for a discussion and analysis of several such algorithms.

\subsection{Null and Alternative Hypotheses}

Consider the hypothesis
\begin{gather*}
    H_{0}: \boldsymbol{r} \in R(G,\mathbf{F},P),
\end{gather*}
where $\boldsymbol{r} \in \mathsf{R}$ is a candidate ranking without ties. The alternative of interest is that $\boldsymbol{r} \notin R(G,\mathbf{F},P)$. By Corollary~\ref{cor:matrix_representation}, the null hypothesis can be expressed as a collection of strict inequalities on the probability matrix ordered according to $\boldsymbol{r}$. Because each off-diagonal entry equals one minus its transpose, it is sufficient to impose restrictions on one triangular half of the matrix. We use the \emph{lower}-triangular entries, which are the probabilities that a worse-ranked team beats a better-ranked team.

For any $(\ell,k)\in E$, let
\begin{gather*}
    p_{\ell,k} \coloneqq \expect_P[Y_{\ell,k}],
\end{gather*}
and let $\boldsymbol{p}\coloneqq\{p_{\ell,k}:\ (\ell,k)\in E\}$ denote the resulting vector of edge probabilities. For the candidate ranking $\boldsymbol{r}$, define
\begin{gather*}
    \tilde{E}_{\boldsymbol{r}} \coloneqq
    \{(j,\ell):\ (j,\ell) \in E,\ r_\ell < r_j \}, \\
    E_{\boldsymbol{r}} \coloneqq
    \{((j,\ell),(i,k)): (j,\ell), (i,k) \in \tilde{E}_{\boldsymbol{r}},
    \ r_{\ell} \le r_k \le r_i \le r_j,\
    (j,\ell) \neq (i,k) \}.
\end{gather*}
The set $\tilde{E}_{\boldsymbol{r}}$ collects the observed worse-versus-better pairs under $\boldsymbol{r}$, while $E_{\boldsymbol{r}}$ indexes pairs of such comparisons that are ordered by the restrictions implied by strong stochastic transitivity. The hypothesis of interest can therefore be written as
\begin{align} \label{eq:H0}
    H_{0}:= \begin{cases}
        p_{j,\ell} - \frac{1}{2} < 0, & \forall (j,\ell) \in \tilde{E}_{\boldsymbol{r}}, \\
        p_{j,\ell} - p_{i,k} < 0, & \forall ((j,\ell),(i,k)) \in E_{\boldsymbol{r}}.
    \end{cases}
\end{align}

The restrictions in \eqref{eq:H0} are strict because the link function is strictly monotone. For inference, however, we work with their closure,
\begin{align} \label{eq:H0c}
    \overline{H}_{0}:= \begin{cases}
        p_{j,\ell} - \tfrac{1}{2} \leq 0, & \forall (j,\ell) \in \tilde{E}_{\boldsymbol{r}}, \\
        p_{j,\ell} - p_{i,k} \leq 0, & \forall ((j,\ell),(i,k)) \in E_{\boldsymbol{r}}.
    \end{cases}
\end{align}
Because $H_0$ is open, distributions on its boundary can be approached arbitrarily closely by distributions satisfying the hypothesis of interest. We therefore base inference on the closed null $\overline{H}_0$, so that size control extends to these boundary distributions as well.

Since every probability vector satisfying \eqref{eq:H0} also satisfies \eqref{eq:H0c}, rejecting $\overline{H}_0$ implies rejecting the hypothesis of interest. The converse need not hold. In particular, a probability vector may satisfy all the weak inequalities in \eqref{eq:H0c} while failing at least one of the corresponding strict inequalities in \eqref{eq:H0}. Replacing $H_0$ by $\overline{H}_0$ therefore enlarges the set of probability vectors under which the test controls size.

Some binding inequalities have a simple interpretation. A direct restriction $p_{j,\ell}\leq1/2$ can bind only when teams $j$ and $\ell$ have equal latent merits. Similarly, if two probabilities being compared correspond to interactions that share the same team, equality requires a tie in the merits of the other two teams. This conclusion does not extend to comparisons between two interactions involving different teams and different opponents. In that case, changes in the two arguments of the link function may offset one another, so that the two probabilities are equal even when all latent merits are distinct.

The relevant question for inference is therefore whether such equalities can allow a candidate ranking that does not belong to the identified set to satisfy the entire collection of weak restrictions in \eqref{eq:H0c}. The following proposition shows that the answer depends on the structure of the tournament graph.

\begin{proposition}\label{prop:closed_null}
    {\normalfont (Open and Closed Ranking Nulls)}
    Let Assumption \ref{as:dgp} hold with $F_0\in\mathbf{F}$, and suppose the latent merits are distinct.
    \begin{enumerate}[label=(\alph*)]
        \item Suppose the tournament graph has diameter at most two, and let $\boldsymbol{r}^*$ denote the ranking generated by the true latent merits. Write $\boldsymbol{p}$ for the probability vector generated by these merits and $F_0$. Then, for every candidate ranking $\boldsymbol{r}$ without ties such that $\boldsymbol{r}\neq\boldsymbol{r}^*$, $\boldsymbol{p}$ violates at least one inequality in \eqref{eq:H0c}. Consequently, among candidate rankings without ties, $\boldsymbol{r}^*$ is the only ranking whose probability vector satisfies the closed restrictions.

        \item Suppose the tournament graph is connected and has diameter greater than two. Then there exist a link function $F_0\in\mathbf{F}$, a vector of distinct latent merits $\boldsymbol{\theta}$, and a candidate ranking $\boldsymbol{r}'$ without ties such that, writing $P$ for the distribution generated by $(\boldsymbol{\theta},F_0)$,
        \begin{gather*}
            \boldsymbol{r}' \notin R(G,\mathbf{F},P),
        \end{gather*}
        while the corresponding probability vector $\boldsymbol{p}$ satisfies all the weak inequalities in \eqref{eq:H0c}.
    \end{enumerate}
\end{proposition}

Proposition~\ref{prop:closed_null} clarifies when replacing the strict restrictions that characterize the hypothesis of interest by their closure changes the population hypothesis. Recall from Theorem~\ref{th:sharp_ranking_general} that a tournament graph with diameter at most two point identifies the ranking under the nonparametric link model. In this case, when latent merits are distinct, replacing the strict restrictions by their closure is innocuous: the true ranking is the only candidate ranking without ties that satisfies the closed restrictions, and every false candidate ranking violates at least one of them strictly. When the graph has diameter greater than two, so that point identification is no longer guaranteed, the closed restrictions may instead admit additional candidate rankings that do not belong to the identified set.

Part (b) also shows that this distinction is not driven by the flexibility of the nonparametric link model. The construction uses the BTL link function, so the discrepancy between $H_0$ and $\overline{H}_0$ can arise even when the data are generated by a standard parametric specification. The source of the discrepancy is the combination of incomplete pairwise comparisons and replacing the strict restrictions implied by the nonparametric characterization with weak inequalities.

For the testing procedure developed below, this distinction does not affect validity: since $H_0$ is contained in $\overline{H}_0$, rejection of the closed null also rejects the hypothesis of interest. It does, however, affect which alternatives lie outside the set over which the test controls size. We return to this distinction in Section~\ref{sec:the_tests} and after Theorem~\ref{th:consistency} in Appendix~\ref{appendix:inference}.

\subsection{Likelihood-based Test Statistics}

The closed null $\overline{H}_0$ is composite and imposes inequality restrictions on the parameters of the Bernoulli random variables $\{Y_{\ell,k}\}_{(\ell,k) \in E}$. This is remarkable: despite a nonparametric link function, the null involves only inequalities on parameters of known distributions. The structure suggests using a likelihood ratio test, comparing the likelihood of the unrestricted model, which does not depend on $\boldsymbol{r}$, with the likelihood under the restrictions implied by $\overline{H}_0$. The larger the gap between these two values, the stronger the evidence that the data reject the null.

To implement this approach, consider the log-likelihood
\begin{gather*}
    l(\boldsymbol{p}) \coloneqq \sum_{(\ell,k) \in E} \sum_{i=1}^{n_{\ell,k}} \Big( Y_{\ell,k,i} \ln( p_{\ell,k}) + (1 - Y_{\ell,k,i}) \ln(1-p_{\ell,k}) \Big),
\end{gather*}
maximized by the sample averages $\hat{p}_{\ell, k}:= \frac{1}{n_{\ell, k}} \sum_{i=1}^{n_{\ell, k}} Y_{\ell, k, i}$. Let $\hat{\boldsymbol{p}} \coloneqq \{ \hat{p}_{\ell, k}, (\ell,k) \in E \}$. The restricted estimator $\hat{\boldsymbol{p}}^{*}$ maximizes $l(\boldsymbol{p})$ subject to the closed restrictions in \eqref{eq:H0c}:
\begin{gather*}
    \hat{\boldsymbol{p}}^*
    \coloneqq
    \arg\max_{\boldsymbol{p}:\ \overline{H}_0}
    l(\boldsymbol{p}).
\end{gather*}
By Proposition~2.4.3 of \citet{silvapulle2011constrained}, this constrained MLE coincides with the weighted Euclidean projection of $\hat{\boldsymbol{p}}$ onto the null polytope,\footnote{\citet{wang2022efficient} propose a fast algorithm, implemented in the R package \texttt{IsotoneOptimization}, for solving the particular optimization problem considered here.}
\begin{gather*}
    \hat{\boldsymbol{p}}^*
    =
    \arg\min_{\boldsymbol{p}:\ \overline{H}_0}
    \sum_{(\ell,k) \in E} n_{\ell,k}\,(\hat{p}_{\ell, k} - p_{\ell, k})^2.
\end{gather*}

The estimator $\hat{\boldsymbol{p}}^{*}$ is defined using the closed restrictions in $\overline{H}_0$. This closure is necessary for the constrained optimization problem to have a solution, since the log-likelihood need not attain its supremum over the open set defined by $H_0$.

The likelihood ratio test statistic is then:
\begin{align*}
    \Lambda =& 2 (l(\hat{\boldsymbol{p}}) - l(\hat{\boldsymbol{p}}^*)) = 2 \sum_{(\ell, k) \in \tilde{E}_{\boldsymbol{r}}} n_{\ell,k} \left( \hat{p}_{\ell,k} \ln \bigg( \frac{\hat{p}_{\ell,k}}{\hat{p}_{\ell,k}^*(\hat{\boldsymbol{p}})} \bigg) + (1 - \hat{p}_{\ell,k}) \ln \bigg( \frac{1 - \hat{p}_{\ell,k}}{1 - \hat{p}_{\ell,k}^*(\hat{\boldsymbol{p}})} \bigg) \right)
\end{align*}
with the standard continuous extension conventions that $0 \ln(0/q) = 0$ for $q \geq 0$, and $p \ln(p/0) = \infty$ for $p > 0$. The statistic is hence a random variable, function of $\hat{\boldsymbol{p}}$ and $\hat{\boldsymbol{p}}^*$: denote by $\lambda$ the realized value of $\Lambda$ in a given application, when $\hat{\boldsymbol{p}}$ and $\hat{\boldsymbol{p}}^*$ are computed using the realized data $\boldsymbol{Y}_{\vn}$. Large values of $\lambda$ give evidence against the null: the statistic equals zero when the restricted and the unrestricted estimators are equal, and hence none of the constraints for $\hat{\boldsymbol{p}}^*$ is binding. The larger the discrepancies between $\hat{\boldsymbol{p}}$ and $\hat{\boldsymbol{p}}^*$, the larger the log-likelihood ratio.

\subsection{Testing the Null Hypothesis}

Given the realized $\lambda$, define the p-value $\pi_\lambda$:
\begin{gather} \label{eq:pvalue}
    \pi_\lambda \coloneqq \sup_{\boldsymbol{p}:\ \overline{H}_0} P\{ \Lambda \ge \lambda \}.
\end{gather}
$\pi_\lambda$ is the largest probability of observing a test statistic greater than $\lambda$ under the closed null hypothesis, which depends on the vector $\boldsymbol{p}$. Taking the supremum over $\overline{H}_0$ is necessary to obtain a test that is valid for any distribution satisfying $\overline{H}_0$, including at the boundary where the least favorable distribution typically lies.

Since $\Lambda$ is a known function of $\hat{\boldsymbol{p}}$, its distribution and hence $P\{ \Lambda \ge \lambda \}$ can be computed exactly or simulated for any $\boldsymbol{p}$. Therefore, if the value of $\boldsymbol{p}$ that satisfies $\overline{H}_0$ and maximizes this probability is known, computing the p-value is straightforward.

Less straightforward is the constrained optimization required to find the least favorable distribution $\boldsymbol{p}$ that, for a given $\lambda$, maximizes $P\{ \Lambda \ge \lambda \}$. We consider two approaches. The first is computationally demanding and feasible only for small tournaments but delivers the exact p-value. The second is simpler and applies in asymptotic settings where the number of games grows large.

\subsubsection{Exact p-value} \label{sec:exact_pv}

The exact p-value is obtained by solving the optimization problem in Equation \ref{eq:pvalue}. For a given test statistic $\lambda$, the rejection probability $P(\Lambda(\boldsymbol{p}) \ge \lambda)$ can be written as a polynomial function of $\boldsymbol{p}$, with the form of the polynomial determined by the graph $G$ and $\boldsymbol{n}$, the number of games between each pair of teams. Appendix \ref{appendix:inference} describes an algorithm to construct this polynomial. This step is the computational bottleneck of the procedure and becomes infeasible even for moderately large tournaments. More efficient methods for this step would greatly improve the practical feasibility of the approach. Once the polynomial is obtained, it can be optimized with respect to $\boldsymbol{p}$ subject to the linear constraints in $\overline{H}_0$. The following example illustrates this p-value computation.

\begin{example}\label{ex:1}
    Consider the tournament graph illustrated in Figure \ref{fig:tournament_pvalue}, where the numbers on the edges indicate the number of games played between each pair of teams.
    \begin{figure}[H]
        \centering
        \begin{tikzpicture}
            \node[shape=circle,draw=black] (A) at (0,0) {A};
            \node[shape=circle,draw=black] (B) at (3,0) {B};
            \node[shape=circle,draw=black] (C) at (6,0) {C};
            \path [-] (A) edge node[anchor=south] {$n_{AB} = 9$} (B);
            \path [-] (B) edge node[anchor=south] {$n_{BC} = 2$} (C);
        \end{tikzpicture}
        \caption{Tournament graph for Example \ref{ex:1}.}
        \label{fig:tournament_pvalue}
    \end{figure}

   Suppose the observed data are $\boldsymbol{Y}_{A,B} = (0, 0, 0, 0, 0, 0, 0, 0, 0)$ and $\boldsymbol{Y}_{C,B} = (0, 1)$. We want to test whether these data are consistent with the team order $(B, A, C)$, that is, whether the ranking $\boldsymbol{r} \coloneqq (r_A, r_B, r_C)' = (2, 1, 3)'$ belongs to the identified set. For this ranking, the realized value of the test statistic equals $\lambda = 3.9294$.

    For convenience, let $p_1 = \expect_P[Y_{A,B}]$ and $p_2 = \expect_P[Y_{C,B}]$. $P(\Lambda \ge \lambda)$ is then
    \begin{equation}\label{eq:cdf_example}
        \begin{split}
            P(\Lambda \ge \lambda) = {} & 9 p_1^8 (1 - p_1) + p_1^9 + (1 - p_1)^9 (2 - p_2) p_2 \\
           & + \left[ 9 p_1 (1 - p_1)^8 + 36 p_1^2 (1 - p_1)^7 + 36 p_1^7 (1 - p_1)^2 \right] p_2^2.
        \end{split}
    \end{equation}
    To compute the p-value, we maximize this polynomial subject to the constraints in $\overline{H}_0$, which for the ranking $(2, 1, 3)$ are $p_2 \le p_1 \le 1/2$. The maximum is attained at $p_1 = p_2 \approx 0.118$ and yields $\pi_\lambda \approx 0.080$.
\end{example}

Example \ref{ex:1} shows that, in general, the least favorable distribution is not $p_{\ell, k} = \frac{1}{2}$ for all $(\ell,k)$, even though in that case all weak inequalities in $\overline{H}_0$ bind. This situation is unusual: when inequalities in $\overline{H}_0$ are tested separately, the least favorable distribution does correspond to a value of $\boldsymbol{p}$ that satisfies the equalities. In such cases the optimization is straightforward, as in Barnard's test for comparing the averages of two binomial distributions \citep{barnard1947significance}. In contrast, testing a ranking requires maximizing a polynomial of the form in \eqref{eq:cdf_example}. Maximizing this polynomial subject to the linear constraints of the null hypothesis is an NP-hard problem and does not scale well as the number of edges in the tournament graph increases.

\subsubsection{Approximate p-value}\label{sec:apprx_pvalue}

As illustrated by Example~\ref{ex:1}, finding the least favorable distribution in finite samples generally requires solving an optimization problem. When this optimization becomes computationally infeasible, an asymptotically valid approximation can be used. Under the asymptotic regime in which the numbers of games on all edges grow to infinity at the same rate, the distribution with $p_{\ell,k}=\tfrac{1}{2}$ for every $(\ell,k)\in E$ is least favorable, as shown in Lemma~\ref{lemma:lfd} in Appendix~\ref{appendix:inference}. Asymptotically valid $p$-values can therefore be obtained by evaluating $P\{\Lambda \geq \lambda\}$ under this distribution.

To do so, we simulate samples from the asymptotically least favorable distribution. Let $\tilde{p}$ denote the distribution under which all inequalities in the closed null hypothesis in \eqref{eq:H0c} bind, so that $\expect_{\tilde{p}}[Y_{\ell,k}] = 0.5$ for every $(\ell,k)\in E$. In each simulated sample, we keep the tournament graph and the numbers of games $\{{n_{\ell,k}}, (\ell,k)\in E \}$ fixed. For every edge $(\ell,k)$, we draw $n_{\ell,k}$ independent Bernoulli outcomes with success probability $1/2$, compute the unrestricted and restricted estimators, and then evaluate the corresponding likelihood ratio statistic.

After repeating this procedure independently many times, the approximate $p$-value is the fraction of simulated likelihood ratio statistics that are at least as large as the observed value $\lambda$. This approximation is expected to perform well when the numbers of games are sufficiently large and do not differ substantially across interacting pairs. Nevertheless, the Monte Carlo evidence in Section~\ref{sec:mc_simulations} shows that empirical rejection frequencies do not exceed the nominal level even in the small and highly unbalanced designs we consider.

\subsubsection{The Tests} \label{sec:the_tests}

Once the p-value, exact or approximate, is computed, the level-$\alpha$ testing procedure rejects the null hypothesis whenever the p-value is no larger than $\alpha$.

In Appendix \ref{appendix:inference}, we formally describe the decision rules and prove finite-sample validity of the exact test, as well as pointwise asymptotic validity of the approximate test. The approximate test is consistent against any fixed permissible alternative whose probability vector violates the closed null $\overline{H}_0$. When the tournament graph has diameter at most two and latent merits are distinct, Proposition~\ref{prop:closed_null} implies that this includes every false candidate ranking without ties. When the diameter is greater than two, a ranking may lie outside the identified set while still satisfying $\overline{H}_0$; the test then does not have power against that distribution. Since only pointwise validity is established, there may exist data generating processes for which the test's size deviates substantially from the nominal level even for large sample sizes. Establishing uniform validity of the test is left for future research. The next section reports extensive simulations, with additional results in Appendix \ref{appendix:mc}, which show that the asymptotic test controls size across a wide range of tournament graphs.

\begin{remark}
    {\normalfont (A Multiple-Testing Alternative)}
    Together, the two procedures provide feasible and informative tests: the exact version when the number of teams and games in the tournament graph is small, and the approximate version when the number of games is large across all edges. When the exact test is computationally infeasible and the number of games is not large, $\overline{H}_0$ is a collection of hypotheses that can be tested separately at levels adjusted to control the family-wise error rate. These methods are generally more conservative than the joint test, but their power can be improved by aggregating the individual sub-hypotheses. The hierarchical structure of the conditions in $\overline{H}_0$, where some inequalities are implied by others through transitivity, can be exploited to design more powerful multiple-testing procedures. The exact implementation depends on the structure of the tournament graph under study. For this reason, we do not pursue this direction here and instead refer the reader to \cite{hochberg1987multiple}.
\end{remark}

\section{Monte Carlo Simulations} \label{sec:mc_simulations}

Section~\ref{sec:inference} establishes finite-sample validity of the exact test and pointwise asymptotic validity of the approximate test. This section studies whether those guarantees are reflected in finite samples. Design 1 provides an illustration of the size and power of both procedures in the setting of Example \ref{example:partial_identification}. We then study the finite sample size of the approximate test across a broader range of distributions satisfying the null. Design 2 considers a small and sparse tournament based on Example \ref{ex:1}, while Design 3 considers a larger and denser tournament motivated by the empirical illustration in Section \ref{sec:empirical_illustration}. Appendix \ref{appendix:mc} reports two additional designs.

\subsection{Design 1: Small Partially Identified Tournament}

Design 1 uses Example \ref{example:partial_identification} to illustrate, in a simple setting, size and power of both the exact and approximate tests as the sample size grows, including power against alternatives that are close to versus far from the identified set. For convenience, Figure \ref{fig:tournament_partial_identification2} reproduces the tournament graph and the corresponding pairwise win probabilities.

\begin{figure}[htbp]
    \centering
    \begin{tikzpicture}
        \node[shape=circle,draw=black] (A) at (0,0) {A};
        \node[shape=circle,draw=black] (B) at (2,0) {B};
        \node[shape=circle,draw=black] (C) at (4,0) {C};
        \node[shape=circle,draw=black] (D) at (6,0) {D};
        \path [-] (A) edge node[anchor=south] {$0.75$} (B);
        \path [-] (B) edge node[anchor=south] {$0.7$} (C);
        \path [-] (C) edge node[anchor=south] {$0.2$} (D);
    \end{tikzpicture}
    \caption{Tournament graph and pairwise win probabilities in Design 1. Each edge label reports the probability that the team on the left defeats the team on the right. For example, $\prob(Y_{A,B}=1)=0.75$.}
    \label{fig:tournament_partial_identification2}
\end{figure}

As shown above, the sharp identified set contains exactly two rankings without ties,
\begin{gather*}
    \boldsymbol{r}^{(1)} = (1, 3, 4, 2)', \qquad \boldsymbol{r}^{(2)} = (2, 3, 4, 1)'.
\end{gather*}
Thus, among rankings without ties, teams $A$ and $D$ occupy the first two positions, but their relative order is not identified. Team $B$ is ranked third, while team $C$ is ranked fourth.

We consider a tournament in which every interaction is repeated the same number of times, denoted by $n$. We test four distinct null hypotheses. The rankings $\{(1, 3, 4, 2)', (2, 3, 4, 1)' \}$ belong to the identified set, so their rejection frequencies measure the size of the test. The rankings $\{(1, 2, 4, 3)', (4, 2, 1, 3)' \}$ do not belong to the identified set, so their rejection frequencies measure power against two different alternatives.

\begin{figure}[htbp]
    \centering
    \includegraphics[width=\textwidth]{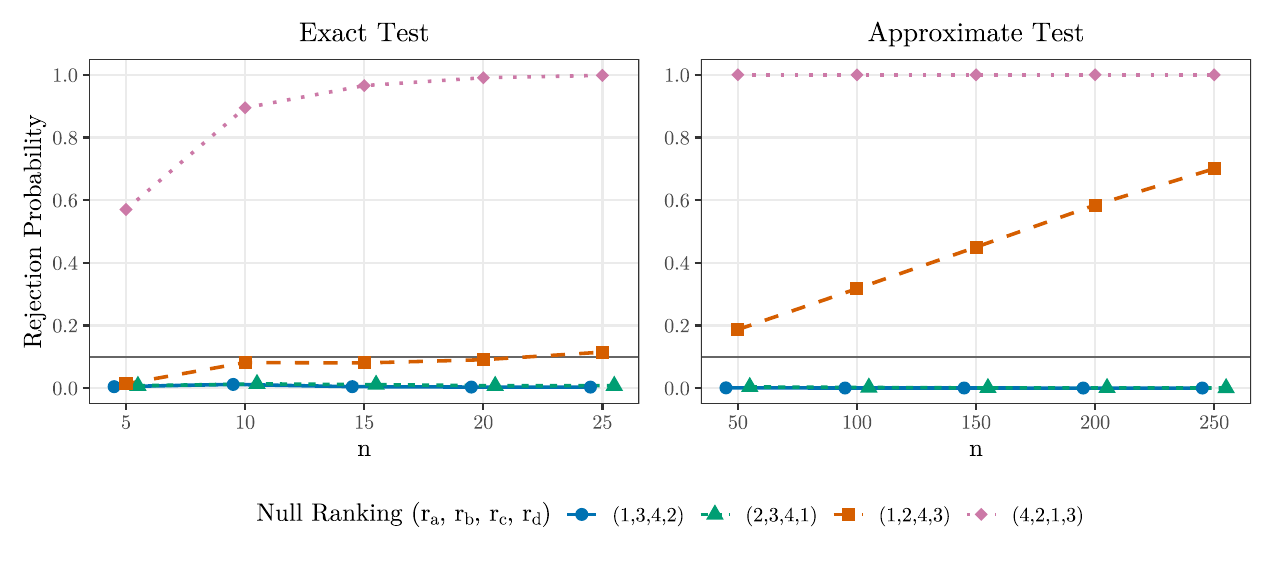}
    \caption{Empirical rejection frequencies in Design 1. The horizontal line denotes the nominal significance level $\alpha=0.10$.}
    \label{fig:mc_design1}
\end{figure}

For $n \in \{5, 10, 15, 20, 25\}$, we compute the exact p-value described in Section \ref{sec:exact_pv}. For $n \in \{50, 100, 150, 200, 250\}$, we compute the approximate p-value described in Section \ref{sec:apprx_pvalue}. For each value of $n$, we generate 10,000 independent samples from the same data generating process,
\begin{gather*}
    p_{A,B}=0.75,\qquad
    p_{B,C}=0.70,\qquad
    p_{C,D}=0.20,
\end{gather*}
and apply the test to each of the four candidate rankings at significance level $\alpha=0.10$. Figure \ref{fig:mc_design1} reports the resulting empirical rejection frequencies.

We first consider the two tie-free rankings that belong to the identified set, shown in blue and green. Their empirical rejection frequencies remain below the nominal significance level for every value of $n$. The exact test controls size by construction. The approximate test also controls size in these designs.

Both tests reject the null with probability strictly below $0.1$. The least favorable calibration protects against distributions on the boundary of the closed null, whereas the probabilities in Design 1 satisfy the relevant inequalities strictly, leading to rejection frequencies below $\alpha=0.10$. The discreteness of the finite sample distribution of the test statistic may further contribute to rejection probabilities below the nominal level.

We next consider the ranking $(1,2,4,3)'$, which lies outside the identified set (shown in orange). This ranking is close to $(1,3,4,2)'$, one of the rankings in the identified set. The two rankings differ only in the relative positions of teams $B$ and $D$. Because these teams do not compete directly, their relative order must be inferred by comparing their performances against team $C$. Under the true data generating process, team $D$ defeats team $C$ with probability $0.8$, whereas team $B$ defeats team $C$ with probability $0.7$. The difference is therefore only $0.1$, making the violation imposed by this ranking difficult to detect when $n$ is small. As $n$ increases, sampling uncertainty declines, and the empirical rejection frequency rises substantially, in line with the consistency of the test.

Finally, consider the ranking $(4,2,1,3)'$, also outside the identified set (shown in pink). This ranking reverses the ordering in $(1,3,4,2)'$, placing team $C$ first and team $A$ last. It conflicts with the direction implied by all three observed comparisons. Evidence against this ranking therefore accumulates across multiple interactions, making the violation easier to detect. Accordingly, the test has substantial power even when each interaction is observed only a small number of times, and its rejection frequency increases rapidly with $n$.

\subsection{Finite-Sample Size of the Approximate Test}
\label{sec:mc_size}

The theoretical results establish pointwise asymptotic size control for the approximate test. That is, for any fixed distribution satisfying the closed null $\overline{H}_0$, the rejection probability is asymptotically bounded by the nominal significance level as the number of repetitions increases. Designs 2 and 3 study whether the approximate test also controls size in finite samples across a broad range of distributions satisfying $\overline{H}_0$.

For a given tournament graph, the data generating process is characterized by one winning probability for each interaction and, therefore, by a vector of $|E|$ probabilities. The closed null $\overline{H}_0$ imposes a collection of weak inequalities on this vector and defines a polytope in the $|E|$-dimensional space of winning probabilities. Evaluating the test at every distribution in this polytope is infeasible, particularly when the tournament graph contains many edges.

We therefore use a two-step procedure to search for distributions under which the approximate test may be most likely to overreject. In the first step, we use Markov chain Monte Carlo to draw winning-probability vectors from the null polytope. For each probability vector, we generate one sample and apply the approximate test. Because the data generating process changes across draws, this exercise is not a standard Monte Carlo simulation. Instead, it provides an exploratory search over the null polytope and identifies regions in which rejections occur relatively frequently.

In the second step, we select representative probability vectors from these regions and conduct standard Monte Carlo simulations. For each selected vector, we hold the data generating process fixed, repeatedly generate samples, and compute the empirical rejection frequency of the approximate test. This second step therefore provides a conventional assessment of size at distributions identified by the exploratory search. Details on the procedure used to draw from the null polytope are provided in Appendix \ref{appendix:mc}.

We apply this procedure to two tournament structures. Design 2 considers a small and sparse graph, while Design 3 considers a larger and denser graph that resembles the tournament observed in the empirical application.

\subsubsection{Design 2: Small Tournament with Unbalanced Repetitions}

Design 2 uses Example \ref{ex:1} to assess finite sample size control of the approximate test in a small tournament with unbalanced repetitions. With only a few games on one edge and a large imbalance across edges, this design is far from the settings in which the asymptotic least-favorable approximation is expected to be accurate, and therefore provides a stringent check of the procedure. Because the null polytope is two-dimensional, the exploratory search over null distributions can also be displayed in full. For convenience, Figure \ref{fig:tournament_design2} reproduces the tournament graph and the number of games on each edge.

\begin{figure}[htbp]
    \centering
    \begin{tikzpicture}
        \node[shape=circle,draw=black] (A) at (0,0) {A};
        \node[shape=circle,draw=black] (B) at (3,0) {B};
        \node[shape=circle,draw=black] (C) at (6,0) {C};
        \path [-] (A) edge node[anchor=south] {$n_{AB} = 9$} (B);
        \path [-] (B) edge node[anchor=south] {$n_{BC} = 2$} (C);
    \end{tikzpicture}
    \caption{Tournament graph and number of games on each edge in Design 2.}
    \label{fig:tournament_design2}
\end{figure}

\begin{figure}[htbp]
    \centering
    \includegraphics[scale=0.6]{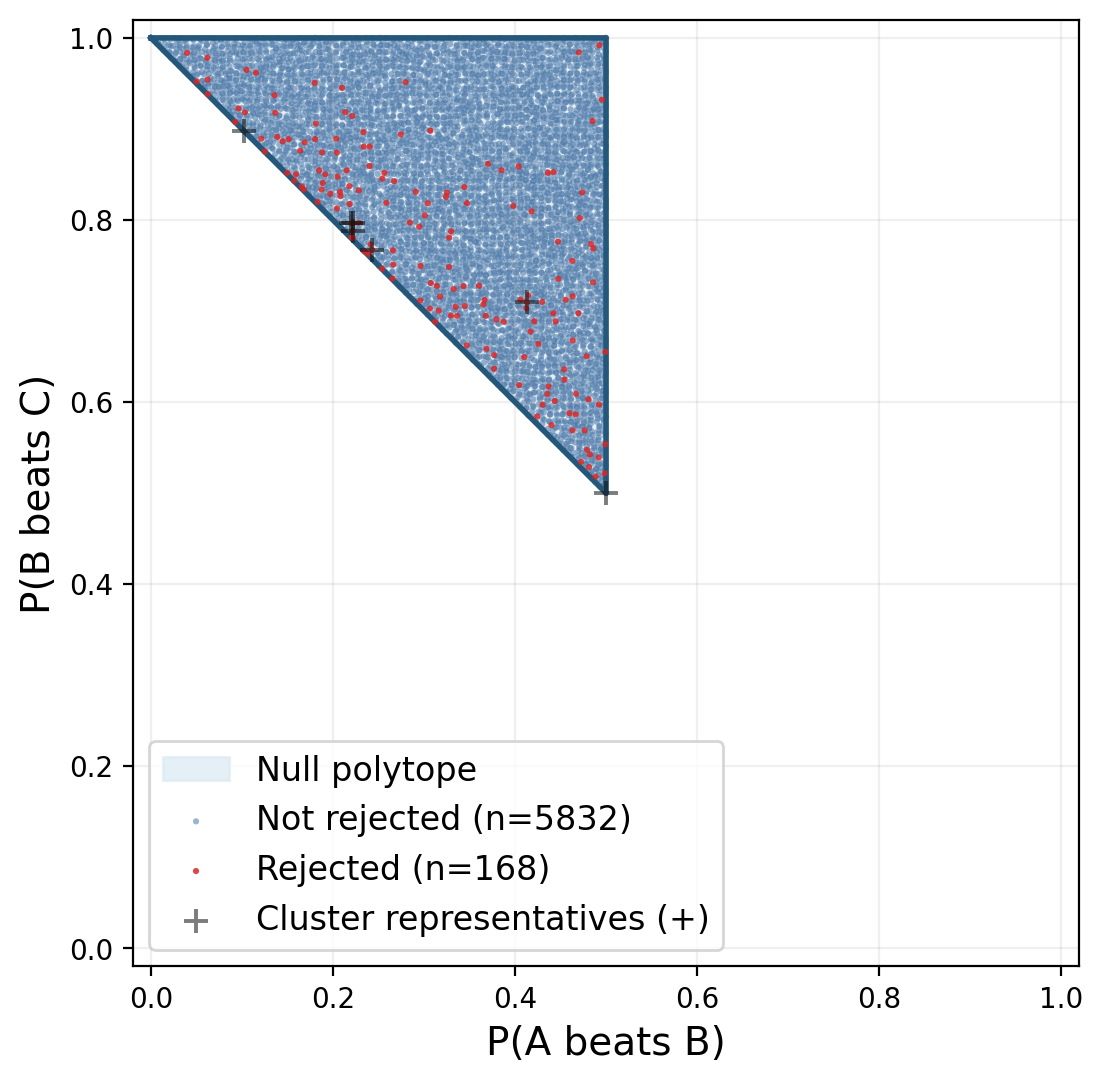}
    \caption{Exploratory draws from the null polytope in Design 2. Blue points correspond to non-rejections, red points to rejections, and ``+'' symbols to the selected representative probability vectors.}
    \label{fig:2dnullproj_des2}
\end{figure}

We test the null hypothesis that the ranking $\boldsymbol{r}=(2,1,3)'$ belongs to the identified set. Under this ranking, team $B$ is ranked first, team $A$ second, and team $C$ third. Let $p_{A,B}=\prob(Y_{A,B}=1)$ and $p_{C,B}=\prob(Y_{C,B}=1)$. The null hypothesis imposes
\begin{gather*}
    p_{C,B}\leq p_{A,B}\leq \frac{1}{2}.
\end{gather*}
It therefore corresponds to a triangular region in the two-dimensional space of winning probabilities, shown in Figure \ref{fig:2dnullproj_des2}. The figure reports on the vertical axis the probability that team $B$ defeats team $C$, $p_{B,C}=1-p_{C,B}$, so that in the displayed coordinates the null polytope is the set of pairs with $p_{A,B}\leq 1/2$ and $p_{B,C}\geq 1-p_{A,B}$.

We first draw 6,000 probability vectors from the null polytope. For each vector, we generate one sample and apply the approximate test at the nominal significance level $\alpha=0.1$. The test rejects for 168 of the 6,000 draws. Figure \ref{fig:2dnullproj_des2} shows the probability vectors for which the test does not reject in blue and those for which it rejects in red.

We then select $k=5$ representative probability vectors at which rejections are locally most frequent. Specifically, among the red points, we select the five points with the largest number of neighboring red points. These probability vectors are marked with ``+'' in Figure \ref{fig:2dnullproj_des2}.

We also consider two benchmark probability vectors. The first is the finite sample least favorable distribution for the exact p-value calculation in Example \ref{ex:1},
\begin{gather*}
    \prob(Y_{A,B}=1)=\prob(Y_{C,B}=1)=0.118.
\end{gather*}
The second is the asymptotically least favorable distribution used to compute the approximate p-value,
\begin{gather*}
    \prob(Y_{A,B}=1)=\prob(Y_{C,B}=1)=0.5.
\end{gather*}

For each of these seven probability vectors, we hold the data generating process fixed, generate $M=100{,}000$ independent samples, and apply the approximate test at the nominal significance level $\alpha=0.1$. Table \ref{tab:cluster-representative-rejections} reports the resulting empirical rejection frequencies. In every case, the empirical rejection frequency is below the nominal significance level, and the largest value is attained at the finite sample least favorable distribution of Example \ref{ex:1}.

\begin{table}[H]
    \centering
    \begin{tabular}{cc}
        Probability vector & Empirical rejection \\
        \hline
        Cluster 1 & 0.06621 \\
        Cluster 2 & 0.02806 \\
        Cluster 3 & 0.06201 \\
        Cluster 4 & 0.06273 \\
        Cluster 5 & 0.06353 \\
        $p = (0.5, 0.5)$ & 0.0610\\
        $p = (0.118, 0.118)$ & 0.0800\\
        \hline
    \end{tabular}
    \caption{Empirical rejection frequencies at the selected probability vectors in Design 2. The nominal significance level is $\alpha=0.1$, and each result is based on $M=100{,}000$ Monte Carlo replications.}
    \label{tab:cluster-representative-rejections}
\end{table}

Even in this design, which is far from the asymptotic regime underlying the approximate p-value, the empirical rejection frequencies remain below the nominal level over a wide range of null distributions. This is consistent with the analytical finding for Example \ref{ex:1} (see Appendix \ref{appendix:inference}) that the finite sample size distortion of the approximate test does not exceed $0.02$ for any level $\alpha$ in this setting.

\subsubsection{Design 3: Empirically Motivated Dense Tournament}

\begin{figure}[htbp]
    \centering
    \includegraphics[scale=0.8]{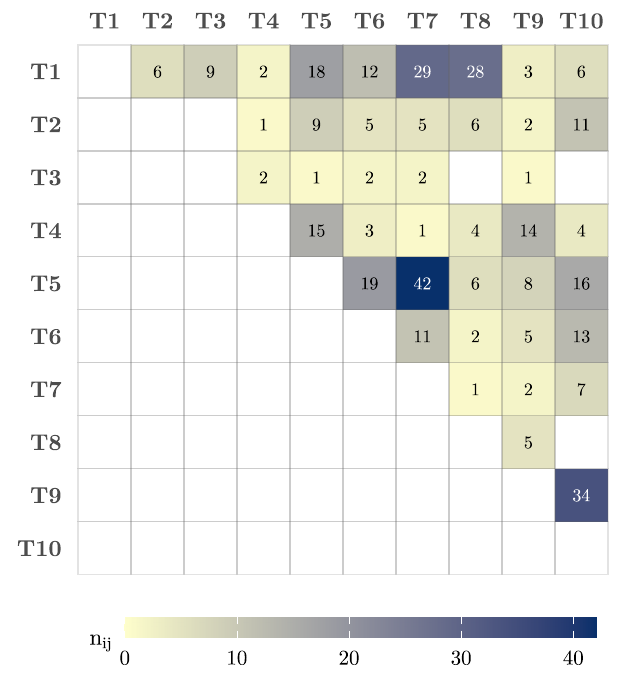}
    \caption{Tournament graph and number of transitions on each edge in Design 3. Firms are ordered according to the PageRank ordering. Entry $(i,j)$ reports the total number of transitions between firms $T_i$ and $T_j$.}
    \label{fig:heatmap_des3}
\end{figure}

Design 3 assesses finite sample size control of the approximate test on a dense, empirically motivated tournament. We use the graph of job-to-job transitions of women across the ten firms studied in Section \ref{sec:empirical_illustration}. Unlike the pooled graph in the main empirical illustration, this women's tournament is not complete: it contains many but not all pairwise interactions, together with unequal numbers of repetitions. We choose this graph because it provides a high-dimensional, empirically relevant null that is denser than Design 2 yet still incomplete. Figure \ref{fig:heatmap_des3} reports the number of transitions on each edge.

We test whether the ranking implied by the PageRank scores belongs to the identified set. PageRank is commonly used in applied work and is described in detail in Section \ref{sec:empirical_illustration}. We order the firms accordingly and denote the resulting ranking by $(T_1,\ldots,T_{10})$. The null hypothesis is that $(T_1,\ldots,T_{10})$ belongs to the identified set.

As in Design 2, we first draw $370{,}000$ winning-probability vectors from the null polytope. For each vector, we generate one sample and apply the approximate test at the nominal significance level $\alpha=0.1$. Appendix \ref{appendix:mc} reports several two-dimensional projections of the null polytope, together with the sampled probability vectors and the corresponding test outcomes.

The approximate test rejects for $1{,}097$ of the $370{,}000$ draws. Among these probability vectors, we select the $10$ points with the largest number of neighboring points that also lead to rejection. These points represent regions of the null polytope in which rejections are locally most frequent.

For each selected probability vector, we then hold the data generating process fixed, generate $M=30{,}000$ independent samples, and apply the approximate test at the nominal significance level $\alpha=0.1$. We also consider the asymptotically least favorable probability vector,
\begin{gather*}
    \boldsymbol{p}=(0.5,\ldots,0.5).
\end{gather*}
Table \ref{tab:cluster-representative-rejections_des3} reports the resulting empirical rejection frequencies. The rejection frequency is below $\alpha=0.1$ at each of the $10$ selected probability vectors and equals the nominal significance level at the asymptotically least favorable distribution. The approximate test therefore controls size at all selected null distributions: Design 3 provides no evidence of size distortion in the dense tournament underlying the empirical illustration.

\begin{table}[htbp]
    \centering
    \begin{tabular}{cc}
        Probability vector & Empirical rejection \\
        \hline
        Cluster 1 & 0.0742 \\
        Cluster 2 & 0.0453 \\
        Cluster 3 & 0.0691 \\
        Cluster 4 & 0.0767 \\
        Cluster 5 & 0.0664 \\
        Cluster 6 & 0.0374\\
        Cluster 7 & 0.0266\\
        Cluster 8 & 0.0312\\
        Cluster 9 & 0.0296\\
        Cluster 10 & 0.0323\\
        $p = (0.5,..., 0.5)$ & 0.100\\
        \hline
    \end{tabular}
    \caption{Empirical rejection frequencies at the selected probability vectors in Design 3. The nominal significance level is $\alpha=0.1$, and each result is based on $M=30{,}000$ Monte Carlo replications.}
    \label{tab:cluster-representative-rejections_des3}
\end{table}

\section{Empirical Illustration}
\label{sec:empirical_illustration}

We illustrate our testing procedure using employer-to-employer transitions from the \textit{Rela\c{c}\~{a}o Anual de Informa\c{c}\~{o}es Sociais} (RAIS), an annual administrative census of formal-sector employment in Brazil. Several studies use job-to-job transitions in RAIS to rank firms and examine labor-market outcomes, including the effects of unions and the gender pay gap \citep{corradini2023collective, lagos2024union, morchio2026gender}. Following \citet{sorkin2018ranking}, this approach begins by constructing a transition matrix whose entries record the frequency of worker movements between firms. Applying the PageRank algorithm \citep{page1999pagerank} to this matrix produces a firm ranking that can be used in subsequent analysis. In our empirical exercise, we take the resulting PageRank ordering as the candidate ranking and test whether it belongs to the identified set under the nonparametric link model.

Following \citet{lavetti2023gender}, we construct a worker-year panel from RAIS for 2015--2017. We retain workers aged 23--65 who hold at least one full-time job, defined as a position with at least 30 contracted hours per week. When a worker holds multiple jobs in the same year, we define the job with the highest estimated annual earnings as the worker's primary employment. Finally, we restrict the sample to firms with at least 20 employees.

For each worker, we use the sequence of primary employers to identify job-to-job transitions. Two firms $\ell$ and $k$ interact if at least one worker moves between them, and $n_{\ell,k}=n_{k,\ell}$ denotes the total number of transitions in either direction. For each $i\in[n_{\ell,k}]$, define
\begin{gather*}
    Y_{\ell,k,i}
    =
    \begin{cases}
        1, & \text{if the worker moves from firm $k$ to firm $\ell$},\\
        0, & \text{if the worker moves from firm $\ell$ to firm $k$}.
    \end{cases}
\end{gather*}
We interpret a transition to firm $\ell$ as a win for $\ell$ over firm $k$. The corresponding sample win probability is
\begin{gather*}
    \hat{p}_{\ell,k}
    =
    \frac{1}{n_{\ell,k}} \sum_{i=1}^{n_{\ell,k}}Y_{\ell,k,i}.
\end{gather*}
Thus, $\hat{p}_{\ell,k}>1/2$ means that, among workers moving between the two firms, more move from $k$ to $\ell$ than from $\ell$ to $k$.

To keep the illustration tractable, we focus on ten firms connected by a dense transition network. We first construct a filtered graph in which two firms are linked whenever more than 20 transitions occur in at least one of the two directions. The largest clique in this graph contains seven firms; among the cliques of seven firms, we retain the one with the highest average degree. We then add firms sequentially, at each step selecting the firm with the largest number of links in the filtered graph to those already included, until the set contains ten firms. Once the firms have been selected, we return to the original data and use all employer-to-employer transitions among them. The resulting tournament graph is complete, so every pair of selected firms interacts.

\begin{figure}[H]
    \centering
    \includegraphics[width=\textwidth]{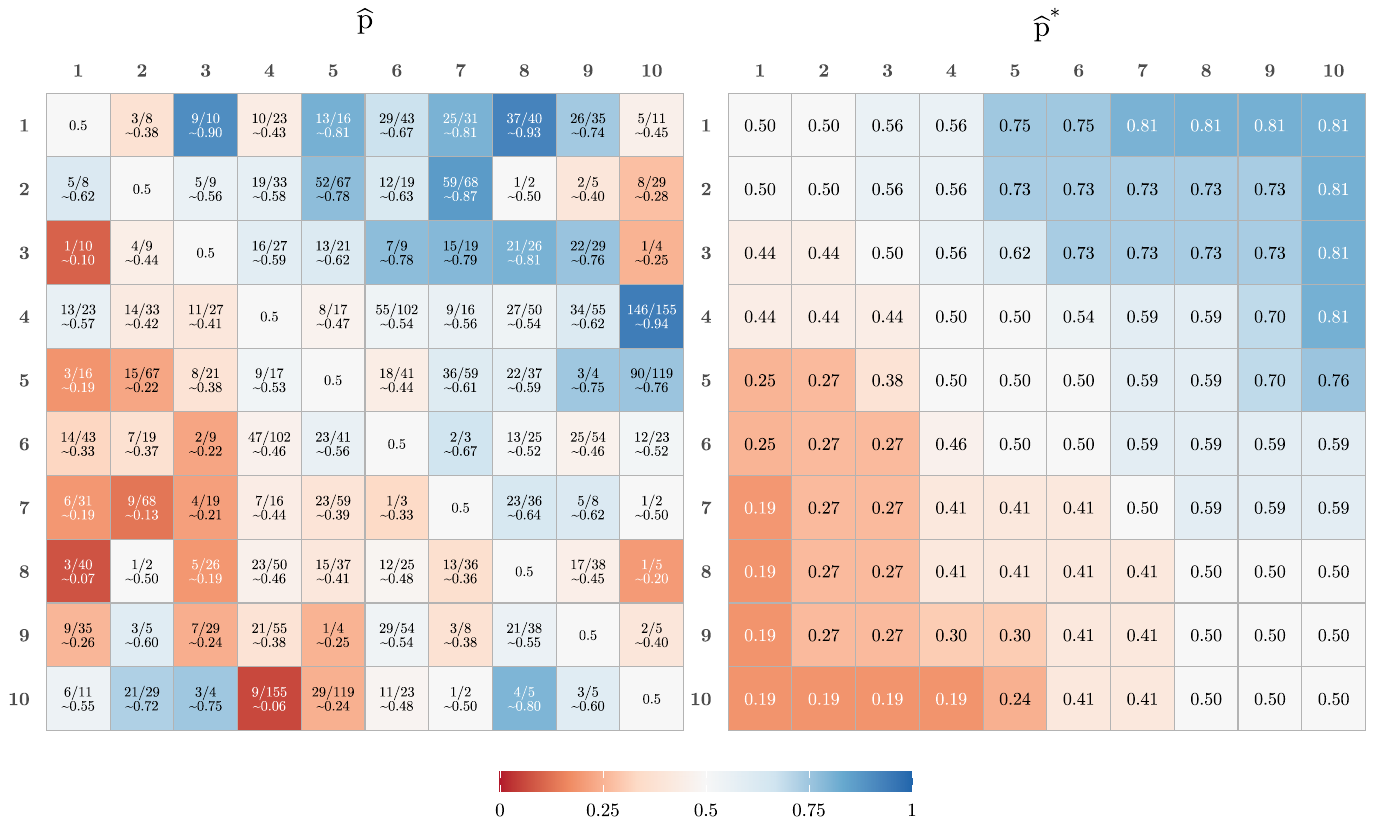}
    \caption{Unrestricted and restricted sample pairwise win probabilities under the PageRank ordering. Each off-diagonal entry reports the estimates. Diagonal entries are fixed at $1/2$.}
    \label{fig:observed_p_matrix}
\end{figure}

Using the transitions among these ten firms, we compute the PageRank ordering following \citet{sorkin2018ranking}. We then calculate the unrestricted sample averages, $\boldsymbol{\hat{p}}$, and the restricted sample averages, $\boldsymbol{\hat{p}^{*}}$, implied by this ordering. Figure~\ref{fig:observed_p_matrix} reports the two probability matrices after arranging firms from best to worst according to the PageRank ordering. Entry $(\ell,k)$ gives the estimated probability that firm $\ell$ wins against firm $k$. Equivalently, it is the probability that a worker moving between firms $\ell$ and $k$ moves from $k$ to $\ell$. In the restricted matrix $\boldsymbol{\hat{p}^{*}}$, shown on the right, the restrictions in Corollary~\ref{cor:matrix_representation} are satisfied: the probabilities weakly increase as one moves northeast through the matrix.

If the PageRank ordering belongs to the identified set, the unrestricted and restricted matrices converge to the same population object. The visible differences between the two matrices therefore suggest a violation of the null restrictions. The likelihood-ratio test developed in Section~\ref{sec:inference} allows us to formally assess this evidence. We approximate the least favorable null distribution, under pairwise win probabilities equal to $1/2$, using $20{,}000$ simulation draws. The test rejects the null hypothesis that the PageRank ordering belongs to the identified set at any conventional significance level. The Monte Carlo $p$-value is $0.000$. Thus, under the strong-stochastic-transitivity restrictions imposed by the nonparametric link model, the observed pairwise win probabilities are inconsistent with the PageRank ordering.

The unrestricted matrix suggests that the incompatibility may not be specific to the PageRank ordering. Among the $\binom{10}{3}=120$ triples of firms, 19 form Condorcet cycles. Specifically, these are triples of firms $\ell$, $k$, and $j$ for which $\hat{p}_{\ell,k}>1/2$, $\hat{p}_{k,j}>1/2$, and $\hat{p}_{j,\ell}>1/2$. If these sample probabilities accurately reflect their population counterparts, no ordering of the three firms can rationalize all pairwise comparisons simultaneously. These cycles are descriptive features of the sample matrix and do not provide a formal test that the population identified set is empty. Nevertheless, they suggest that the rejection is not driven solely by the particular ordering produced by PageRank.

Appendix~\ref{appendix:empirical_robustness} reports a series of robustness exercises. We continue to reject the PageRank ordering when we separate its construction from its evaluation through sample splitting, use the alternative PageRank procedure proposed by \citet{corradini2023collective}, conduct the analysis separately by gender, and alter or substantially expand the set of firms. These results are relevant for applied work that interprets PageRank orderings derived from job-to-job transitions as revealed-preference rankings of latent firm amenities. Such an interpretation requires the PageRank ordering to rationalize pairwise win probabilities under strong stochastic transitivity, the structure imposed by the nonparametric link model. Our procedure tests this restriction directly and rejects it for every ranking we consider. PageRank may still summarize the transition network usefully, but in these data it cannot be read as an amenity ranking consistent with the observed flows.

\section{Conclusion} \label{sec:conclusion}

In this paper, we studied models for ranking objects based on latent merits using data from their pairwise interactions. Unlike much of the existing literature, we did not assume that all interactions were observed, and we accounted for potential non-randomness in the observed interactions. This approach reflects real-world applications in economics, where the tournament graph is sparse, and the edges are not randomly assigned.

We explored what can be inferred about the true population ranking depending on the structure of the tournament and the assumptions made about the link function. Under the nonparametric link model, which encodes strong stochastic transitivity, we showed that informative inference remains possible and sharply characterized the identified set for the ranking. We also demonstrated how this characterization can be used to conduct formal statistical tests to determine whether a ranking is consistent with the observed data. Illustrating the procedure with Brazilian job-to-job transitions, we rejected the hypothesis that the PageRank firm ordering belongs to the identified set.

Some problems remain open for future research. First, it would be interesting to expand the model by incorporating covariates or other available data in practical scenarios, to study their influence on the ranking. Second, it seems important to develop testing procedures that target more specific null hypotheses, ensuring both computational feasibility and statistical power. Our identification results, along with the potential for partial identification, provide a foundation for these future advancements.

\newpage
\bibliography{references}

\newpage
\appendix

\section{Proofs}

\setcounter{proposition}{0}
\setcounter{theorem}{0}
\setcounter{lemma}{0}
\setcounter{corollary}{0}

\subsection*{Theorem \ref{th:ident_param}}

\begin{theorem}
    {\normalfont (Point Identification in the Linear Parametric Model)}
    Let Assumption \ref{as:dgp} hold with $F_0 \in \mathbf{F}_L$ known. For a tournament graph $G$ and $F_0$, the ranking is point identified if and only if $G$ is connected.
\end{theorem}

\begin{proof}
    By Definition \ref{def:linear}, there exists a known, strictly increasing function
    $f:\mathbb{R}\mapsto(0,1)$ such that
    \begin{gather}
        F_0(x,y)=f(y-x),
        \qquad
        \forall\ (x,y)\in\mathbb{R}^2.
        \label{eq:proof_parametric_link}
    \end{gather}
    We prove sufficiency and necessity separately.

    \medskip
    \noindent
    \textit{Sufficiency.}
    Suppose that $G$ is connected. Fix any permissible distribution
    $P\in\mathbf{P}(G,\{F_0\})$, and let
    $\boldsymbol{\theta},\widetilde{\boldsymbol{\theta}}
    \in\Theta(G,\{F_0\},P)$ be two observationally equivalent collections
    of merits. By the definition of observational equivalence, for every
    $(\ell,k)\in E$,
    \begin{align}
        F_0(\theta_\ell,\theta_k)
        &=
        P(Y_{\ell,k}=1)
        =
        F_0(\widetilde{\theta}_\ell,\widetilde{\theta}_k).
        \label{eq:proof_parametric_equal_prob}
    \end{align}
    Using \eqref{eq:proof_parametric_link}, Equation
    \eqref{eq:proof_parametric_equal_prob} becomes
    \begin{align}
        f(\theta_k-\theta_\ell)
        &=
        f(\widetilde{\theta}_k-\widetilde{\theta}_\ell),
        \qquad
        \forall\ (\ell,k)\in E.
        \label{eq:proof_parametric_equal_f}
    \end{align}
    Since $f$ is strictly increasing and therefore injective,
    \begin{align}
        \theta_k-\theta_\ell
        &=
        \widetilde{\theta}_k-\widetilde{\theta}_\ell,
        \qquad
        \forall\ (\ell,k)\in E.
        \label{eq:proof_parametric_equal_differences}
    \end{align}

    Define
    \begin{gather*}
        d_\ell
        \coloneqq
        \widetilde{\theta}_\ell-\theta_\ell,
        \qquad
        \ell\in[q].
    \end{gather*}
    Equation \eqref{eq:proof_parametric_equal_differences} implies that
    \begin{align}
        d_k-d_\ell
        &=
        (\widetilde{\theta}_k-\widetilde{\theta}_\ell)
        -
        (\theta_k-\theta_\ell)
        =
        0,
        \qquad
        \forall\ (\ell,k)\in E.
        \label{eq:proof_parametric_equal_shifts}
    \end{align}
    Hence $d_\ell=d_k$ whenever $\ell$ and $k$ are connected by an edge.

    Because $G$ is connected, for every pair of teams $\ell,k\in[q]$ there
    exists a path
    \begin{gather*}
        \ell=i_0,i_1,\ldots,i_m=k
    \end{gather*}
    such that $(i_{s-1},i_s)\in E$ for every $s=1,\ldots,m$. Applying
    \eqref{eq:proof_parametric_equal_shifts} along this path gives
    \begin{gather*}
        d_\ell=d_{i_1}=\cdots=d_{i_m}=d_k.
    \end{gather*}
    Therefore, there exists some constant $c\in\mathbb{R}$ such that
    \begin{gather}
        \widetilde{\boldsymbol{\theta}}
        =
        \boldsymbol{\theta}
        +
        c\boldsymbol{1},
        \label{eq:proof_parametric_location}
    \end{gather}
    where $\boldsymbol{1}$ denotes the $q$-dimensional vector of ones.

    Adding a common constant to all latent merits does not change their
    ordering. Thus,
    \begin{gather*}
        \vr(\widetilde{\boldsymbol{\theta}})
        =
        \vr(\boldsymbol{\theta}).
    \end{gather*}
    Since the two observationally equivalent merit vectors were arbitrary,
    every element of $\Theta(G,\{F_0\},P)$ induces the same ranking.
    Consequently,
    \begin{gather*}
        R(G,\{F_0\},P)
    \end{gather*}
    is a singleton. Since this holds for every
    $P\in\mathbf{P}(G,\{F_0\})$, the ranking is point identified.

    \medskip
    \noindent
    \textit{Necessity.}
    Necessity is straightforward: if there are two components in the tournament graph that are not connected, then nothing can be learnt about the ranking between any two teams from the different components.

    Combining the sufficiency and necessity arguments proves the result.
\end{proof}

\subsection*{Theorem \ref{th:sharp_ranking_general}}

\begin{theorem}
    {\normalfont (Identification in the Nonparametric Link Model)}
    Let Assumption \ref{as:dgp} hold with $F_0 \in \mathbf{F}$, where $F_0$ is unknown, and consider some $\boldsymbol{r}:=(r_1,\dots,r_q)'\in\mathsf{R}$. Then:
    \begin{enumerate}[label=(\alph*)]
        \item $\boldsymbol{r}\in R(G,\mathbf{F},P)$ if and only if the following three conditions hold:
        \begin{enumerate}[label=(\roman*)]
            \item For all $(\ell,k)\in E$,
            \begin{gather*}
                \operatorname{sign}\left(
                    \expect_P\left[Y_{\ell,k}-\frac{1}{2}\right]
                \right)
                =
                \operatorname{sign}(r_k-r_\ell).
            \end{gather*}

            \item For all $(i,k),(j,\ell)\in E$ with $r_i=r_j$,
            \begin{gather*}
                \operatorname{sign}\left(
                    \expect_P[Y_{i,k}-Y_{j,\ell}]
                \right)
                =
                \operatorname{sign}(r_k-r_\ell).
            \end{gather*}

            \item For all $(i,k),(j,\ell)\in E$ with $r_i>r_j$,
            \begin{gather*}
                r_\ell\geq r_k
                \quad\Longrightarrow\quad
                \expect_P[Y_{i,k}]
                <
                \expect_P[Y_{j,\ell}].
            \end{gather*}
        \end{enumerate}

        \item The ranking is point identified with respect to $G$ and $\mathbf{F}$ if and only if there is a path of length at most two between any two teams in the tournament graph.
    \end{enumerate}
\end{theorem}

\begin{proof}

\noindent
\textbf{Part (a).}

\medskip
\noindent
\textit{Necessity.}
Suppose that $\boldsymbol{r}\in R(G,\mathbf{F},P)$. Then there exist $\boldsymbol{\theta}\in\Theta(G,\mathbf{F},P)$ and
$F\in\mathbf{F}$ such that $\vr(\boldsymbol{\theta})=\boldsymbol{r}$ and, for all $(\ell,k)\in E$, $\expect_P[Y_{\ell,k}] = F(\theta_\ell,\theta_k)$.
For any $(\ell,k)\in E$, symmetry implies $F(\theta_\ell,\theta_\ell)=\frac{1}{2}$.
Since $F$ is strictly increasing in its second argument,
\begin{align*}
    \operatorname{sign}\left(
        \expect_P\left[Y_{\ell,k}-\frac{1}{2}\right]
    \right)
    &=
    \operatorname{sign}\left(
        F(\theta_\ell,\theta_k)
        -
        F(\theta_\ell,\theta_\ell)
    \right) \\
    &=
    \operatorname{sign}(\theta_k-\theta_\ell) \\
    &=
    \operatorname{sign}(r_k-r_\ell).
\end{align*}
Therefore, \eqref{eq:dir_games} holds.
Next, consider $(i,k),(j,\ell)\in E$ such that $r_i=r_j$. Then
$\theta_i=\theta_j$, and hence
\begin{align*}
    \operatorname{sign}\left(
        \expect_P[Y_{i,k}-Y_{j,\ell}]
    \right)
    &=
    \operatorname{sign}\left(
        F(\theta_i,\theta_k)
        -
        F(\theta_i,\theta_\ell)
    \right) \\
    &=
    \operatorname{sign}(\theta_k-\theta_\ell) \\
    &=
    \operatorname{sign}(r_k-r_\ell).
\end{align*}
Therefore, \eqref{eq:indir_games_tie} holds.
Finally, consider $(i,k),(j,\ell)\in E$ such that $r_i>r_j$ and $r_\ell\geq r_k$. Then $\theta_i>\theta_j$ and $\theta_\ell\geq\theta_k$.
By strict monotonicity in the first argument and monotonicity in the second,
\begin{align*}
    \expect_P[Y_{i,k}]
    &=
    F(\theta_i,\theta_k) \\
    &<
    F(\theta_j,\theta_k) \\
    &\leq
    F(\theta_j,\theta_\ell) \\
    &=
    \expect_P[Y_{j,\ell}].
\end{align*}
Therefore, \eqref{eq:indir_games} holds.

\medskip
\noindent
\textit{Sufficiency.}
Suppose that $\boldsymbol{r}\in\mathsf{R}$ satisfies
\eqref{eq:dir_games}--\eqref{eq:indir_games}. For every observed interaction, let
$p_{\ell,k}\coloneqq\expect_P[Y_{\ell,k}]$, and define the probability for the reverse orientation as $p_{k,\ell}\coloneqq1-p_{\ell,k}$.

Consider the merit vector defined by $\theta_\ell\coloneqq r_\ell$ for every $\ell\in[q]$. Since $\boldsymbol{r}\in\mathsf{R}$, $\vr(\boldsymbol{\theta})=\boldsymbol{r}$.

Let
\[
  D \coloneqq \bigl\{ (r_\ell, r_k) \ :\ (\ell,k) \in E \bigr\},
\]
where each unordered edge contributes \emph{both} orientations $(r_\ell,r_k)$ and $(r_k,r_\ell)$, and
define $\widetilde F : D \to (0,1)$ by $\widetilde F(r_\ell, r_k) \coloneqq p_{\ell,k}$.
$\widetilde F$ is well defined. If $(i,k),(j,\ell) \in E$ generate the same point, i.e.\ $r_i = r_j$ and
$r_k = r_\ell$, then \eqref{eq:indir_games_tie} gives
\begin{gather*}
\text{sign}(p_{i,k} - p_{j,\ell}) = \text{sign}(r_k - r_\ell) = 0,
\end{gather*}
hence $p_{i,k} = p_{j,\ell}$; the same argument applied to the oriented pairs $(i,k)$ and $(\ell,j)$
covers the case in which the two orientations of two different edges coincide.
By construction $\widetilde F$ is antisymmetric on $D$:
\begin{equation}\label{eq:antisym-D}
  \widetilde F(x,y) + \widetilde F(y,x) = 1 \qquad \text{for every } (x,y) \in D ,
\end{equation}
and, in particular, $\widetilde F(a,a) = 1/2$ whenever $(a,a) \in D$, which is also implied directly by
\eqref{eq:dir_games}.



Conditions \eqref{eq:dir_games}--\eqref{eq:indir_games} also imply that
$\widetilde{F}$ is strictly decreasing in its first argument and strictly
increasing in its second argument on its domain. In particular, for any
two points $(x,y)$ and $(x',y')$ in its domain,
\begin{gather}
    x\geq x',
    \qquad
    y\leq y',
    \qquad
    (x,y)\neq(x',y')
    \quad\Longrightarrow\quad
    \widetilde{F}(x,y)
    <
    \widetilde{F}(x',y').
\label{eq:proof_monotone_partial_F}
\end{gather}

Set $\widetilde G(x,y) \coloneqq \widetilde F(-x,y)$ on $-D \coloneqq \{(-x,y) : (x,y) \in D\}$. By \eqref{eq:proof_monotone_partial_F},
$\widetilde G$ is strictly increasing with respect to the componentwise order on $\mathbb{R}^2$. Since ranks take
values in $\{1,\dots,q\}$, the set $D$, and hence $-D$, is finite, so it is compact and $\widetilde G$ is
trivially continuous on it.
Let $h : (0,1) \to \mathbb{R}$ be an increasing homeomorphism. Then $h \circ \widetilde G : -D \to \mathbb{R}$ is
continuous and strictly increasing, so Corollary~2 of \citet{husseinov2010monotonic} provides a continuous,
strictly increasing $H : \mathbb{R}^2 \to \mathbb{R}$ with $H = h \circ \widetilde G$ on $-D$. Define
$G \coloneqq h^{-1} \circ H$ and
\[
  F^{\mathrm{e}}(x,y) \coloneqq G(-x,y), \qquad (x,y) \in \mathbb{R}^2 .
\]
Then $F^{\mathrm{e}}$ is continuous, takes values in $(0,1)$, is strictly decreasing in its first argument
and strictly increasing in its second, and satisfies $F^{\mathrm{e}} = \widetilde F$ on $D$.


Since $F^{\mathrm{e}} > 0$ everywhere, we may define
\begin{equation}\label{eq:symmetrization}
  F(x,y) \coloneqq \frac{F^{\mathrm{e}}(x,y)}{F^{\mathrm{e}}(x,y) + F^{\mathrm{e}}(y,x)},
  \qquad (x,y) \in \mathbb{R}^2 .
\end{equation}
We verify that $F \in \mathbf{F}$ and that $F$ reproduces the observed probabilities.
\emph{(i) Range.} $F^{\mathrm{e}}(x,y), F^{\mathrm{e}}(y,x) \in (0,1)$ implies $F(x,y) \in (0,1)$.
\emph{(ii) Condition 1 of Definition \ref{def:func}.} Directly from \eqref{eq:symmetrization},
$F(x,y) + F(y,x) = 1$ for all $(x,y) \in \mathbb{R}^2$; in particular $F(x,x) = 1/2$.
\emph{(iii) Condition 2 of Definition \ref{def:func}.} Fix $x$ and take $y_1 < y_2$. Write
$A_m \coloneqq F^{\mathrm{e}}(x,y_m)$ and $B_m \coloneqq F^{\mathrm{e}}(y_m,x)$. Strict monotonicity of
$F^{\mathrm{e}}$ gives $A_1 < A_2$ and $B_1 > B_2$, all four numbers being strictly positive, so
$A_1 B_2 < A_2 B_1$ and therefore
\[
  F(x,y_1) = \frac{A_1}{A_1 + B_1} < \frac{A_2}{A_2 + B_2} = F(x,y_2).
\]
Hence $F$ is strictly increasing in its second argument on all of $\mathbb{R}^2$. The same computation with
$x_1 < x_2$ and $y$ fixed gives $A_1 > A_2$, $B_1 < B_2$, hence $A_1 B_2 > A_2 B_1$ and
$F(x_1,y) > F(x_2,y)$, so $F$ is strictly decreasing in its first argument.
\emph{(iv) Fit.} For $(\ell,k) \in E$ we have $(r_\ell,r_k), (r_k,r_\ell) \in D$, so by
\eqref{eq:antisym-D},
\[
  F(r_\ell, r_k)
  = \frac{\widetilde F(r_\ell,r_k)}{\widetilde F(r_\ell,r_k) +  \widetilde F(r_k,r_\ell)}
  = \frac{p_{\ell,k}}{p_{\ell,k} + (1 - p_{\ell,k})}
  = p_{\ell,k} = E_P[Y_{\ell,k}] .
\]
Therefore, $\boldsymbol{\theta}\in\Theta(G,\mathbf{F},P)$. Since
$\vr(\boldsymbol{\theta})=\boldsymbol{r}$,
\begin{gather*}
    \boldsymbol{r}\in R(G,\mathbf{F},P).
\end{gather*}
This proves Part (a).

\bigskip
\noindent
\textbf{Part (b).}

\medskip
\noindent
\textit{Sufficiency.}
Suppose that there is a path of length at most two between every pair of
teams. Fix any $P\in\mathbf{P}(G,\mathbf{F})$ and any two teams
$\ell,k\in[q]$.

If $(\ell,k)\in E$, their relative ranking is determined by
\begin{gather*}
    \operatorname{sign}\left(
    \expect_P\left[Y_{\ell,k}-\frac{1}{2}\right]
    \right).
\end{gather*}

If $(\ell,k)\notin E$, there exists a team $i$ such that
$(i,\ell),(i,k)\in E$. For any observationally equivalent representation,
\begin{align*}
    \operatorname{sign}\left(
    \expect_P[Y_{i,\ell}-Y_{i,k}]
    \right)
    &=
    \operatorname{sign}\left(
    F(\theta_i,\theta_\ell)
    -
    F(\theta_i,\theta_k)
    \right) \\
    &=
    \operatorname{sign}(\theta_\ell-\theta_k) \\
    &=
    \operatorname{sign}(r_\ell-r_k).
\end{align*}
Thus, the distribution of the observed data determines the relative
ranking of every pair of teams and therefore determines the full ranking.
Hence $R(G,\mathbf{F},P)$ is a singleton.

\medskip
\noindent
\textit{Necessity.}
Suppose that there exist two teams $\ell$ and $k$ for which there is no
path of length at most two. Therefore, $(\ell,k)\notin E$, and $\ell$ and
$k$ have no common opponent.

Choose any $F_0\in\mathbf{F}_L\subseteq\mathbf{F}$ and a merit vector
$\boldsymbol{\theta}_0$ such that
$\theta_{0,\ell}=\theta_{0,k}=0$, while the merits of all other teams are
distinct and strictly positive. Let $P$ be the distribution generated by
$(F_0,\boldsymbol{\theta}_0)$.

Consider the two rankings obtained by breaking the tie between $\ell$ and
$k$ in either direction, while leaving the ordering of all other teams
unchanged. We show that both rankings satisfy the conditions in Part (a).

Condition \eqref{eq:dir_games} holds under both rankings because $\ell$ and
$k$ do not interact and have the same ordering relative to every other
team. Condition \eqref{eq:indir_games_tie} also holds. In particular,
distinguishing between $\ell$ and $k$ as opponents of the same team would
require them to have a common opponent, which they do not.

Finally, consider condition \eqref{eq:indir_games}. Any restriction affected
by breaking the tie compares observed interactions involving $\ell$ and
$k$. If they appear as the first arguments, their merits are equal, and
the required strict inequality follows from the distinct ordering of
their opponents. If they appear as the second arguments, the required
strict inequality follows from the distinct ordering of the first teams.
The absence of a common opponent ensures that the remaining teams are
different. All other restrictions are unchanged and follow from the
strict monotonicity of $F_0$.

Therefore, both rankings satisfy
\eqref{eq:dir_games}--\eqref{eq:indir_games}. By Part (a), both belong to
$R(G,\mathbf{F},P)$. Since the two rankings are different, the ranking is
not point identified.

Combining the two directions proves Part (b).

\end{proof}

\subsection*{Corollary \ref{cor:matrix_representation}}

\begin{corollary}
    {\normalfont (Matrix Representation of Identification Conditions)}
    Under the conditions of Theorem \ref{th:sharp_ranking_general}, fix
    $\boldsymbol{r}\in\mathsf{R}$ without ties, and form the $q\times q$
    matrix $A_{\boldsymbol{r}}$ whose rows and columns are ordered from best
    to worst under $\boldsymbol{r}$. Set every diagonal entry equal to
    $1/2$. For $i\neq j$, set
    $a_{i,j}=\expect_P[Y_{\ell,k}]$ if $(\ell,k)\in E$ and
    $(r_\ell,r_k)=(i,j)$, and leave $a_{i,j}$ empty otherwise. Then
    $\boldsymbol{r}\in R(G,\mathbf{F},P)$ if and only if, for all nonempty
    entries,
    \begin{align*}
        i\geq i',
        \quad
        j\leq j',
        \quad
        (i,j)\neq(i',j')
        \quad\Longrightarrow\quad
        a_{i,j}<a_{i',j'}.
    \end{align*}
\end{corollary}

\begin{proof}
    \textit{Necessity.}
    Suppose that $\boldsymbol{r}\in R(G,\mathbf{F},P)$. Consider two
    nonempty entries $a_{i,j}$ and $a_{i',j'}$ such that $i\geq i'$,
    $j\leq j'$, and at least one inequality is strict.

    Suppose first that both entries are off the diagonal. If $i=i'$, the
    absence of ties implies that the two entries correspond to the same
    team facing two differently ranked opponents. Since $j<j'$,
    \eqref{eq:indir_games_tie} implies $a_{i,j}<a_{i',j'}$. If $i>i'$,
    \eqref{eq:indir_games} and $j'\geq j$ imply
    $a_{i,j}<a_{i',j'}$.

    If $a_{i,j}$ is on the diagonal, then $a_{i,j}=1/2$ and $i'<j'$.
    By \eqref{eq:dir_games}, $a_{i',j'}>1/2$. Similarly, if
    $a_{i',j'}$ is on the diagonal, then $i>j$ and
    $a_{i,j}<1/2=a_{i',j'}$. Therefore, the matrix inequalities hold.

    \medskip
    \noindent
    \textit{Sufficiency.}
    Suppose that the matrix inequalities hold. We verify the three
    conditions in Theorem \ref{th:sharp_ranking_general}.

    For any $(\ell,k)\in E$, if $r_\ell>r_k$, comparison of
    $a_{r_\ell,r_k}$ with $a_{r_k,r_k}=1/2$ gives
    $\expect_P[Y_{\ell,k}]<1/2$. If $r_\ell<r_k$, comparison of
    $a_{r_\ell,r_\ell}=1/2$ with $a_{r_\ell,r_k}$ gives
    $\expect_P[Y_{\ell,k}]>1/2$. Thus, \eqref{eq:dir_games} holds.

    Next, since $\boldsymbol{r}$ has no ties, $r_i=r_j$ implies $i=j$.
    Comparing $a_{r_i,r_k}$ and $a_{r_i,r_\ell}$ gives
    \begin{align*}
        \operatorname{sign}\left(
            \expect_P[Y_{i,k}-Y_{j,\ell}]
        \right)
        &=
        \operatorname{sign}(r_k-r_\ell),
    \end{align*}
    so \eqref{eq:indir_games_tie} holds.

    Finally, if $r_i>r_j$ and $r_\ell\geq r_k$, the matrix inequalities
    directly imply
    \begin{align*}
        \expect_P[Y_{i,k}]
        &=
        a_{r_i,r_k} \\
        &<
        a_{r_j,r_\ell} \\
        &=
        \expect_P[Y_{j,\ell}].
    \end{align*}
    Therefore, \eqref{eq:indir_games} holds. Theorem
    \ref{th:sharp_ranking_general} then implies that
    $\boldsymbol{r}\in R(G,\mathbf{F},P)$.
\end{proof}

\subsection*{Proposition \ref{prop:closed_null}}

\begin{proposition}
    {\normalfont (Open and Closed Ranking Nulls)}
    Let Assumption \ref{as:dgp} hold with $F_0\in\mathbf{F}$, and suppose the latent merits are distinct.
    \begin{enumerate}[label=(\alph*)]
        \item Suppose the tournament graph has diameter at most two, and let $\boldsymbol{r}^*$ denote the ranking generated by the true latent merits. Write $\boldsymbol{p}$ for the probability vector generated by these merits and $F_0$. Then, for every candidate ranking $\boldsymbol{r}$ without ties such that $\boldsymbol{r}\neq\boldsymbol{r}^*$, $\boldsymbol{p}$ violates at least one inequality in \eqref{eq:H0c}. Consequently, among candidate rankings without ties, $\boldsymbol{r}^*$ is the only ranking whose probability vector satisfies the closed restrictions.

        \item Suppose the tournament graph is connected and has diameter greater than two. Then there exist a link function $F_0\in\mathbf{F}$, a vector of distinct latent merits $\boldsymbol{\theta}$, and a candidate ranking $\boldsymbol{r}'$ without ties such that, writing $P$ for the distribution generated by $(\boldsymbol{\theta},F_0)$,
        \begin{gather*}
            \boldsymbol{r}' \notin R(G,\mathbf{F},P),
        \end{gather*}
        while the corresponding probability vector $\boldsymbol{p}$ satisfies all the weak inequalities in \eqref{eq:H0c}.
    \end{enumerate}
\end{proposition}

\begin{proof}
   \textit{Part (a).}
    Consider any candidate ranking $\boldsymbol{r}\neq\boldsymbol{r}^*$. List the teams according to $\boldsymbol{r}$, from best to worst, and consider their ranks under $\boldsymbol{r}^*$. Since $\boldsymbol{r}\neq\boldsymbol{r}^*$ and merits are distinct, this sequence is a permutation of $\{1,\dots,q\}$ other than the identity and therefore contains an adjacent inversion. Hence, there exist teams $a$ and $b$ that are consecutive under $\boldsymbol{r}$ (no team is ranked between them under $\boldsymbol{r}$) and ordered in opposite directions under $\boldsymbol{r}$ and $\boldsymbol{r}^*$.

    Without loss of generality, let
    \begin{gather*}
        \theta_a < \theta_b,
        \qquad
        r_b < r_a.
    \end{gather*}
    Thus, the true ranking places $a$ above $b$, while the candidate ranking places $b$ above $a$. Since the tournament graph has diameter at most two, either $(a,b)\in E$, or $a$ and $b$ have a common neighbor.

    Suppose first that $(a,b)\in E$. Since $r_b<r_a$, $(a,b)\in\tilde{E}_{\boldsymbol{r}}$, so \eqref{eq:H0c} requires
    \begin{gather*}
        p_{a,b}\leq\frac{1}{2}.
    \end{gather*}
    However, $\theta_a<\theta_b$, and strict monotonicity of $F_0$ in its second argument implies
    \begin{gather*}
        p_{a,b}
        =
        F_0(\theta_a,\theta_b)
        >
        F_0(\theta_a,\theta_a)
        =
        \frac{1}{2}.
    \end{gather*}
    Hence the closed restrictions are violated.

    Suppose instead that $(a,b)\notin E$, and let $c$ be a common neighbor of $a$ and $b$. Since $a$ and $b$ are consecutive under $\boldsymbol{r}$, team $c$ must be ranked either above both or below both.

    If
    \begin{gather*}
        r_c < r_b < r_a,
    \end{gather*}
    then $(a,c),(b,c)\in\tilde{E}_{\boldsymbol{r}}$, and \eqref{eq:H0c} requires
    \begin{gather*}
        p_{a,c}\leq p_{b,c}.
    \end{gather*}
    But $\theta_a<\theta_b$, and strict monotonicity of $F_0$ in its first argument implies
    \begin{gather*}
        p_{a,c}
        =
        F_0(\theta_a,\theta_c)
        >
        F_0(\theta_b,\theta_c)
        =
        p_{b,c}.
    \end{gather*}
    Hence the restriction is violated.

    If
    \begin{gather*}
        r_b < r_a < r_c,
    \end{gather*}
    then $(c,a),(c,b)\in\tilde{E}_{\boldsymbol{r}}$, and \eqref{eq:H0c} requires
    \begin{gather*}
        p_{c,b}\leq p_{c,a}.
    \end{gather*}
    Since $\theta_a<\theta_b$, strict monotonicity of $F_0$ in its second argument implies
    \begin{gather*}
        p_{c,b}
        =
        F_0(\theta_c,\theta_b)
        >
        F_0(\theta_c,\theta_a)
        =
        p_{c,a}.
    \end{gather*}
    Thus, the restriction is again violated.

    In every case, any candidate ranking different from $\boldsymbol{r}^*$ violates at least one inequality in \eqref{eq:H0c}. Since $\boldsymbol{r}^*\in R(G,\mathbf{F},P)$, its probability vector satisfies the strict restrictions in \eqref{eq:H0}, and therefore also the weak restrictions in \eqref{eq:H0c}. This proves Part (a).

    \medskip
    \noindent
    \textit{Part (b).}
    Since the tournament graph is connected and has diameter greater than two, there exist teams $a$ and $b$ such that their graph distance is greater than two. Therefore, $(a,b)\notin E$, and $a$ and $b$ have no common neighbor. Moreover, connectedness implies that both $N_G(a)$ and $N_G(b)$ are nonempty.

    Consider the BTL link
    \begin{gather*}
        F_0(x,y)
        =
        f(y-x),
        \qquad
        f(t)
        =
        \frac{\exp(t)}{1+\exp(t)}.
    \end{gather*}
    Fix some $\varepsilon\in(0,1/2)$ and set
    \begin{gather*}
        \theta_a=0,
        \qquad
        \theta_b=\varepsilon.
    \end{gather*}
    Since $N_G(a)$ and $N_G(b)$ are disjoint, assign distinct merits to the teams in $N_G(a)$ inside the interval $(1,2]$, with some team $k\in N_G(a)$ satisfying $\theta_k=2$. Assign distinct merits to the teams in $N_G(b)$ inside $[2+\varepsilon,3)$, with some team $\ell\in N_G(b)$ satisfying
    \begin{gather*}
        \theta_\ell=2+\varepsilon.
    \end{gather*}
    Assign all remaining teams distinct merits larger than $3$. The resulting merit vector has no ties.

    Let $\boldsymbol{r}^*$ be the ranking induced by $\boldsymbol{\theta}$. Teams $a$ and $b$ occupy the first two positions, with $a$ ranked above $b$. Let $\boldsymbol{r}'$ be the candidate ranking obtained by swapping only $a$ and $b$, so that $b$ is ranked first and $a$ second.

    Since $a$ and $b$ do not interact and have no common neighbor, every direct restriction under $\boldsymbol{r}'$ is the same as under $\boldsymbol{r}^*$. Moreover, because $a$ and $b$ remain ranked above every other team, swapping their first two positions does not change the ordering restrictions between an edge involving only one of $a$ or $b$ and an edge that involves neither. The only restrictions that can change compare an interaction involving $a$ with an interaction involving $b$.

    Consider any $u\in N_G(a)$ and $v\in N_G(b)$. By construction,
    \begin{gather*}
        \theta_u\leq2,
        \qquad
        \theta_v\geq2+\varepsilon,
    \end{gather*}
    and hence
    \begin{gather*}
        \theta_v-\theta_u\geq\varepsilon.
    \end{gather*}
    Under $\boldsymbol{r}'$, the relevant weak restriction in \eqref{eq:H0c} is
    \begin{gather*}
        p_{v,b}\leq p_{u,a}.
    \end{gather*}
    Using the BTL link,
    \begin{align*}
        p_{v,b}
        &=
        f(\theta_b-\theta_v) \\
        &=
        f(\varepsilon-\theta_v) \\
        &\leq
        f(-\theta_u) \\
        &=
        f(\theta_a-\theta_u) \\
        &=
        p_{u,a},
    \end{align*}
    where the inequality follows from $\theta_v-\theta_u\geq\varepsilon$ and strict monotonicity of $f$. Thus, all the weak restrictions created by the swap are satisfied.

    For the particular teams $k$ and $\ell$ selected above,
    \begin{gather*}
        \theta_\ell-\theta_k=\varepsilon,
    \end{gather*}
    and therefore
    \begin{align*}
        p_{\ell,b}
        &=
        f(\theta_b-\theta_\ell) \\
        &=
        f\big(\varepsilon-(2+\varepsilon)\big) \\
        &=
        f(-2) \\
        &=
        f(\theta_a-\theta_k) \\
        &=
        p_{k,a}.
    \end{align*}
    Hence one of the inequalities in \eqref{eq:H0c} binds. The corresponding strict inequality in \eqref{eq:H0} fails, so Corollary~\ref{cor:matrix_representation} implies
    \begin{gather*}
        \boldsymbol{r}'\notin R(G,\mathbf{F},P).
    \end{gather*}
    At the same time, all the weak restrictions in \eqref{eq:H0c} are satisfied. This proves Part (b).
\end{proof}

\section{Inference for Moment Inequalities Literature} \label{appendix:moment_inequ}

The null hypothesis of our test (the closed version $\overline{H}_0$ of $H_0:\ \boldsymbol{r}\in R(G,\mathbf{F},P)$) is equivalent to a collection of moment inequalities, reported in \eqref{eq:H0c}. A natural question is how this null hypothesis relates to the null hypotheses considered in the literature on testing moment inequalities. In this appendix, we will show that they differ, and there is no one-to-one mapping between the two problems.

We will consider the problem of testing a set of moment inequalities as presented by \cite{canay2017practical}. They indicate the identified set by $\Theta_0(P) = \{ \theta \in \Theta: \expect_P[m(W_i,\theta)] \leq 0 \}$, and then, for any value of $\theta$, they consider the test for the null $H_\theta: \expect_P[m(W_i,\theta)] \leq 0$. $\theta$ is the parameter of interest, and $W_i$ is $k$-dimensional random vector where it is typically assumed that $\{W_i\}_{i=1}^n$ is an iid collection.

In our case, the parameter of interest is $\boldsymbol{r}$ instead of $\theta$, and the data are $\{\{Y_{\ell,k,i}\}_{i=1}^{n_{\ell, k}} \}_{(\ell, k) \in E}$. Recall that $E$ is the set of all pairs $(\ell,k)$ of interacting teams. Define the set $E^2 \coloneqq \{ ((\ell,k),(u,v)): (\ell,k), (u,v) \in E \}$.

Let $\boldsymbol{Y}_{\ell, k} \coloneqq (Y_{\ell, k, i})_{i=1}^{n_{\ell, k}}$; define the functions $m_{\ell,k}$ and $m_{\ell,k,u,v}$:
\begin{align*}
    m_{\ell,k}(\boldsymbol{Y}_{\ell,k}, \boldsymbol{r}) \coloneqq & \bigg( \frac{1}{n_{\ell,k}} \sum_{i=1}^{n_{\ell,k}} Y_{\ell,k,i} - \frac{1}{2} \bigg) \ind \{ r_k \leq r_\ell \} \\
    m_{\ell,k,u,v}(\boldsymbol{Y}_{\ell,k}, \boldsymbol{Y}_{u,v}, \boldsymbol{r}) \coloneqq & \bigg( \frac{1}{n_{\ell,k}} \sum_{i=1}^{n_{\ell,k}} Y_{\ell,k,i} - \frac{1}{n_{u,v}} \sum_{j=1}^{n_{u,v}} Y_{u,v,j} \bigg) \ind \{ r_k \leq r_v \leq r_u \leq r_\ell \}.
\end{align*}

For any $\boldsymbol{r}$, the null hypothesis collects the inequalities
\begin{align*}
    H_{\boldsymbol{r}}: \ & \expect_P[m_{\ell,k}(\boldsymbol{Y}_{\ell,k}, \boldsymbol{r})] \leq 0 \\
    & \expect_P[m_{\ell,k,u,v}(\boldsymbol{Y}_{\ell,k}, \boldsymbol{Y}_{u,v}, \boldsymbol{r})] \leq 0
\end{align*}
for any $(\ell,k) \in E$ and for any $((\ell,k), (u,v)) \in E^2$.
This formulation of the null hypothesis is similar to that in \cite{canay2017practical}, but with a key distinction. Our data, $\{\{Y_{\ell,k,i}\}_{i=1}^{n_{\ell, k}} \}_{(\ell, k) \in E}$, cannot be viewed as an iid collection of $n$-element $k$-dimensional random vectors unless $n_{\ell, k} = n_{u, v}$ for all $(\ell, k), (u, v) \in E$. The issue is that we do not assume each pair of interacting teams plays the same number of games. Without this condition, the results presented in Section 4.1.1 by \cite{canay2017practical} regarding the least favorable distribution no longer hold.

\section{Additional Results for Inference}\label{appendix:inference}

Here, we extend and formalize the discussion of likelihood-based inference presented in Section \ref{sec:inference}. Define

\begin{gather*}
    \boldsymbol{P}_0(\boldsymbol{r}):= \{P:\ \text{the probability vector }\boldsymbol{p}\text{ of }P\text{ satisfies }\overline{H}_0\text{ in }\eqref{eq:H0c}\}, \\
    \boldsymbol{P}_1(\boldsymbol{r}):= \{P \in \mathbf{P}(G, \mathbf{F}):\ \text{the probability vector }\boldsymbol{p}\text{ of }P\text{ violates }\overline{H}_0\text{ in }\eqref{eq:H0c}\}.
\end{gather*}

Thus, $\boldsymbol{P}_0(\boldsymbol{r})$ is the closed null used for size control, while $\boldsymbol{P}_1(\boldsymbol{r})$ collects permissible distributions that violate at least one of the weak restrictions in \eqref{eq:H0c}. Since the hypothesis of interest $H_0$ implies $\overline{H}_0$, validity over $\boldsymbol{P}_0(\boldsymbol{r})$ implies validity for $ H_0$.

To test the hypothesis of interest $H_0$, we work with the closed null $\overline{H}_0$ and construct a test function $\phi^{FS}_{\alpha}(\boldsymbol{Y}_{\vn})$,
\begin{gather*}
    \phi^{FS}_{\alpha}(\boldsymbol{Y}_{\vn}) = \begin{cases}
        1, \text{ if rejects}\\
        0, \text{ else}
    \end{cases}
\end{gather*}
that delivers a test valid in finite samples (Theorem \ref{th:validity}):
\begin{align}\label{eq:test_coverage}
    E_P[\phi^{FS}_{\alpha}(\boldsymbol{Y}_{\vn})] \le \alpha,\ \forall\ P \in \boldsymbol{P}_0(\boldsymbol{r}),\ \forall\ \vn.
\end{align}
However, the proposed construction of $\phi_{\alpha}^{FS}(\cdot)$ can become computationally hard when the tournament graph is large. For the latter case, we propose another test function $\phi^{AS}_{\alpha}(\cdot)$ that is easy to compute and that delivers a pointwise asymptotically valid test (Theorem \ref{th:asympt_validity}):
\begin{align*}
    \limsup_{N \to \infty} E_P[\phi_{\alpha}^{AS}(\boldsymbol{Y}_{\vn})] \le \alpha,\ \forall\ P \in \boldsymbol{P}_0(\boldsymbol{r})
\end{align*}
where $N = \sum_{(\ell, k) \in E} n_{\ell, k}$, and $\forall\ (\ell, k) \in E:\ n_{\ell, k}/N \to a_{\ell, k} \in (0, 1)$.

Additionally, we show that the test based on $\phi^{AS}_\alpha(\cdot)$ is consistent against any fixed permissible alternative whose probability vector violates the closed null (Theorem \ref{th:consistency}):
\begin{align}\label{eq:test_consistency}
    \liminf_{N \to \infty} E_P[\phi^{AS}_{\alpha}(\boldsymbol{Y}_{\vn})] = 1,\ \forall\ P \in \boldsymbol{P}_1(\boldsymbol{r}),
\end{align}
where $N = \sum_{(\ell, k) \in E} n_{\ell, k}$, and $\forall\ (\ell, k) \in E:\ n_{\ell, k}/N \to a_{\ell, k} \in (0, 1).$

\subsection{p-values}

The algorithm to compute $\pi_{\lambda}$ is provided below.

\begin{algorithm}[H]
    \caption{Compute p-value $\pi_{\lambda}$}
    \label{alg:p_value}
    \begin{algorithmic}[1]
        \STATE \textbf{Input:} $E$, $\{ n_{\ell,k} \}_{(\ell,k) \in E}$, $\lambda$
        \STATE \texttt{$\pi_{\lambda}(\boldsymbol{p}):= 0$}
        \FOR{$(i_j\:\ j \in E)$ in $\prod_{j \in E}\ \{0,\dots, n_j\}$}
        \STATE \texttt{$\hat{\boldsymbol{p}}:= (i_j/n_j\:\ j \in E)$}
        \IF{$\Lambda(\hat{\boldsymbol{p}}) \ge \lambda:$}
        \STATE \texttt{$\pi_{\lambda}(\boldsymbol{p}):= \pi_{\lambda}(\boldsymbol{p}) + \prod_{j \in E}\binom{n_j}{i_j}p_j^{i_j}(1-p_j)^{n_j-i_j}$}
        \ENDIF
        \ENDFOR
        \STATE \texttt{$\pi_{\lambda}:= \max\{\pi_{\lambda}(\boldsymbol{p}) \ s.t.\ \boldsymbol{p}\text{ satisfies }\eqref{eq:H0c}\}$}
        \STATE \textbf{Output:} p-value $\pi_{\lambda}$
    \end{algorithmic}
\end{algorithm}

As discussed in the main text, for a given value of the test statistic $\lambda$, the rejection probability $P(\Lambda(\boldsymbol{p}) \ge \lambda)$ can be expressed as a polynomial function of $\boldsymbol{p}$. The specific form of this polynomial depends on the number of games between each interacting pair of teams. Steps 2-8 of Algorithm \ref{alg:p_value} construct this polynomial by enumerating all possible interaction outcomes that result in larger realizations of the statistic. We use the Python package \verb|SymPy| for symbolic computation to execute steps 2-8, although any symbolic computation software could be used. In step 9, the resulting symbolic expression is converted into a numeric function, which is then optimized with respect to $\boldsymbol{p}$ subject to a set of linear constraints that define the closed null $\overline{H}_0$ for the tested ranking.

The approximate p-values discussed in Section \ref{sec:apprx_pvalue} can be computed by simulations, as illustrated by the following Algorithm \ref{alg:example}, where $\tilde{p}$ is the distribution defined in Section \ref{sec:apprx_pvalue} under which every inequality in \eqref{eq:H0c} binds.

\begin{algorithm}[H]
    \caption{Compute p-value $\tilde{\pi}_{\lambda}$}
    \label{alg:example}
    \begin{algorithmic}[1]
        \STATE \textbf{Input:} $m$, $E$, $\{ n_{\ell,k} \}_{(\ell,k) \in E}$, $\lambda$
        \FOR{$i = 1$ to $m$}
        \STATE For any $(\ell,k) \in E$, draw $n_{\ell,k}$ independent Bernoulli with probability $\tilde{p}_{\ell, k} = 0.5$. Store the simulated data in $\boldsymbol{Y_i}$
        \STATE Compute $\hat{\boldsymbol{p}}_i$
        \STATE Compute $\Lambda(\hat{\boldsymbol{p}}_i)$
        \ENDFOR
        \STATE Compute $\tilde{\pi}_{\lambda} = \frac{1}{m} \sum_{i=1}^m \ind \{ \Lambda(\hat{\boldsymbol{p}}_i) \ge \lambda \} $
        \STATE \textbf{Output:} p-value $\tilde{\pi}_{\lambda}$
    \end{algorithmic}
\end{algorithm}

\subsection{The Tests}
We consider two tests:
\begin{gather}
    \phi^{FS}_\alpha (\boldsymbol{Y}_{\vn}) = I\{ \pi_{\Lambda(\hat{\boldsymbol{p}})} \leq \alpha \}, \label{eq:test}\\
    \phi^{AS}_\alpha (\boldsymbol{Y}_{\vn}) = I\{ \tilde{\pi}_{\Lambda(\hat{\boldsymbol{p}})} \leq \alpha \}. \label{eq:test_asympt}
\end{gather}
Both tests reject the null hypothesis whenever the significance level $\alpha$ is larger than the p-value, which is computed based on the observed value of the test statistic $\Lambda(\hat{\boldsymbol{p}})$. The difference between the two tests lies in the p-value they use. Test $\phi^{FS}_\alpha(\cdot)$ computes p-value $\pi_{\Lambda(\hat{\boldsymbol{p}})}$ using Algorithm \ref{alg:p_value}, whereas test $\phi^{AS}(\cdot)$ uses $\tilde{\pi}_{\Lambda(\hat{\boldsymbol{p}})}$ computed by Algorithm \ref{alg:example}.

Test $\phi^{FS}_\alpha(\cdot)$ is finite sample valid: this follows directly from the definition of $\pi_\lambda$, and is formally stated in the following theorem. Recall that $\boldsymbol{P}_0(\boldsymbol{r})$ is the set of distributions satisfying the closed null $\overline{H}_0$ in \eqref{eq:H0c}.

\begin{theorem} \label{th:validity}
    {\normalfont (Finite-Sample Validity of the Exact Test)}
    Let Assumption \ref{as:dgp} hold with $F_0\in\mathbf{F}$. For any $\alpha\in(0,1)$, the test $\phi^{FS}_{\alpha}(\cdot)$ defined in \eqref{eq:test} satisfies \eqref{eq:test_coverage}, and is therefore finite-sample valid.
\end{theorem}

\begin{proof}
    By the definition of the test,
    \begin{align*}
        \sup_{P\in\boldsymbol{P}_0(\boldsymbol{r})}
        E_P\left[
            \phi^{FS}_{\alpha}(\boldsymbol{Y}_{\vn})
        \right]
        &=
        \sup_{P\in\boldsymbol{P}_0(\boldsymbol{r})}
        E_P\left[
            I\left\{
                \pi_{\Lambda(\hat{\boldsymbol{p}})}
                \leq\alpha
            \right\}
        \right] \\
        &=
        \sup_{P\in\boldsymbol{P}_0(\boldsymbol{r})}
        P\left\{
            \pi_{\Lambda(\hat{\boldsymbol{p}})}
            \leq\alpha
        \right\} \\
        &\leq
        \alpha.
    \end{align*}
    The last inequality follows from the standard property of p-values, since $\pi_{\lambda}$ is defined in \eqref{eq:pvalue} as the supremum of $P\{\Lambda\geq\lambda\}$ over all $P\in\boldsymbol{P}_0(\boldsymbol{r})$; see, for example, Lemma 3.3.1 in \citet{lehmann2022testing}. Therefore, \eqref{eq:test_coverage} holds for every $\vn$.
\end{proof}

As shown in the continuation of Example \ref{ex:1} below, test $\phi^{AS}_\alpha(\cdot)$ in general does not control size in finite samples, however, as we demonstrate in the following theorem, it is pointwise asymptotically valid when $N:= \sum_{(\ell, k) \in E} n_{\ell, k} \to \infty$, and
$n_{\ell, k}/N \to a_{\ell, k} \in (0, 1),\ \forall\ (\ell, k) \in E$.

\begin{theorem}\label{th:asympt_validity}

    {\normalfont (Pointwise Asymptotic Validity of the Approximate Test)}

    Let Assumption \ref{as:dgp} hold with $F_0 \in \mathbf{F}$. The test $\phi^{AS}_\alpha(\cdot)$ defined in \eqref{eq:test_asympt} is pointwise asymptotically valid:
    \begin{align*}
        \limsup_{N \to \infty} E_P[\phi^{AS}_\alpha (\boldsymbol{Y}_{\vn})] \le \alpha,\ \forall\ P \in \boldsymbol{P}_0(\boldsymbol{r})
    \end{align*}
    where $N = \sum_{(\ell, k) \in E } n_{\ell, k}$, and $\frac{n_{\ell, k}}{N} \to a_{\ell, k} \in (0, 1),\ \forall\ (\ell, k) \in E.$
\end{theorem}
\begin{proof}
    To prove the theorem, we will show that the distribution with $\boldsymbol{p}=0.5$ is the least favorable distribution when $N \to \infty$ (Lemma \ref{lemma:lfd}). To prove the lemma, we first need to introduce additional notation and definitions, and derive some intermediary results.

    It is useful to consider a different notation for the closed null hypothesis. Numerate elements in $\tilde{E}_{\boldsymbol{r}}$ arbitrarily, that is consider any one-to-one mapping $q: \tilde{E}_{\boldsymbol{r}} \mapsto \{1, 2, \dots, K\}, K = |\tilde{E}_{\boldsymbol{r}}|,$ and let $q(E_{\boldsymbol{r}}):= \{(q((j, \ell)), q((i, k))):\ ((j, \ell), (i, k)) \in E_{\boldsymbol{r}} \}.$ The closed null hypothesis can be equivalently stated as
    \begin{align}
        \overline{H}_0:\ &p_i - p_j \le 0,\ \forall\ (i, j) \in q(E_{\boldsymbol{r}}) \label{eq:isotone} \\
        &p_i \le \frac{1}{2},\ \forall\ i = 1, \dots, K. \label{eq:boundary}
    \end{align}
    where $Y_i \coloneqq Y_{q^{-1}(i)}, i=1, \dots, K$, and $p_i \coloneqq E_P[Y_i], i=1, \dots, K$.

    With this notation, if one observes independent realizations $\{\{Y_{i, j} \}_{j=1}^{n_i}\}_{i=1}^K$, the restricted ML estimator of $\boldsymbol{p}:= (p_1, \dots, p_K)'$ is defined as:
    \begin{gather*}
        \hat{\boldsymbol{p}}^*:= \underset{\substack{\tilde{p}_i - \tilde{p}_j \le 0,\ \forall\ (i, j) \in q(E_{\boldsymbol{r}})\\ \tilde{p}_i \le 1/2,\ \forall\ i \in [K]}}{\arg \max}\ \sum_{i=1}^K \sum_{j=1}^{n_i} \left( Y_{i, j} \log \tilde{p}_i + (1-Y_{i, j})\log(1 - \tilde{p}_i) \right),
    \end{gather*}
    and the unrestricted MLE is
    \begin{gather*}
        \hat{p}_i = \frac{1}{n_i}\sum_{j=1}^{n_i} Y_{i, j},\ \forall\ i = 1, \dots, K.
    \end{gather*}
    The test statistic $\Lambda$ is given by:
    \begin{align*}
        \Lambda = 2 \sum_{i=1}^K w_i \left(\hat{p}_i (\log \hat{p}_i - \log \hat{p}^*_i) + (1-\hat{p}_i) (\log (1-\hat{p}_i) - \log (1-\hat{p}^*_i))\right).
    \end{align*}

    It is also useful to introduce the definition of projection on closed convex set.
    \begin{definition}
        {\normalfont (Projection on closed convex set)}
        Let $D \subseteq \mathbb{R}^k$ be a closed convex subset of $\mathbb{R}^k$, and $\boldsymbol{y} \in \mathbb{R}^k$ be an arbitrary vector, $\boldsymbol{w} \in \mathbb{R}^k$ be a positive vector of weights. Let
        \begin{align*}
            \boldsymbol{y}^* = \underset{\tilde{\boldsymbol{y}} \in D}{\arg \min} \sum_{i=1}^k w_i(\tilde{y}_i - y_i)^2,
        \end{align*}
        whenever $\boldsymbol{y}^*$ exists it is called a projection of $\boldsymbol{y}$ on $D$ with weights $\boldsymbol{w}$. We denote $\boldsymbol{y}^* \equiv P_{\boldsymbol{w}}(\boldsymbol{y}|D).$
    \end{definition}

    Let $C$ be a collection of $K$ dimensional vectors that are isotonic with respect to a quasi-order induced by \eqref{eq:isotone}, that is $C:= \{\boldsymbol{x} \in \mathbb{R}^K:\ (i, j) \in q(E_{\boldsymbol{r}}) \implies x_i \le x_j \}$. Let $B_{\boldsymbol{b}}:= \{\boldsymbol{x} \in \mathbb{R}^K:\ x_i \le b_i,\ \forall\ i \in [K] \}$. Note that $B_{\boldsymbol{b}} \subseteq B_{\boldsymbol{b}'}$ whenever $0 \le \boldsymbol{b} \le \boldsymbol{b}'.$
    \cite{hu1997maximum} established the uniqueness and existence of $P_{\boldsymbol{w}}(\hat{\boldsymbol{p}}|C\ \bigcap\ B_{1/2})$ and that $\hat{\boldsymbol{p}}^* = P_{\boldsymbol{w}}(\hat{\boldsymbol{p}}|C\ \bigcap\ B_{1/2})$, where $\boldsymbol{w} = (n_1, \dots, n_K)'.$

    Before proving Lemma \ref{lemma:lfd}, consider the following results.
    \begin{lemma}
        Let $Y_{i, j} \sim Bern(p_i)$, a sample of independent realizations $\{\{Y_{i, j}\}_{j=1}^{n_i}\}_{i=1}^{k}$ is observed, and $\hat{p}_i:= \frac{1}{n_i} \sum_{j=1}^{n_i} Y_{i, j}.$ Then
        \begin{align*}
            \begin{pmatrix}
                \sqrt{n_1} (\hat{p}_1 - p_1)\\
                \dots\\
                \dots\\
                \dots\\
                \sqrt{n_k} (\hat{p}_k - p_k)
            \end{pmatrix} \overset{d}{\to} \mathcal{N} \left( \begin{pmatrix}
                0\\
                \dots\\
                \dots\\
                \dots\\
                0
            \end{pmatrix}, \begin{bmatrix}
                p_1(1-p_1) & 0 & \dots & \dots & 0\\
                 0 & p_2(1-p_2) & \dots & \dots & 0\\
                \dots & \dots & \dots & \dots & \dots \\
                \dots & \dots & \dots & \dots & \dots \\
                0 & 0 & 0 & 0 & p_k(1-p_k) \\
            \end{bmatrix} \right),\\
            \text{as }n_1 \to \infty, \dots, n_k \to \infty.
        \end{align*}
    \end{lemma}
    \begin{proof}
        By CLT for each $i = 1, \dots, k$ it follows that

        \begin{gather*}
            \sqrt{n_i}(\hat{p}_i - p_i) \overset{d}{\to} \mathcal{N}(0, p_i(1-p_i)).
        \end{gather*}
        Then we show that for any $(\lambda_1, \dots, \lambda_k) \in \mathbb{R}^k:$
        \begin{align*}
            \sum_{i=1}^k \lambda_i \sqrt{n}_i(\hat{p}_i - p_i) \overset{d}{\to} \sum_{i=1}^k \lambda_i Z_i, \text{as }n_1 \to \infty, \dots, n_k \to \infty.
        \end{align*}
        where $Z_i \sim \mathcal{N}(0, p_i(1-p_i))$ are jointly independent. The characteristic function of $\lambda_i \sqrt{n}_i(\hat{p}_i - p_i)$ is
        \begin{align*}
            \phi_i(t) = \left(1 - p_i + p_i e^{i \lambda_i \frac{t}{\sqrt{n_i}}} \right)^{n_i} e^{-it \lambda_i \sqrt{n_i}p_i},
        \end{align*}
        by independence, the characteristic function of $\sum_{i=1}^k \lambda_i \sqrt{n}_i(\hat{p}_i - p_i)$ is
        \begin{align*}
            \phi(t) = \prod_{i=1}^k \phi_i(t) = \prod_{i=1}^k \left(1 - p_i + p_i e^{i \lambda_i \frac{t}{\sqrt{n_i}}} \right)^{n_i} e^{-it \lambda_i \sqrt{n_i}p_i},
        \end{align*}
        then
        \begin{align*}
            \lim_{n_1 \to \infty, \dots, n_k \to \infty} \ln \phi(t) =\lim_{n_1 \to \infty, \dots, n_k \to \infty} \sum_{i=1}^k \left( n_i \ln \left(1 - p_i + p_i e^{i \lambda_i \frac{t}{\sqrt{n_i}}} \right) - i t \lambda_i \sqrt{n_i}p_i \right) = \\ =
            \sum_{i=1}^k \lim_{n_i \to \infty} \left( n_i \ln \left(1 - p_i + p_i e^{i \lambda_i \frac{t}{\sqrt{n_i}}} \right) - i t \lambda_i \sqrt{n_i}p_i \right).
        \end{align*}
        By Taylor expansion,
        \begin{align*}
            n_i \ln \left(1 - p_i + p_i e^{i \lambda_i \frac{t}{\sqrt{n_i}}} \right) - i t \lambda_i \sqrt{n_i}p_i = n_i \ln \left(1 - p_i + p_i \left(1 + i\lambda_i \frac{t}{\sqrt{n_i}} - \lambda_i^2\frac{t^2}{2 n_i} + o \left( \frac{t^2}{n_i} \right) \right) \right)-\\
            -i t \lambda_i \sqrt{n_i}p_i
            = n_i \ln \left( 1 + i \lambda_i p_i \frac{t}{\sqrt{n_i}} - p_i \lambda_i^2 \frac{t^2}{2 n_i} + o\left(\frac{t^2}{n_i} \right) \right)- i t \lambda_i \sqrt{n_i}p_i = \\ = n_i \left( i \lambda_i p_i \frac{t}{\sqrt{n_i}} - p_i \lambda_i^2 \frac{t^2}{2 n_i} + \lambda_i^2p_i^2\frac{t^2}{2 n_i} + o\left(\frac{t^2}{n_i} \right) \right)- i t \lambda_i \sqrt{n_i}p_i=\\=
            -p_i(1-p_i)\lambda_i^2 \frac{t^2}{2} + o(1), \text{ as } n_i \to \infty,
        \end{align*}
        which implies that
        \begin{align*}
            \lim_{n_i \to \infty} \left(1 - p_i + p_i e^{i \lambda_i \frac{t}{\sqrt{n_i}}} \right)^{n_i} e^{- i t \lambda_i \sqrt{n_i}p_i} = e^{-p_i(1-p_i)\lambda_i^2 \frac{t^2}{2}}.
        \end{align*}
        Notice that $e^{-p_i(1-p_i)\lambda_i^2 \frac{t^2}{2}}$ is a characteristic function of $\lambda_i\mathcal{N}(0, p_i(1-p_i))$, from this the result follows.
    \end{proof}
    \begin{corollary}\label{corol:as_normality}
        Let $N = \sum_{i=1}^k n_i$, and suppose that for each $i = 1, \dots, k:\ \frac{n_i}{N} \to a_i \in (0, 1)$ as $n_1 \to \infty, \dots, n_k \to \infty.$ Then
        \begin{align*}
            \sqrt{N}\begin{pmatrix}
                \hat{p}_1 - p_1\\
                \dots\\
                \dots\\
                \dots\\
                \hat{p}_k - p_k
            \end{pmatrix} \overset{d}{\to} \mathcal{N} \left( \begin{pmatrix}
                0\\
                \dots\\
                \dots\\
                \dots\\
                0
            \end{pmatrix}, \begin{bmatrix}
                \frac{p_1(1-p_1)}{a_1} & 0 & \dots & \dots & 0\\
                0 & \frac{p_2(1-p_2)}{a_2} & \dots & \dots & 0\\
                \dots & \dots & \dots & \dots & \dots \\
                \dots & \dots & \dots & \dots & \dots \\
                0 & 0 & 0 & 0 & \frac{p_k(1-p_k)}{a_k} \\
            \end{bmatrix} \right),\\
            \text{as }n_1 \to \infty, \dots, n_k \to \infty,
        \end{align*}
    \end{corollary}
    \begin{proof}
        This follows from:
        \begin{align*}
            \sqrt{N}\begin{pmatrix}
                \hat{p}_1 - p_1\\
                \dots\\
                \dots\\
                \dots\\
                \hat{p}_k - p_k
            \end{pmatrix} = \begin{pmatrix}
                \frac{\sqrt{n}}{\sqrt{n_1}}\\
                \dots \\
                \dots\\
                \dots\\
                \frac{\sqrt{n}}{\sqrt{n_k}}
            \end{pmatrix} \begin{pmatrix}
                \sqrt{n_1} (\hat{p}_1 - p_1)\\
                \dots\\
                \dots\\
                \dots\\
                \sqrt{n_k} (\hat{p}_k - p_k)
            \end{pmatrix} \overset{d}{\to} \\ \overset{d}{\to} \begin{pmatrix}
                \sqrt{\frac{1}{a_1}}\\
                \dots \\
                \dots\\
                \dots\\
                \sqrt{\frac{1}{a_k}}
            \end{pmatrix} \mathcal{N} \left( \begin{pmatrix}
                0\\
                \dots\\
                \dots\\
                \dots\\
                0
            \end{pmatrix}, \begin{bmatrix}
                p_1(1-p_1) & 0 & \dots & \dots & 0\\
                0 & p_2(1-p_2) & \dots & \dots & 0\\
                \dots & \dots & \dots & \dots & \dots \\
                \dots & \dots & \dots & \dots & \dots \\
                0 & 0 & 0 & 0 & p_k(1-p_k) \\
            \end{bmatrix} \right),\\
            \text{as }n_1 \to \infty, \dots, n_k \to \infty,
        \end{align*}
        where the last line follows from Slutsky's theorem.
    \end{proof}

    We can now prove that the distribution with $\boldsymbol{p}=0.5$ is the least favorable distribution when $N \to \infty$.
    \begin{lemma} \label{lemma:lfd}
        Let $\{Y_{i, j} \}_{j=1}^{n_i}$ for $i=1, \dots, K$ be independent samples from $Bern(p_i)$, let $n_i/N \to a_i \in (0, 1), i = 1, \dots, K$ where $N = \sum_{i=1}^K n_i$, then $\forall\ t > 0$ and for all $\boldsymbol{p}$ that satisfies \eqref{eq:isotone}, \eqref{eq:boundary}:
        \begin{align*}
            \limsup_{N \to \infty} P_{\boldsymbol{p}}(\Lambda \ge t) \le \lim_{N \to \infty} P_{1/2}(\Lambda \ge t)
        \end{align*}
    \end{lemma}
    \begin{proof}
        Expanding $\Lambda$ about $\hat{p}_i$ with a second-degree remainder term one obtains:
        \begin{align*}
            \Lambda = \sum_{i=1}^K n_i \left[ \frac{\hat{p}_i}{\alpha_i^2} + \frac{1-\hat{p}_i}{(1-\gamma_i)^2}\right] (\hat{p}_i^* - \hat{p}_i)^2,
        \end{align*}
        where $\alpha_i$ and $\gamma_i$ are between $\hat{p}_i^*$ and $\hat{p}$. Under the null hypothesis, $\hat{p}_i^* \overset{a.s.}{\to} p_i$, $\hat{p}_i \overset{a.s.}{\to} p_i$, hence $\alpha_i\overset{a.s.}{\to} p_i, \gamma_i \overset{a.s.}{\to} p_i.$ Consider the set $C_{\boldsymbol{p}}:= \{\boldsymbol{x} \in \mathbb{R}^K:\ (i, j) \in q(E_{\boldsymbol{r}})\ \text{and } p_i = p_j \implies x_i \le x_j \}$. Note that $C \subseteq C_{\boldsymbol{p}}$. Let $\eta_1 < \eta_2 < \dots < \eta_h < 1/2$ be the distinct values of $p_1, \dots, p_k$ and set $S_i = \{j:\ p_j = \eta_i \}$ for $i=1, \dots, h$, and $S_{1/2} = \{j:\ p_j = 1/2 \}$, with $S_{1/2}$ can potentially be empty. Since $\hat{p}_i \overset{a.s.}{\to} p_i$ for almost all $\omega$ in the underlying probability space and for sufficiently large $N$,
        \begin{align*}
            \underset{j \in S_1}{\max}\ \hat{p}_j < \underset{j \in S_2}{\min}\ \hat{p}_j \le \underset{j \in S_2}{\max}\ \hat{p}_j < \dots < \underset{j \in S_h}{\min}\ \hat{p}_j \le \underset{j \in S_h}{\max}\ \hat{p}_j < \min \left\{\frac{1}{2}, \underset{j \in S_{1/2}}{\min}\ \hat{p}_j \right\}
        \end{align*}
        with the convention that $\underset{\varnothing}{\min} \equiv 1.$

        Let $\bar{\boldsymbol{p}}:= P_{\boldsymbol{w}}(\hat{\boldsymbol{p}}|C_{\boldsymbol{p}} \bigcap B_{1/2})$, then for almost all $\omega$ in the underlying probability space and for sufficiently large $N$ it follows that $\bar{\boldsymbol{p}} \in C \bigcap B_{1/2}$. To see this, first, by definition $\bar{\boldsymbol{p}} \in B_{1/2}$; for $\bar{\boldsymbol{p}} \in C$ suppose the true hypothesis is $p_i \le p_j$ for some $i, j \in [K]$, then if $p_i < p_j < 1/2$ then $i \in S_u, j \in S_v$ with $u < v, u \neq 1/2, v \neq 1/2$ and hence $\bar{p}_i \le \underset{j \in S_{u}}{\max}\ \hat{p}_j < \underset{j \in S_{v}}{\min}\ \hat{p}_j \le \bar{p}_j;$ if $p_i < p_j = 1/2$ then $i \in S_u, j \in S_{1/2} \implies \bar{p}_i \le \underset{j \in S_{u}}{\max}\ \hat{p}_j \le \underset{j \in S_{h}}{\max}\ \hat{p}_j,$ further if $\underset{j \in S_{1/2}}{\min}\ \hat{p}_j < 1/2$ then $\underset{j \in S_{1/2}}{\min}\ \hat{p}_j \le \bar{p}_j \le 1/2$ and $\bar{p}_i \le \underset{j \in S_{u}}{\max}\ \hat{p}_j \le \underset{j \in S_{h}}{\max}\ \hat{p}_j < \underset{j \in S_{1/2}}{\min}\ \hat{p}_j \le \bar{p}_j$, if $\underset{j \in S_{1/2}}{\min}\ \hat{p}_j \ge 1/2$ then $\bar{p}_j = 1/2$, and again one has $\bar{p}_i \le \underset{j \in S_{u}}{\max}\ \hat{p}_j \le \underset{j \in S_{h}}{\max}\ \hat{p}_j < \frac{1}{2} = \bar{p}_j$; if $p_i = p_j$ then $i, j \in S_u$ for some $u = 1/2, 1, \dots, h$ $\implies \bar{p}_i \le \bar{p}_j$ by definition of projection on $C_{\boldsymbol{p}}$.

        Then, since $C\ \bigcap\ B_{1/2} \subseteq C_{\boldsymbol{p}}\ \bigcap B_{1/2}$ it follows that for almost all $\omega$ and sufficiently large $N$, $\bar{\boldsymbol{p}} = \hat{\boldsymbol{p}}^*$. Further,
        \begin{align*}
            \underset{\substack{\tilde{\boldsymbol{p}} \le 1/2,\\ i, j \in S_u\ \&\
            \hat{p}_i \le \hat{p}_j \implies \tilde{p}_i \le \tilde{p}_j} }{\min} \sum_{i=1}^K w_i(\tilde{p}_i - \hat{p}_i)^2 = \\
            =\underset{\substack{\tilde{\boldsymbol{p}} \le 1/2,\\ i, j \in S_u\ \&\
            \hat{p}_i \le \hat{p}_j \implies \tilde{p}_i \le \tilde{p}_j}}{\min} \sum_{S_j}\sum_{i \in S_j} w_i(\tilde{p}_i - \hat{p}_i)^2 =\\= \sum_{S_j} \underset{\substack{\tilde{\boldsymbol{p}} \le 1/2, \hat{p}_i \le \hat{p}_j\\ \implies \tilde{p}_i \le \tilde{p}_j}}{\min}\ \sum_{i \in S_j} w_i(\tilde{p}_i - \hat{p}_i)^2,
        \end{align*}
        hence subtracting a constant (on each $S_i$) or multiplying by a constant (on each $S_i$) does not change the projection. Thus,
        $P_{\boldsymbol{w}}(\hat{\boldsymbol{p}}|C_{\boldsymbol{p}}\ \bigcap\ B_{1/2}) - \boldsymbol{p} = P_{\boldsymbol{w}}\left(\hat{\boldsymbol{p}} - \boldsymbol{p}|C_{\boldsymbol{p}}\ \bigcap\ B_{1/2-\boldsymbol{p}}\right)$ and
        \begin{align*}
            \sqrt{\frac{N}{\boldsymbol{p}(1-\boldsymbol{p})}} \left( P_{\boldsymbol{w}}(\hat{\boldsymbol{p}}|C_{\boldsymbol{p}}\ \bigcap\ B_{1/2}) - \boldsymbol{p} \right) = P_{\boldsymbol{w}/N} \left(\sqrt{\frac{N}{\boldsymbol{p}(1-\boldsymbol{p})}}(\hat{\boldsymbol{p}} - \boldsymbol{p}) \bigg|\ C_{\boldsymbol{p}}\ \bigcap\ B_{\sqrt{\frac{N}{\boldsymbol{p}(1-\boldsymbol{p})}}\left(\frac{1}{2} - \boldsymbol{p} \right)} \right),
        \end{align*}
        where $\sqrt{\frac{N}{\boldsymbol{p}(1-\boldsymbol{p})}}:= \left( \sqrt{\frac{N}{p_1(1-p_1)}},\dots, \sqrt{\frac{N}{p_K(1-p_K)}}\right)'$.
        Denoting $\hat{\boldsymbol{g}} = \sqrt{\frac{N}{\boldsymbol{p}(1-\boldsymbol{p})}}(\hat{\boldsymbol{p}} - \boldsymbol{p})$,
        \begin{gather*}
            \hat{\boldsymbol{g}}^* = P_{\boldsymbol{w}/N}\left(\hat{\boldsymbol{g}}\ \bigg|\ C_{\boldsymbol{p}}\ \bigcap\ B_{\sqrt{\frac{N}{\boldsymbol{p}(1-\boldsymbol{p})}}\left(\frac{1}{2} - \boldsymbol{p} \right)}\right),
        \end{gather*}
        as shown by \cite{robertson1988order} and \cite{hu1997maximum}:
        \begin{align*}
            \hat{g}^*_i = \min\left\{ \underset{U: \hat{g}_i \in U}{\max}\ \underset{L:\ \hat{g}_i \in L}{\min}\ Av_{\boldsymbol{w}/N}\left(L\ \bigcap\ U\right),\ \sqrt{\frac{N}{p_i(1-p_i)}}\left(\frac{1}{2} - p_i \right) \right\},
        \end{align*}
        where $U$ is an upper set, that is a set $U$ such that $\hat{g}_i \in U, (i, j) \in q(E_{\boldsymbol{r}})\ \&\ p_i = p_j \implies \hat{g}_j \in U$, and $L$ is a lower set: $\hat{g}_i \in L,\ (j, i) \in q(E_{\boldsymbol{r}})\ \&\ p_i = p_j \implies \hat{g}_j \in L$, and
        \begin{gather*}
            Av_{\boldsymbol{w}/N}(A) = \frac{\sum_{i: \hat{g}_i \in A} w_i \hat{g}_i}{\sum_{i: \hat{g}_i \in A} w_i}.
        \end{gather*}
Denote
\[
  b^{(N)}_i \coloneqq \sqrt{\tfrac{N}{p_i(1-p_i)}}\Bigl(\tfrac{1}{2} - p_i\Bigr),
  \qquad
  b^{\infty}_i \coloneqq
  \begin{cases}
    0,       & \text{if } p_i = \tfrac{1}{2},\\[2pt]
    +\infty, & \text{if } p_i < \tfrac{1}{2},
  \end{cases}
\]
$b^{(N)}_i \to b^{\infty}_i$ as $N \to \infty$: the bound is identically zero for every $N$ when $p_i = 1/2$, and diverges to
$+\infty$ when $0 < p_i < 1/2$. Denote
\[
  D_N \coloneqq C_{\boldsymbol{p}} \cap B_{\boldsymbol{b}^{(N)}},
  \qquad
  D_\infty \coloneqq C_{\boldsymbol{p}} \cap B_{\boldsymbol{b}^{\infty}},
\]
and, for $\boldsymbol{y} \in \mathbb{R}^K$ and a weight vector $\boldsymbol{v} \in (0,\infty)^K$,
\[
  \Psi_N(\boldsymbol{y}, \boldsymbol{v}) \coloneqq P_{\boldsymbol{v}}\bigl( \boldsymbol{y} \,\big|\, D_N \bigr),
  \qquad
  \Psi_\infty(\boldsymbol{y}, \boldsymbol{v}) \coloneqq P_{\boldsymbol{v}}\bigl( \boldsymbol{y} \,\big|\, D_\infty \bigr).
\]
The projection does not change when all weights are multiplied by a common positive
constant, so $P_{\boldsymbol{w}} (\cdot|\cdot) = P_{\boldsymbol{w}/N}(\cdot | \cdot)$ and $\hat{\boldsymbol{g}}^{*} = \Psi_N(\hat{\boldsymbol{g}}, \boldsymbol{w}/N)$. Next we show that

\begin{align}\label{eq:aseq}
P\Bigl( \Psi_N\bigl(\hat{\boldsymbol{g}}, \boldsymbol{w}/N\bigr) = \Psi_\infty\bigl(\hat{\boldsymbol{g}}, \boldsymbol{w}/N\bigr) \Bigr) \to 1,\ N \to \infty,
\end{align}
which together with continuity of $\Psi_N(\cdot, \cdot)$ in both arguments for each $N$, and the continuous mapping theorem will imply:
\[
\Psi_N\bigl(\hat{\boldsymbol{g}}, \boldsymbol{w}/N\bigr) \overset{d}{\to} \Psi_{\infty}(\boldsymbol{U}, \boldsymbol{a}),\ N \to \infty.
\]
The event in \eqref{eq:aseq} is equivalent to
\[
\Psi_{\infty}(\hat{\boldsymbol{g}}, \boldsymbol{w}/N) \in D_N,
\]
because the minimizer over the superset remains that over a subset, hence \eqref{eq:aseq} is equivalent to
\begin{align}\label{eq:aseq2}
P\Bigl( \Psi_{\infty}(\hat{\boldsymbol{g}}, \boldsymbol{w}/N) \not \in D_N \Bigr) \to 0,\ N \to \infty.
\end{align}
The event in \eqref{eq:aseq2} is equivalent to the finite union of events:
\begin{align}\label{eq:aseq3}
\bigcup_{i \,:\, p_i < 1/2}
       \Bigl\{ \Psi_{\infty,i}\bigl(\hat{\boldsymbol{g}},\boldsymbol{w}/N\bigr) > b^{(N)}_i \Bigr\}.
\end{align}
Each event in the union above has the limiting probability 0 as $N \to \infty$ because $b^{(N)}_i \to \infty$ deterministically, whereas $\Psi_{\infty,i}\bigl(\hat{\boldsymbol{g}},\boldsymbol{w}/N\bigr) = O_p (1)$. To see the latter note that we have $\boldsymbol{0} \in D_\infty$, so $\boldsymbol{0}$ is
feasible in the minimization. Hence, for every $i \in [K]$,
\begin{align}\label{eq:aseq4}
  \Bigl(\min_{j \in [K]} v_j\Bigr) \bigl(z^{*}_i - y_i\bigr)^2
  \ \le\ \sum_{j=1}^{K} v_j \bigl(z^{*}_j - y_j\bigr)^2
  \ \le\ \sum_{j=1}^{K} v_j y_j^2
  \ \le\ K \Bigl(\max_{j \in [K]} v_j\Bigr) \|\boldsymbol{y}\|_\infty^2 ,
\end{align}
the second inequality follows by feasibility of $\boldsymbol{0}$. By the triangular inequality, using $|y_i| \le \|\boldsymbol{y}\|_\infty$, and rearranging and taking square
roots in \eqref{eq:aseq4}, for every $i \in [K]$,
\[
  \bigl| z^{*}_i \bigr|
  \ \le\ |y_i| + \bigl| z^{*}_i - y_i \bigr|
  \ \le\ \|\boldsymbol{y}\|_\infty \Biggl( 1 + \sqrt{\,K \frac{\max_{j} v_j}{\min_{j} v_j}\,} \Biggr) ,
\]
then taking the maximum over $i \in [K]$,
\begin{equation}\label{eq:Psibound}
  \bigl\| \Psi_\infty(\boldsymbol{y}, \boldsymbol{v}) \bigr\|_\infty
  \ \le\ \|\boldsymbol{y}\|_\infty \Biggl( 1 + \sqrt{\,K \frac{\max_{j} v_j}{\min_{j} v_j}\,} \Biggr)
  \qquad \text{for all } \boldsymbol{y} \in \mathbb{R}^K,\ \boldsymbol{v} \in (0,\infty)^K .
\end{equation}
Applying
\eqref{eq:Psibound} at $\boldsymbol{y} = \hat{\boldsymbol{g}}$ and $\boldsymbol{v} = \boldsymbol{w}/N$: since $n_j/N \to a_j \in (0,1)$ for every
$j$, the ratio $\max_j (n_j/N) / \min_j (n_j/N)$ converges to a finite limit and is therefore bounded, and
$\hat{\boldsymbol{g}} = O_p(1)$ by Corollary \ref{corol:as_normality}, so
\[
  \bigl\| \Psi_\infty\bigl(\hat{\boldsymbol{g}}, \boldsymbol{w}/N\bigr) \bigr\|_\infty = O_p(1),
\]
which implies that the limiting probability of the event in \eqref{eq:aseq3} is zero, therefore \eqref{eq:aseq} holds.

Since $(\hat{\boldsymbol{g}}, \boldsymbol{w} / N) \overset{d}{\to} (\boldsymbol{U}, \boldsymbol{a})$, $\Psi_{\infty}(\cdot, \cdot)$ is continuous, the continuous mapping theorem implies:
\[
\Psi_{\infty}(\hat{\boldsymbol{g}}, \boldsymbol{w} / N) \overset{d}{\to} \Psi_{\infty}(\boldsymbol{U}, \boldsymbol{a}),
\]
where $\boldsymbol{a} = (a_1, \dots, a_K)'$.
\eqref{eq:aseq} implies that the same holds for $\hat{\boldsymbol{g}}^* = \Psi_N (\hat{\boldsymbol{g}}, \boldsymbol{w} / N)$:
\[
\Psi_{N}(\hat{\boldsymbol{g}}, \boldsymbol{w} / N) \overset{d}{\to} \Psi_{\infty}(\boldsymbol{U}, \boldsymbol{a}).
\]

        Rewrite:
        \begin{align*}
            \Lambda = \sum_{i=1}^K n_i \frac{p_i(1-p_i)}{N} \left[ \frac{\hat{p}_i}{\alpha_i^2} + \frac{1-\hat{p}_i}{(1-\gamma_i)^2}\right] \left(\sqrt{\frac{N}{p_i(1-p_i)}}(\hat{p}_i^* - p_i) - \sqrt{\frac{N}{p_i(1-p_i)}}(\hat{p}_i - p_i)\right)^2,
        \end{align*}
        since $\left[ \frac{\hat{p}_i}{\alpha_i^2} + \frac{1-\hat{p}_i}{(1-\gamma_i)^2}\right] \overset{a.s.}{\to} \frac{1}{p_i(1-p_i)}, n_i/N \to a_i,\ \forall\ i=1, \dots, K$, Slutsky's theorem implies:
        \begin{align}\label{eq:as_distribution}
            \Lambda \overset{d}{\to} \sum_{i=1}^K a_i \left(\Psi_{\infty, i}(\boldsymbol{U}, \boldsymbol{a}) - U_i \right)^2 =: LD_{\boldsymbol{p}}.
        \end{align}
        Notice that for any $\boldsymbol{p}$ satisfying $\overline{H}_0$ we have $\{ C \cap B_{0} \} \subseteq \{ C_{\boldsymbol{p}} \cap B_{\boldsymbol{b}^{\infty}} \}$ and $\boldsymbol{p} = 1/2 \implies \{ C_{\boldsymbol{p}} \cap B_{\boldsymbol{b}^{\infty}} \} = \{ C \cap B_{0} \}$. Therefore, since the minimum of a function over a subset can only be weakly larger than that over a superset, we have:
        \begin{align*}
            \forall\ \omega: LD_{\boldsymbol p}(\omega)
        = \min_{\tilde{\boldsymbol y} \in C_{\boldsymbol{p}} \cap B_{\boldsymbol{b}^{\infty}}} \sum_{i=1}^K a_i \left(\tilde y_i - U_i(\omega)\right)^2
        \ \le\
        \min_{\tilde{\boldsymbol y} \in C \cap B_0} \sum_{i=1}^K a_i \left(\tilde y_i - U_i(\omega)\right)^2
        = LD_{1/2}(\omega)
        \end{align*}
        Further, following similar steps as in the proof of Theorem 2.3.1 in \cite{robertson1988order} one can show that the distribution of
        \begin{align*}
            LD_{1/2}:= \sum_{i=1}^K a_i \left(P_{\boldsymbol{a}} \left(\boldsymbol{U}\ \bigg|\ C\ \bigcap\ B_0 \right)_i - U_i \right)^2
        \end{align*}
        is equal to the distribution of the mixture:
        \begin{align*}
            \chi^2_{K-\ell} + \sum_{i=1}^{\ell} (\max\{0, \mathcal{N}(0, 1)\})^2,\ \text{w.p. } P(\ell; K; \boldsymbol{a}),\ \ell = 1, \dots, K,
        \end{align*}
        where $\chi^2_0 \equiv 0$, $\sum_{i=1}^{\ell} (\max\{0, \mathcal{N}(0, 1)\})^2$ denotes the sum of independent truncated squared standard normal random variables, and $P(\ell;K;\boldsymbol{a})$ are weights defined in Chapter 2 in \cite{robertson1988order}. That is
        \begin{align} \label{eq:limit_distribution}
            LD_{1/2} \sim \sum_{\ell=1}^K P(\ell; K;\boldsymbol{a}) \left(\chi^2_{K-\ell} + \sum_{i=1}^{\ell} (\max\{0, \mathcal{N}(0, 1)\})^2 \right)
        \end{align}
        This mixture is absolutely continuous for all $t > 0$ (and has an atom at $t=0$), hence for $\boldsymbol{p}=1/2:$
        \begin{align*}
            \lim_{N \to \infty} P(\Lambda \ge t) = P(LD_{1/2} \ge t),\ \forall\ t > 0.
        \end{align*}
        Hence, for any $\boldsymbol{p}$ that satisfies the null hypothesis,

        \begin{align*}
    \limsup_{N \to \infty} P_{\boldsymbol{p}}(\Lambda \ge t)
    \le P(LD_{\boldsymbol{p}} \ge t)
    \le P(LD_{1/2} \ge t)
    = \lim_{N \to \infty} P_{1/2}(\Lambda \ge t),
    \end{align*}

    where the first inequality is the Portmanteau theorem applied to the closed set $[t,\infty)$, using $\Lambda \overset{d}{\to} LD_{\boldsymbol{p}}$ from \eqref{eq:as_distribution}; the second follows from the pointwise domination $LD_{\boldsymbol{p}} \le LD_{1/2}$ shown above; and the equality holds because $t>0$ is a continuity point of the distribution of $LD_{1/2}$, so $\Lambda \overset{d}{\to} LD_{1/2}$ when $\boldsymbol{p}=1/2$ gives convergence of $P(\Lambda \ge t)$ to $P(LD_{1/2}\ge t)$ directly.
    \end{proof}

\end{proof}

The asymptotic framework considered in Theorem \ref{th:asympt_validity} reflects a finite sample situation where all pairs of teams in $E$ interact many times, and the number of games for each interacting pair is relatively balanced. In practice, for example for the tournament graph illustrated in Figure \ref{fig:tournament_pvalue}, we expect the approximation to be accurate when the numbers of games are 20 and 50, and considerably less accurate when they are 20 and 20,000. Intuitively, in case of imbalances, we expect the least favorable distribution to deviate from $\tilde{p}$, leading the test based on $\tilde{\pi}_\lambda$ to reject the null hypothesis more frequently than the nominal level.

Even when the numbers of games are small and somewhat imbalanced, the excess type-I error that $\phi^{AS}_{\alpha}(\cdot)$ incurs over $\alpha$, because $\tilde{p}$ is not the least favorable distribution, need not be large. The following example quantifies it for the tournament graph in Figure \ref{fig:tournament_pvalue}.

\addtocounter{example}{-1}
\begin{example}[Continued]
    In this example, although $\tilde{p}$ is not the least favorable distribution, direct computation gives
    \begin{align*}
        \underset{\alpha \in (0, 1)}{\sup}\ \underset{P \in \boldsymbol{P}_0(\boldsymbol{r})}{\sup}\ \left( E_P[\phi^{AS}_{\alpha}(\boldsymbol{Y}_{\vn})] - \alpha \right) \approx 0.0195,
    \end{align*}
    which is achieved at $\alpha \approx 0.061$ and at $p_1 = p_2 \approx 0.118$.
\end{example}

\begin{remark}
    {\normalfont (Asymptotic Distribution of $\Lambda$)}
    Under $\tilde{p}$, the asymptotic distribution of $\Lambda$ is presented in the proof of Theorem \ref{th:asympt_validity} that can be derived following the same steps as in Theorem 2.3.1 in \cite{robertson1988order}. However, the direct usage of the asymptotic distribution requires the computation of mixing weights $P(\ell;K; \boldsymbol{w})$ whose expressions are unavailable in closed form (see Section 2.4 in \cite{robertson1988order} for more discussion).
\end{remark}

We now state the consistency result. Since the test is formulated for the closed null $\overline{H}_0$, consistency is established against fixed alternatives whose probability vectors lie outside this set. That is, we consider distributions for which at least one of the weak inequalities in \eqref{eq:H0c} is strictly violated. The relationship between this class of alternatives and the hypothesis of interest $H_0$ depends on the structure of the tournament graph, as discussed below.

\begin{theorem}\label{th:consistency}
    {\normalfont (Test Consistency)}
    Let Assumption \ref{as:dgp} hold with $F_0 \in \mathbf{F}$. The test $\phi^{AS}_\alpha (\cdot)$ is consistent against alternatives that violate the closed null:
    \begin{align*}
        \liminf_{N \to \infty} E_P[\phi^{AS}_\alpha (\boldsymbol{Y}_{\vn})] = 1,\ \forall\ P \in \boldsymbol{P}_1(\boldsymbol{r})
    \end{align*}
    where $N = \sum_{(\ell, k) \in E } n_{\ell, k}$, and $\frac{n_{\ell, k}}{N} \to a_{\ell, k} \in (0, 1),\ \forall\ (\ell, k) \in E.$
\end{theorem}
\begin{proof}
    To prove consistency of the test, we will first show that there exists a finite number $c$ such that
    \begin{align*}
        \limsup_{N \to \infty} P\{ \Lambda(\hat{\boldsymbol{p}}) \ge c \} \leq \alpha,\ \forall\ P \in \boldsymbol{P}_0(\boldsymbol{r}).
    \end{align*}
    Then, we will show that $\Lambda(\hat{\boldsymbol{p}}) \to \infty$ as $N \to \infty$, for any $P \in \boldsymbol{P}_1(\boldsymbol{r})$. This implies
    \begin{align*}
        \lim_{N \to \infty} P\{ \Lambda(\hat{\boldsymbol{p}}) \ge c \} = 1,\ \forall\ P \in \boldsymbol{P}_1(\boldsymbol{r})
    \end{align*}
    and hence proves the theorem.

    In Theorem \ref{th:asympt_validity}, we derived the limiting distribution for the statistic, in the least favorable case. Fix $c$ as the $1-\alpha$ quantile of the distribution described in Equation \ref{eq:limit_distribution}, and note that it satisfies
    \begin{align*}
        \limsup_{N \to \infty} P\{ \Lambda(\hat{\boldsymbol{p}}) \ge c \} \leq \alpha,\ \forall\ P \in \boldsymbol{P}_0(\boldsymbol{r}).
    \end{align*}
    Then, note that under any alternative hypothesis $P \in \boldsymbol{P}_1(\boldsymbol{r})$ there exists at least one pair of $\hat{p}_{\ell, k}$ and $\hat{p}^*_{\ell, k}$ such that $\hat{p}_{\ell, k} \to^p p_{\ell,k}$, $\hat{p}^*_{\ell, k} \to^p p^*_{\ell,k}$ and $p_{\ell,k} \neq p^*_{\ell,k}$. Let $\hat{p}_{i,j}$ and $\hat{p}^*_{i,j}$ be that pair, and note that $\Lambda(\hat{\boldsymbol{p}}) \geq n_{i,j} \left( \hat{p}_{i,j} \ln \bigg( \frac{\hat{p}_{i,j}}{\hat{p}_{i,j}^*} \bigg) + (1 - \hat{p}_{i,j}) \ln \bigg( \frac{1 - \hat{p}_{i,j}}{1 - \hat{p}_{i,j}^*} \bigg) \right)$. Slutsky's theorem guarantees that, as $N \to \infty$, $\left( \hat{p}_{i,j} \ln \bigg( \frac{\hat{p}_{i,j}}{\hat{p}_{i,j}^*} \bigg) + (1 - \hat{p}_{i,j}) \ln \bigg( \frac{1 - \hat{p}_{i,j}}{1 - \hat{p}_{i,j}^*} \bigg) \right) \to^p \left( p_{i,j} \ln \bigg( \frac{p_{i,j}}{p_{i,j}^*} \bigg) + (1 - p_{i,j}) \ln \bigg( \frac{1 - p_{i,j}}{1 - p_{i,j}^*} \bigg) \right) > 0$. This implies that,
    \begin{gather*}
        \forall\ \epsilon>0,\ \exists\ N_0>0:\ \forall\ N \ge N_0:\ P\left(n_{i,j} \left( \hat{p}_{i,j} \ln \bigg( \frac{\hat{p}_{i,j}}{\hat{p}_{i,j}^*} \bigg) + (1 - \hat{p}_{i,j}) \ln \bigg( \frac{1 - \hat{p}_{i,j}}{1 - \hat{p}_{i,j}^*} \bigg) \right) > c \right) > 1 - \epsilon,
    \end{gather*}
    and hence
    \begin{align*}
        \forall\ \epsilon > 0,\ \exists\ N_0:\ \forall\ N \ge N_0:\ \inf_{m \ge N} P(\Lambda(\hat{\boldsymbol{p}}) \ge c) > 1 - \epsilon \implies \liminf_{N \to \infty} P(\Lambda(\hat{\boldsymbol{p}}) \ge c) = 1,\ \forall\ P \in \boldsymbol{P}_1(\boldsymbol{r}),
    \end{align*}
    and proves consistency of $\phi^{AS}_\alpha(\cdot)$.
\end{proof}

Theorem \ref{th:consistency} establishes consistency against the distributions in $\boldsymbol{P}_1(\boldsymbol{r})$, which violate the closed null, rather than against every distribution for which $\boldsymbol{r}\notin R(G,\mathbf{F},P)$. Proposition \ref{prop:closed_null} clarifies when the two classes differ. When the tournament graph has diameter at most two and the latent merits are distinct, every candidate ranking without ties other than the true one violates at least one of the weak restrictions in \eqref{eq:H0c}, so the theorem delivers consistency against all of them. When the diameter is greater than two, a candidate ranking may fail to belong to the identified set while its probability vector still satisfies every weak restriction. Such a distribution is an alternative to the hypothesis of interest, but it belongs to $\boldsymbol{P}_0(\boldsymbol{r})$, the set over which Theorem \ref{th:asympt_validity} controls size. Its rejection probability is therefore asymptotically bounded by $\alpha$, and no test constructed on the closed null can be consistent against it. Intuitively, a ranking that fails the strict characterization only because one or more restrictions hold with equality is indistinguishable from a boundary point of $\overline{H}_0$.

The set of such alternatives is small. A candidate ranking outside the identified set can satisfy the closed null only if at least one of the restrictions in \eqref{eq:H0c} holds with equality, and each restriction binds only on a hyperplane in the space of win probabilities: either a single probability equals one half, or two probabilities coincide. Since the restrictions are finite in number, the probability vectors at which the test loses power are contained in a finite union of hyperplanes, and hence form a Lebesgue-null subset of the win probabilities compatible with the tournament graph. Almost any perturbation of the win probabilities removes the equality and places the distribution either inside the hypothesis of interest or among the alternatives against which the test is consistent.

\section{Additional Monte Carlo Simulations}
\label{appendix:mc}

This Appendix provides additional details and results for the Monte Carlo analysis in Section \ref{sec:mc_simulations}. Section \ref{sec:mc_details} describes the procedures used to sample winning-probability vectors from the null polytope and to select representative probability vectors for the fixed DGP simulations. Section \ref{sec:mc_design3_additional_fig} reports additional graphical results for Design 3. Finally, Section \ref{sec:mc_additional_designs} presents two additional simulation designs that examine the finite sample size of the approximate test under different combinations of tournament size and graph density.

\subsection{Implementation of the Simulation Procedure}
\label{sec:mc_details}

\subsubsection{Sampling Winning-Probability Vectors}

For a tournament graph with edge set $E$, the data generating process is characterized by one winning probability for each interaction. The null hypothesis therefore defines a polytope in the $|E|$-dimensional space of winning-probability vectors. Our objective is to sample vectors from this polytope while obtaining reasonably broad coverage.

A standard method for sampling from a high-dimensional polytope is the hit-and-run algorithm. Starting from an interior point, the algorithm selects a random direction, determines the segment of the corresponding line that lies within the polytope, and draws the next point uniformly from that segment. Under regularity conditions, the resulting Markov chain converges to the uniform distribution over the polytope.

In finite simulations, however, consecutive hit-and-run draws may be strongly dependent, and the algorithm may explore elongated or irregularly shaped regions slowly. Moreover, a sample that is approximately uniform with respect to the volume of a high-dimensional set need not provide broad coverage of its boundary or of other low-volume regions. We therefore use a hybrid modification of the hit-and-run algorithm. In addition to the interior points generated by the Markov chain, we construct candidate points on faces of the feasible set by optimizing random linear objective functions and move some of these points toward the interior. We then select the final winning-probability vectors using a maximin-distance criterion, which discourages near-duplicate points and promotes broader coverage of the feasible parameter space.

For larger tournaments, we supplement this procedure with projection-stratified sampling. We repeatedly select pairs of probability coordinates and sample points from the corresponding two-dimensional projections of the null polytope. We then complete the remaining coordinates subject to the requirement that the full probability vector belongs to the null polytope. To assess the coverage of the resulting sample, we report selected two-dimensional projections of the null polytope together with the sampled winning-probability vectors.

\subsubsection{Selecting Representative Probability Vectors}

Cluster representatives are selected from the simulated winning-probability vectors for which the approximate test rejects. For each rejected vector, we identify its $15$ nearest neighbors in the full probability-vector space and compute the fraction of these neighbors for which the test also rejects. We then select the rejected vectors with the highest local rejection fractions, retaining either five or ten representatives depending on the design. These representatives are used as fixed winning-probability vectors in the subsequent Monte Carlo simulations. Their two-dimensional projections are also marked in the figures to show where the regions with relatively frequent rejections are located.

\subsection{Additional Graphical Results for Design 3}
\label{sec:mc_design3_additional_fig}

\begin{figure}[H]
    \centering
    \includegraphics[scale=0.35]{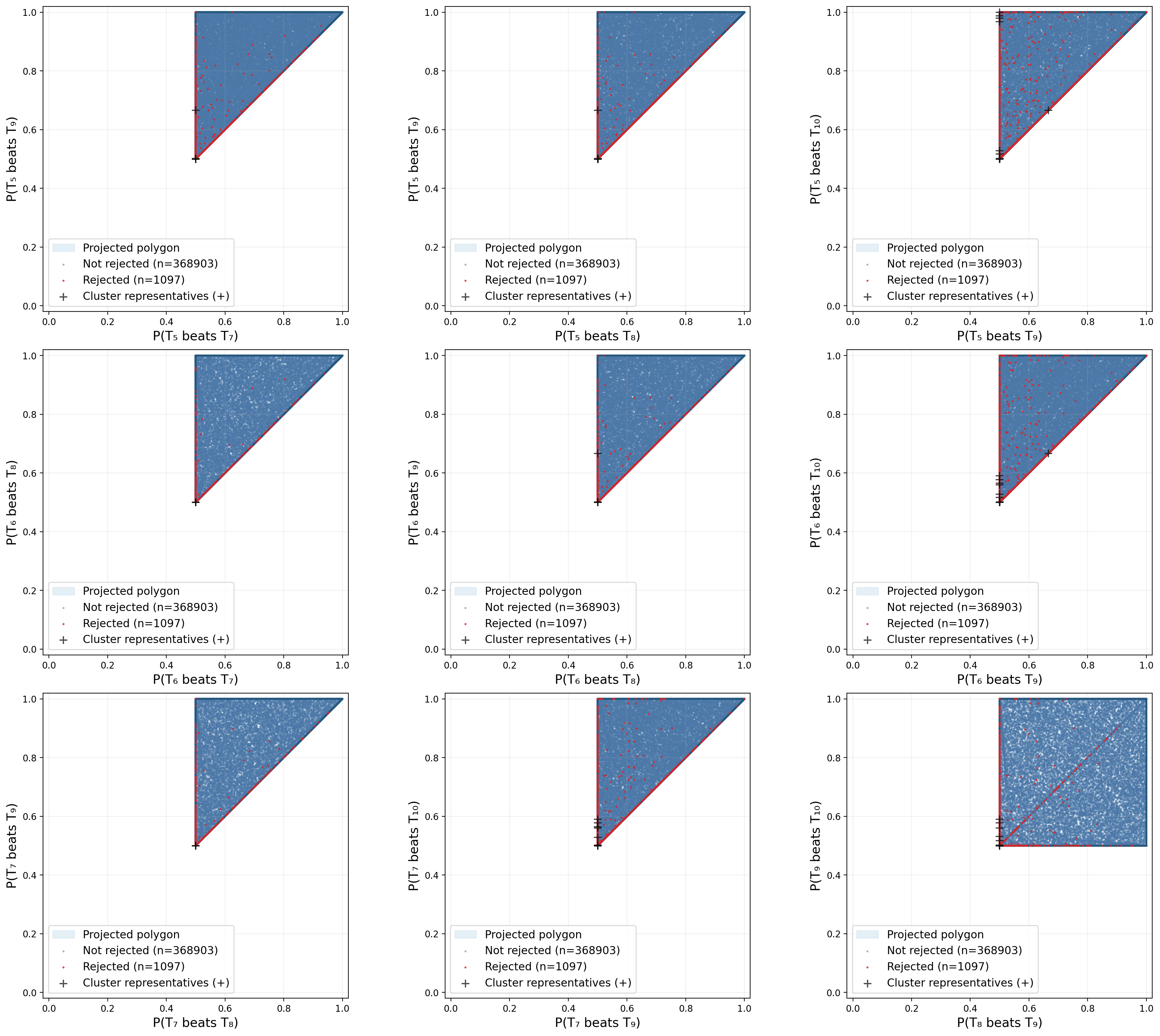}
    \caption{Selected two-dimensional projections of the null polytope in Design 3. Blue points correspond to non-rejections, red points to rejections, and ``+'' symbols to the selected representative probability vectors.}
    \label{fig:2dnullproj_des3_1}
\end{figure}

\begin{figure}[H]
    \centering
    \includegraphics[scale=0.35]{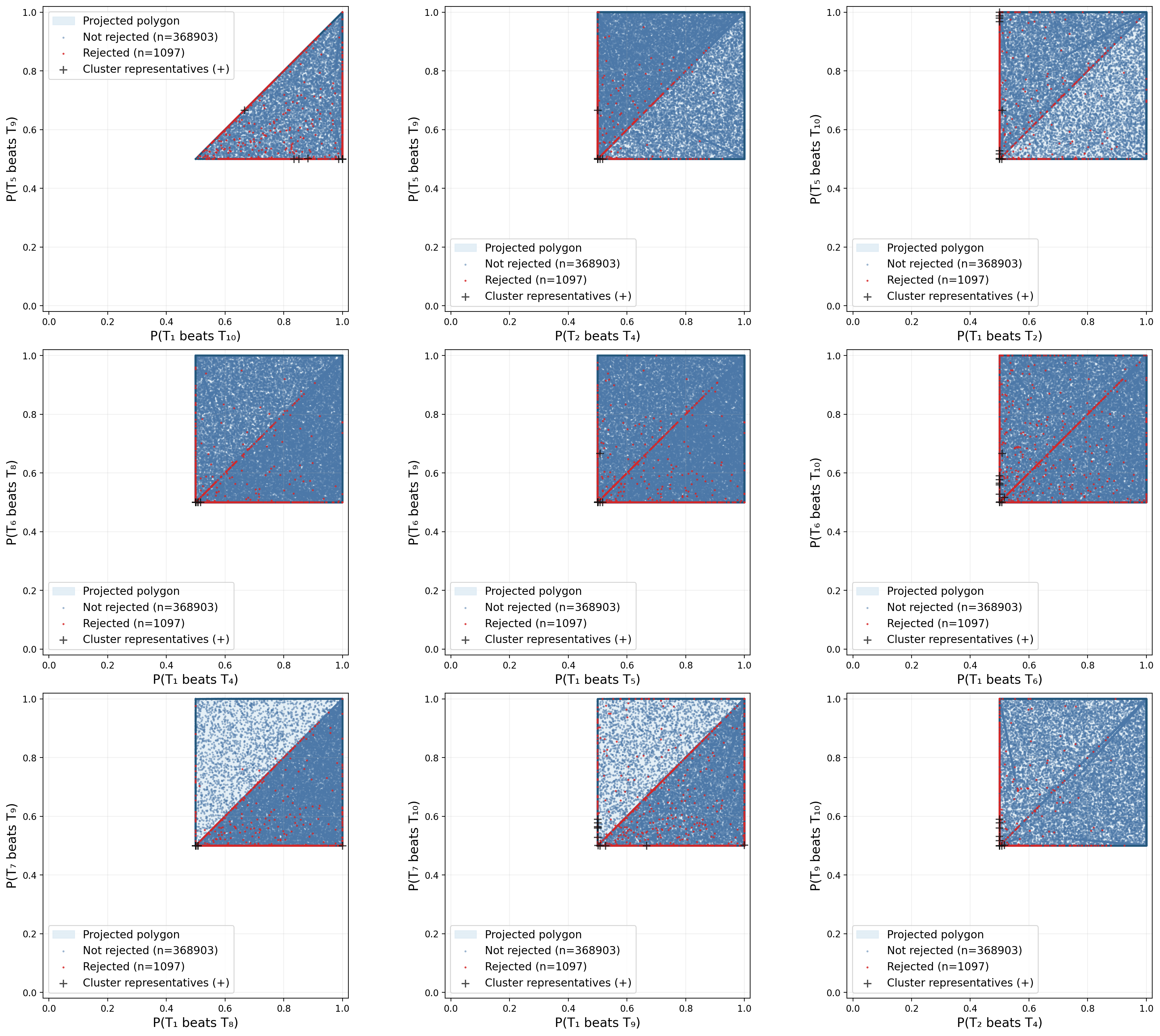}
    \caption{Additional two-dimensional projections of the null polytope in Design 3. Blue points correspond to non-rejections, red points to rejections, and ``+'' symbols to the selected representative probability vectors.}
    \label{fig:2dnullproj_des3_2}
\end{figure}

We report additional graphical results for Design 3. Figures \ref{fig:2dnullproj_des3_1} and \ref{fig:2dnullproj_des3_2} show selected two-dimensional projections of the $41$-dimensional null polytope. Each figure overlays the winning-probability vectors drawn in the exploratory exercise with the corresponding outcomes of the approximate test. Blue points indicate draws for which the test does not reject, red points indicate draws for which it rejects, and ``+'' symbols mark the representative probability vectors selected for the fixed-DGP Monte Carlo simulations.

\subsection{Additional Simulation Designs}
\label{sec:mc_additional_designs}

\subsubsection{Design 4: Small Complete Tournament}
\label{sec:mc_design4}

\begin{figure}[H]
    \centering
    \includegraphics[scale=0.5]{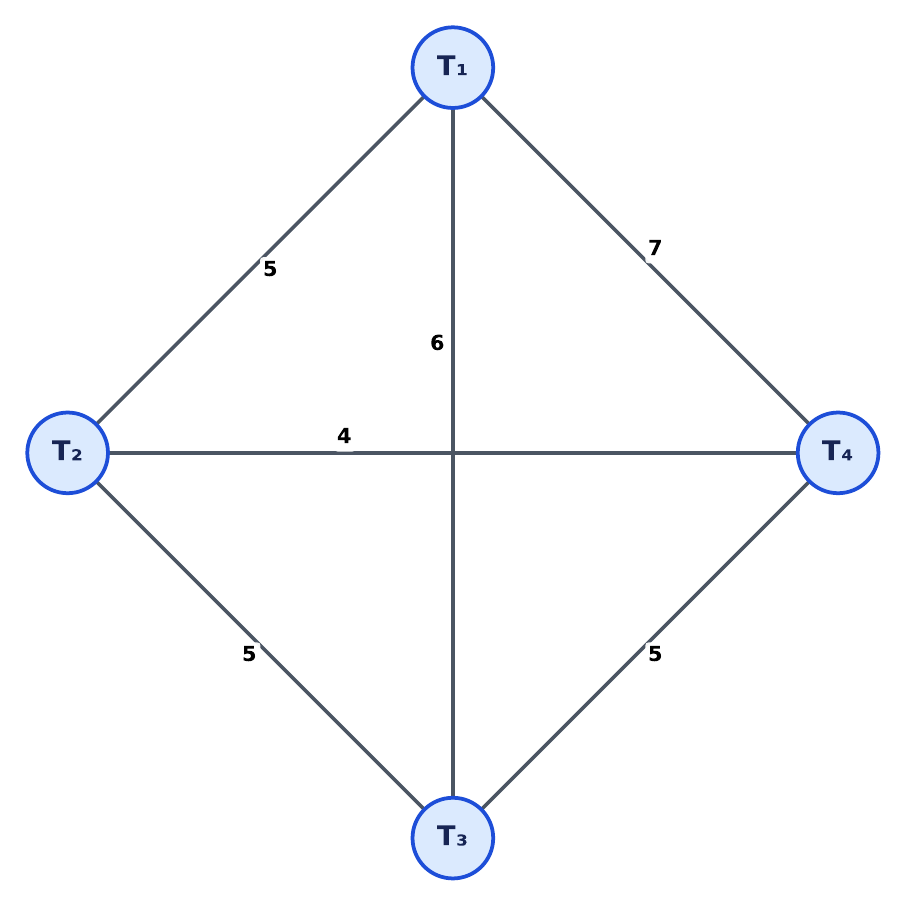}
    \caption{Tournament graph and number of games on each edge in Design 4.}
    \label{fig:tourn_graph_des4}
\end{figure}

Design 4 considers a small tournament with four teams and a complete tournament graph. Figure \ref{fig:tourn_graph_des4} reports the number of games played on each edge.

We test the null hypothesis that the ranking $\boldsymbol{r}=(1,2,3,4)'$ belongs to the identified set. Because the graph contains six interacting pairs, the null hypothesis defines a polytope in the six-dimensional space of winning-probability vectors.

We first draw $24{,}000$ winning-probability vectors from the null polytope. For each vector, we generate one sample and apply the approximate test at the nominal significance level $\alpha=0.1$. The test rejects for $305$ of the $24{,}000$ draws.

\begin{figure}[H]
    \centering
    \includegraphics[scale=0.35]{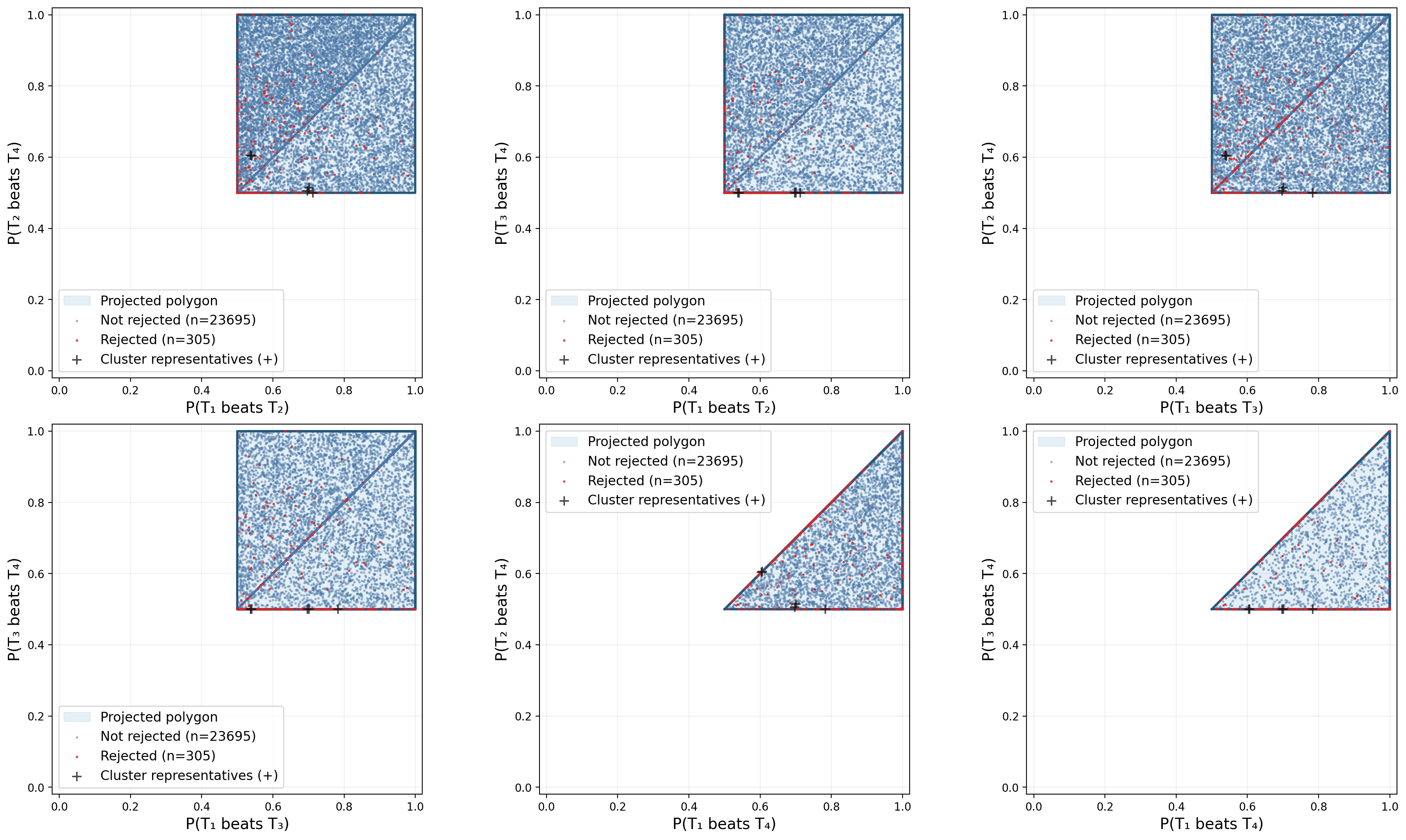}
    \caption{Selected two-dimensional projections of the null polytope in Design 4. Blue points correspond to non-rejections, red points to rejections, and ``+'' symbols to the selected representative probability vectors.}
    \label{fig:2dnullproj_des4}
\end{figure}

Figure \ref{fig:2dnullproj_des4} reports six of the $\binom{6}{2}=15$ possible two-dimensional projections of the null polytope. Blue points correspond to draws for which the test does not reject, red points to draws for which it rejects, and ``+'' symbols to the five selected representative probability vectors.

For each representative probability vector, we hold the data generating process fixed, generate $M=10{,}000$ independent samples, and apply the approximate test at the nominal significance level $\alpha=0.1$. Table \ref{tab:cluster-representative-rejections_des4} reports the resulting empirical rejection frequencies. In all five cases, the empirical rejection frequency is below the nominal significance level.

\begin{table}[ht]
    \centering
    \begin{tabular}{cc}
        Cluster & Empirical rejection \\
        \hline
        1 & 0.0536 \\
        2 & 0.0551 \\
        3 & 0.0398 \\
        4 & 0.0376 \\
        5 & 0.0331 \\
        \hline
    \end{tabular}
    \caption{Empirical rejection frequencies at the selected probability vectors in Design 4. The nominal significance level is $\alpha=0.1$, and each result is based on $M=10{,}000$ Monte Carlo replications.}
    \label{tab:cluster-representative-rejections_des4}
\end{table}

\subsubsection{Design 5: Larger Sparse Tournament}
\label{sec:mc_design5}

\begin{figure}[H]
    \centering
    \includegraphics[scale=0.5]{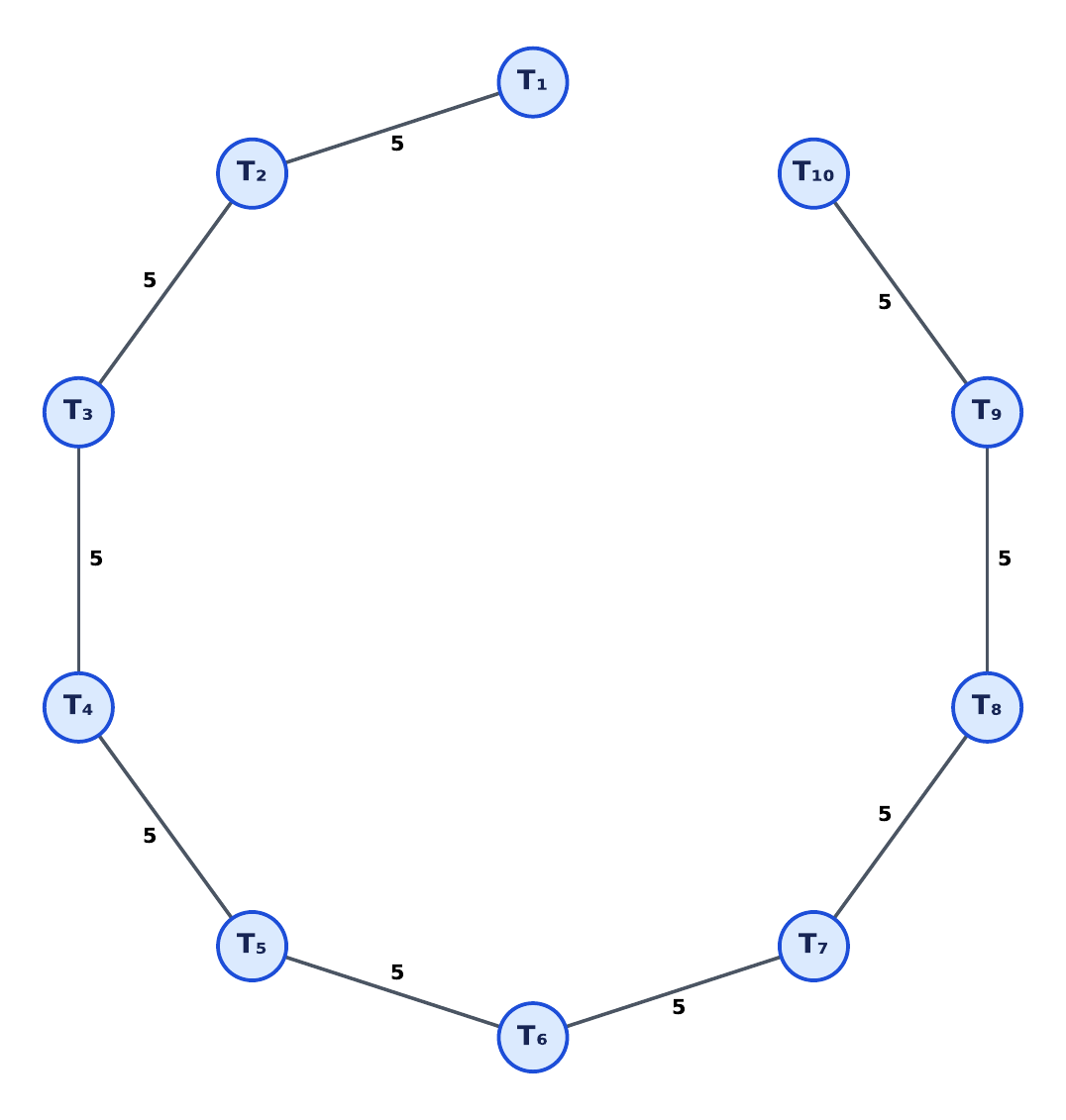}
    \caption{Tournament graph and number of games on each edge in Design 5.}
    \label{fig:tourn_graph_des5}
\end{figure}

Design 5 considers ten teams connected by a sparse, minimally connected tournament graph. Figure \ref{fig:tourn_graph_des5} reports the number of games played on each edge.

We test the null hypothesis that the ranking $\boldsymbol{r}=(1,2,\ldots,10)'$ belongs to the identified set. Because the graph contains nine interacting pairs, the null hypothesis defines a polytope in the nine-dimensional space of winning-probability vectors.

We first draw $100{,}000$ winning-probability vectors from the null polytope. For each vector, we generate one sample and apply the approximate test at the nominal significance level $\alpha=0.1$. The test rejects for $994$ of the $100{,}000$ draws.

\begin{figure}[H]
    \centering
    \includegraphics[scale=0.35]{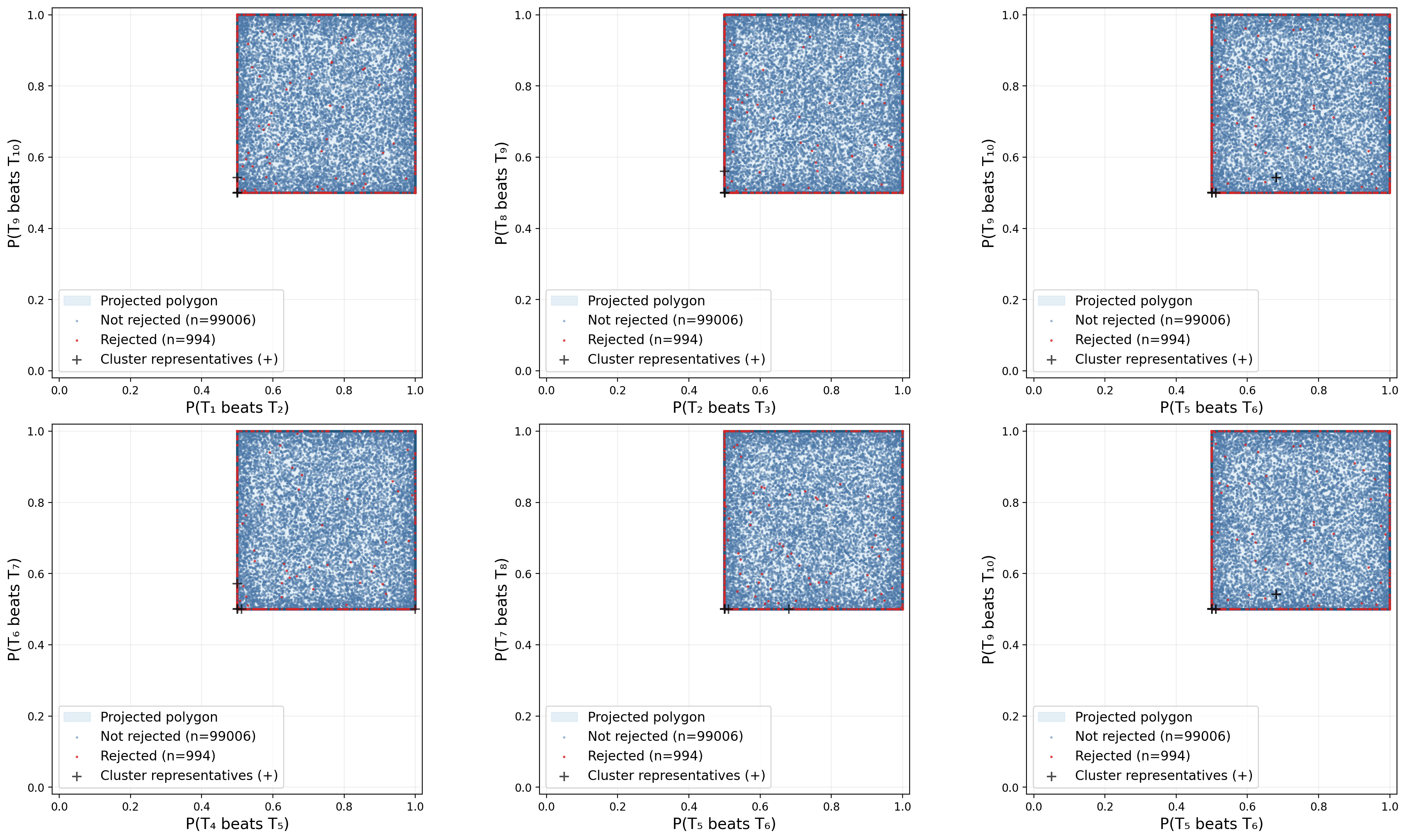}
    \caption{Selected two-dimensional projections of the null polytope in Design 5. Blue points correspond to non-rejections, red points to rejections, and ``+'' symbols to the selected representative probability vectors.}
    \label{fig:2dnullproj_des5}
\end{figure}

Figure \ref{fig:2dnullproj_des5} reports six of the $\binom{9}{2}=36$ possible two-dimensional projections of the null polytope. Blue points correspond to draws for which the test does not reject, red points to draws for which it rejects, and ``+'' symbols to the five selected representative probability vectors.

Following the same procedure as in Design 2, we hold each representative probability vector fixed, generate $M=10{,}000$ independent samples, and apply the approximate test at $\alpha=0.1$. Table \ref{tab:cluster-representative-rejections_des5} reports the empirical rejection frequencies, which do not exceed the nominal level at any of the five vectors.

\begin{table}[ht]
    \centering
    \begin{tabular}{cc}
        Cluster & Empirical rejection \\
        \hline
        1 & 0.0916 \\
        2 & 0.0989 \\
        3 & 0.0900 \\
        4 & 0.0670 \\
        5 & 0.0188 \\
        \hline
    \end{tabular}
    \caption{Empirical rejection frequencies at the selected probability vectors in Design 5. The nominal significance level is $\alpha=0.1$, and each result is based on $M=10{,}000$ Monte Carlo replications.}
    \label{tab:cluster-representative-rejections_des5}
\end{table}

\section{Robustness of the Empirical Illustration}
\label{appendix:empirical_robustness}

This appendix examines three possible explanations for the rejection in the main empirical illustration. First, we consider whether the result is driven by using the same observations to construct and test the candidate ranking or by the particular PageRank specification. Second, we examine whether pooling workers with different firm rankings can account for the rejection by conducting the analysis separately by gender. Third, we assess sensitivity to the set of firms by considering both a disjoint set of ten firms and a substantially larger transition network. Unless otherwise noted, all tests use $20{,}000$ Monte Carlo draws to approximate the reference distribution under the null.

\subsection{Robustness to the Construction of the Candidate Ranking}

\paragraph{Sample splitting.}
The main analysis uses the same transition data to construct the PageRank ordering and test whether it belongs to the identified set. Although this mirrors the common empirical practice of treating the estimated PageRank ordering as the ranking used in subsequent analysis, its data-dependent selection could in principle contribute to the rejection. To separate ranking construction from evaluation, we randomly divide the transition observations into two nonoverlapping subsamples. We compute the PageRank ordering using only the first subsample and test whether it belongs to the identified set using only the second. The null hypothesis is again rejected, with an approximate $p$-value of $0.001$. Thus, the rejection persists when the candidate ranking is constructed from observations that are not used to evaluate it.

\paragraph{Alternative PageRank specification.}
We next consider the damped PageRank specification of \citet{corradini2023collective} in place of the undamped procedure of \citet{sorkin2018ranking}. For the pooled transitions among these ten firms, the two procedures return the same ordering. The candidate ranking evaluated by the test is therefore unchanged, and the damped PageRank ordering is rejected with the same approximate $p$-value of $0.000$. The rejection does not depend on the choice of damping.

\subsection{Worker Heterogeneity: Gender-Specific Rankings} \label{sec:gender_rankings}

One possible explanation for the pooled rejection is that the transition matrix combines groups of workers with different firm rankings. Applied work documents gender differences in job ladders and firm rankings \citep{corradini2023collective,morchio2026gender}, and aggregating groups whose pairwise probabilities are individually consistent with different rankings need not preserve strong stochastic transitivity.

We examine this explanation by retaining the same ten firms but conducting the ranking and testing procedures separately by gender. We construct the PageRank ordering using women's transitions and test it against their transition matrix. We then repeat the same procedure using men's transitions.

The test rejects the ranking constructed from women's transitions with an approximate $p$-value of $0.019$. It also rejects the ranking constructed from men's transitions with an approximate $p$-value of $0.000$. Pooling transitions across genders may contribute to the patterns observed in the full sample, but it cannot by itself explain the rejection of the PageRank ordering.

\subsection{Alternative Sets of Firms}

\paragraph{A disjoint set of firms.}
We first repeat the original firm-selection procedure after excluding the ten firms used in the main illustration. We begin with a highly connected clique of five firms and then add firms sequentially, at each step selecting a firm with many connections to those already included, until the set contains ten firms. This yields a disjoint set of highly connected firms. We construct the PageRank ordering for this alternative set and test whether it belongs to the corresponding identified set. The null hypothesis is again rejected, with an approximate $p$-value of $0.001$. The main result is therefore not specific to the identities of the ten firms selected for the primary illustration.

\paragraph{A larger transition network.}
We next consider a substantially broader firm network. Starting from the ten core firms used in the main analysis, we add every firm that has at least ten transitions, counting moves in both directions, with at least one core firm. This procedure produces a set of 119 firms. We then construct a directed transition matrix using all employer-to-employer moves among these firms.

To compute undamped PageRank, we restrict the directed graph to its largest strongly connected component. A strongly connected component is a set of firms such that each firm can be reached from every other firm through a directed sequence of employer-to-employer transitions. This restriction removes eight peripheral firms and leaves a network of 111 firms.

We compute the PageRank ordering using all transitions within this component and test whether the resulting ordering belongs to the identified set. Because of the larger network, we approximate the null distribution using $2{,}000$ Monte Carlo draws. The null hypothesis is again rejected, with an approximate $p$-value of $0.000$. The rejection therefore persists in a much larger firm network and is not an artifact of restricting the analysis to a small, densely connected set of firms.

\end{document}